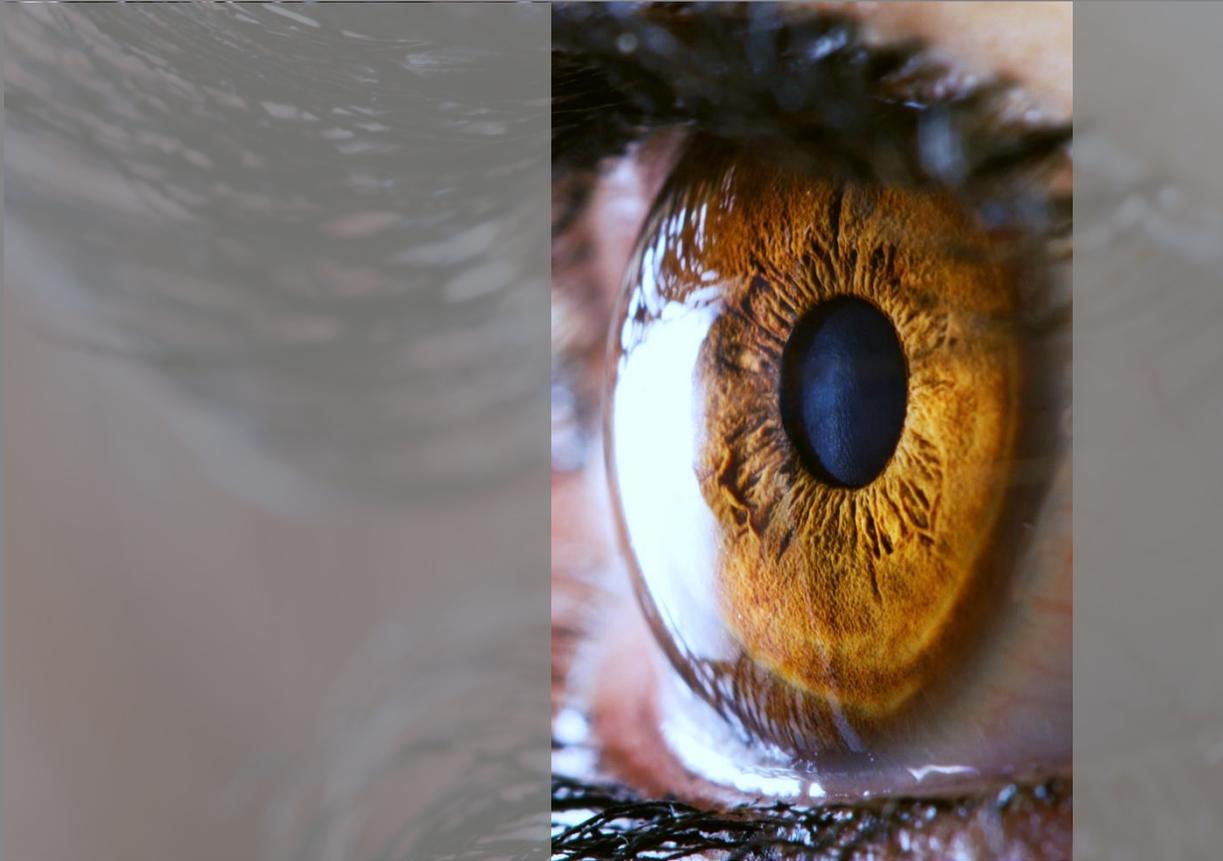

# Assessment of Biomechanical Properties for Corneal Post Refractive Surgery

by

Hassan Mostafa Ahmed Hassan Fahmy Gbr

PhD. Biomedical Engineering (2022)

<u>**Under supervision of**</u>

Prof. Nancy Mustafa Ahmed Salem

Helwan University

Prof. Walid I. Al-Atabany

Helwan University

2022

## Disclaimer

The information contained within this eBook is strictly for research purposes. If you wish to apply ideas contained in this eBook, you are taking full responsibility for your actions. The author has made every effort to ensure the accuracy of the information within this book was correct at time of publication. The author does not assume and hereby disclaims any liability to any party for any loss, damage, or disruption caused by errors or omissions, whether such errors or omissions result from accident, negligence, or any other cause.

No part of this eBook may be reproduced or transmitted in any form or by any means, electronic or mechanical, recording or by any information storage and retrieval system, without written permission from the author.

## Copyright

# Acknowledgement

In the beginning of my acknowledgment, I would like to dedicate this research concluded from this PhD, every research I have concluded during my life and any upcoming research that will be concluded to the man who devoted himself, his life and his wealth so that I reach this day and stand in front you; my father and his soul; I would not have accomplished any of these researches. For my mother and her endless support to me, I dedicate this research to you.

For my supervisors, prof. Nancy and prof. Walid, I thank you for your guidance through the past four years to achieve better quality research. For prof. Nancy whom supported me since my master degree, I cannot find any words describing my overwhelming gratitude.

Finally, I present this thesis as a first stone in my intellectual heritage regarding ophthalmic domain that I aim to build up during my long journey.

# Publications

# Abstract

A stable shape for corneas experiencing refractive surgery has to be sustained so as to elude post-refractive surgery de-compensation. This de-compensation leads to visual complications and unsatisfactory procedure recovery. Variation in corneal lamellae and collagen fibres is induced by recent LASER refractive surgical procedures utilizing LASER ablation and disruption techniques. Conserving a steady response of central apex flattening and peripheral steepening in an elastic cornea pre- and post- procedure is the ultimate purpose of successful refractive surgery. Early diagnosis of ectatic corneal disorders and better understanding of corneal pathogenesis is achieved by assessment of corneal biomechanical properties.

The ultimate objective of this research is to estimate the biomechanical properties for both normal and pathogenic corneal tissue pre- and post-operative refractive surgery. This achieved using ultrasonic acoustic radiation force impulse as a non-invasive method accounting for its high localization. Induced displacement tracking methods will be utilized for assessment of soft tissue biomechanical properties related to the investigated soft tissue. Ultrasound probe simulations will be carried out to optimize the probe design. FEM simulations will take place to precisely estimate in-situ corneal tissue biomechanics.

In this research, corneal biomechanical properties are studied and estimated using acoustic radiation force impulse. This is achieved either by estimating the focal peak axial deformation value or by estimating the shear wave speed for the resulting propagating deformation wave. This is accomplished by means of three steps. Firstly, a mathematical model is derived based on standard linear solid model to study the viscoelastic behavior of corneal tissue in response to an acoustic radiation force in either loading situation of constant or transient loading situation.

Secondly, based on the mathematical model outcomes a 3D FEM simulation is performed to study the corneal tissue in 3D and in response to an acoustic radiation force impulse. FEM simulations are carried out for two major case studies, the pre-refractive and the post-refractive case study. The corresponding corneal biomechanical properties for each case study from these two major case studies are obtained from relevant experimental research studies. The availability to estimate corneal biomechanics using estimated shear wave speed resulting from application of acoustic radiation force impulse is investigated by eleven different elastic moduli 3D FEMs. FEMs' biomechanical properties are chosen to cover a wide range of human corneal biomechanics in both normal and pathogenic states. For simulating normal post-refractive surgery cornea, six different FEMs are investigated with elastic moduli of 3, 30, 140, 300, 600 and 800 KPa respectively. For simulating the pathogenic pre-refractive surgery cornea, five different FEMs are investigated with elastic moduli of 1, 1.5, 2, 2.5 and 3 MPa respectively. At this stage of research, two ultrasonic tissue deformation estimation methods are utilized, the radial shear wave speed (rSWS) method and the focal peak axial deformation method (FPAD). For rSWS method, two B-mode frame rates are used, 10 KHz and 100 KHz, while for the FPAD method only 100 KHz frame rate is used. For FPAD method, two mathematical curve fitting formulae are used in order to include all non-involved elastic moduli values in this experiment.

Thirdly, ultrasound transducer model in conjunction with two other models, namely, the FEM and the scatterrers model, is used to study the behavior of corneal tissue deformation in response to acoustic radiation force impulse as imaged by high-frequency B-mode ultrasound probe. At this step of research, the ultrasound transducer model is simulated to generate acoustic radiation force map to be applied on the 3D FEM and is simulated to image the resulting deformation wave using high-frequency B-mode imaging as well. Nine FEMs are used to be investigated and to cover both normal and pathogenic cornea states for both pre- and post-refractive surgeries. For post-refractive surgery cornea four elastic moduli are used, 140KPa, 300KPa, 600KPa and 800KPa respectively. For pre-refractive surgery cornea five elastic moduli are used, 1MPa, 1.5MPa, 2MPa, 2.5MPa and 3MPa respectively. Time-To-Peak (TTP) deformation, deformation amplitude (DA), deformation amplitude ratio at 2 mm (DA at 2 mm) and shear wave speed (SWS) are all estimated for the involved different nine FEMs. These parameters are used as metrics to assess corneal biomechanics before and after corneal refractive surgeries.

Simulation results show that, rSWS is optimum for high-frequency ultrasound imaging. Maximum accuracy of 99.8% with a range of 2.4% is achieved with 100 KHz frame rate while for 10 KHz frame rate the achieved maximum accuracy is 95.5% with a range of 29%, with a nearly stable accuracy at 100 KHz on contrary to highly fluctuating accuracy at 10 KHz. Also, FPAD show that logarithmic formula is optimum with mean square value of 0.006 compared to 0.09 for power formula. Results have shown that, FPAD is optimum with low frame rate transducers. TTP results have shown decreasing values for increasing corneal elastic modulus. DA values are decreasing with increasing elastic modulus as well. Difficulty of corneal tissue to deform uniformly at higher elastic moduli is reflected by decreasing DA ratio values at increasing elastic moduli. Obtained SWS shows an estimation accuracy of 98%.

# Abbreviations

| | |
|---|---|
| AFM | Atomic Force Microscopy |
| AK | Astigmatic Keratectomy |
| ARF | Acoustic Radiation Force |
| ARFI | Acoustic Radiation Force Impulse |
| csv | Comma separated value |
| CXL | Corneal Cross-Linking |
| FEA | Finite Element Analysis |
| FEM | Finite Element Modelling |
| FLEX | femtosecond lenticule extraction |
| FNM | Fast Near Field Method |
| FPAD | Focal Peak Axial Deformation |
| HMI | Harmonic Motion Imaging |
| I-CXL | assisted trans-epithelial corneal collagen cross-linking |
| IOP | Intraocular pressure |
| LASEK | LASER Assisted Sub-Epithelium Keratomileusis |
| LASIK | Laser-assisted in situ keratomileusis |
| OPE | Ocular Pulse Elastography |
| OPI | Ocular Pulse Imaging |
| ORA | Ocular Response Analyzer |
| PIOL | Phakic Intra-Ocular Lens |
| PRK | Photorefractive keratectomy |
| RK | Radial Keratectomy |
| RLE | Refractive Lens Exchange |
| rSWS | radial Shear Wave Speed |
| SD-OCT | Spectral-domain optical coherence tomography |
| SMILE | Small incision lenticule extraction |
| SSI | Supersonic Shear Imaging |
| SWS | Shear Wave Speed |
| TOF | Time-Of-Flight |
| TTP | Time-To-Peak |
| US | Ultrasound |
| UVA CXL | UltraViolet-A Corneal Cross-Linking |

# Notations & Greek symbols

| | |
|---|---|
| $F$ | Acoustic Radiation Force |
| α | Attenuation coefficient |
| $f_0$ | Central frequency |
| Δ | Delta (difference) |
| ρ | Density |
| $e$ | Exponential function |
| $σ_0$ | Initial stress |
| ϑ | Poisson's ratio |
| μ | Shear modulus |
| $C$ | Shear wave speed |
| E | Strain |
| σ | Stress |
| τ | tau |
| $I$ | Temporal average intensity of the ultrasound beam |
| $T$ | time |
| η | Viscosity |
| $E$ | Young's modulus of elasticity |

# TABLE OF CONTENTS

# LIST OF FIGURES

# LIST OF TABLES

# Chapter 1 | Introduction

## 1.1. INTRODUCTION

The cornea is the first and most powerful refractive surface of the optical system of the eye. The production of an accurate image in the retinal receptors requires the cornea to be transparent and have a suitable refractive power [1]–[3]. The structural integrity of the cornea can be altered in the refractive surgery modifying its refractive properties. These procedures have been developed empirically without detailed knowledge of corneal behavior. Measuring the change in corneal shape has been available in the past decade through computerized analysis of the reflection of photokeratoscopic ring surface of the cornea (corneal topographer). Little is known about the behavior of the internal structure of the cornea. Although the ultrastructure was analyzed by electron microscopy, the role of each one of the layers has not been examined in detail. The tools to measure and understand these processes arise from mechanical engineering and have been used in other medical specialties, such as orthopedics, which evaluates quantitatively the requirements of the prosthesis. Many of these methods are not suitable for soft tissue such as the cornea, but general principles can be applied.

The refractive power of the cornea depends on the curvature at its anterior and posterior surfaces, its thickness, and the refractive index difference between the air and the aqueous humor. The radius of curvature on the anterior surface of the cornea varies from the center (apex) toward the periphery. It is steeper in the central and somewhat flattened at the periphery. Average values of 7.8 and 6.7 mm from anterior to the posterior surface.

At the apex, a wide range of variation of the radius of curvature of 7–8.5 mm is compatible with good visual function, and in pathological conditions increases even more the spectrum. Short radii result in refractive power and high myopia. Conversely, large rays give a low dioptric power and farsightedness. Furthermore, the central corneal curvature is often not equal in all meridians. The focus of the rays reflected by a point object does not form a point image on the retina. In these cases, astigmatism occurs (a, without; stigma, point). The curvature of the cornea changes somewhat with age. It is more spherical in childhood and changes to the rule astigmatism (more curved horizontal meridian) in adolescence. Then again it becomes more spherical in adulthood to turn against the rule astigmatism (steepest vertical meridian) with senility. The thickness also varies from the center to the periphery. This causes the difference in curvatures of the anterior and posterior surfaces. It has an average value of 0.5 mm center, increasing toward the periphery at an average of 1.2 mm at the limbus (limit of the cornea with sclera). The corneal thickness is determined largely by corneal hydration that increases with increasing hydration and slightly with the age.

Gradual loss of the spherical shape of cornea happens in the Keratoconus disease process yielding a conical shape of cornea[4], [5]. The Lack of sufficient in vivo clinical data as well as lack of proper clinical instrument monitoring the corneal pathological changes results in opaque nature of the Keratoconus disease etiology.

Accurate measurement of intra-ocular pressure as well as early detection of corneal diseases can be obtained from information held by corneal viscoelastic properties. Amongst these properties, the corneal shape is essential for visual acuity. Cornea shape is affected by biomechanical properties where it is made from load bearing collagen and is exposed to mechanical forces such as fluctuating intraocular pressure arising from blood circulation.

Noninvasive estimation of these properties leads to local cornea stiffness estimation and results in assessing more corneal pathologies like Keratoconus. In addition, it could highlight the mechanical effects induced by corneal transplant using femtosecond laser or therapy based on Riboflavin/UltraViolet-A Corneal Cross Linking (UVA CXL). In terms of glaucoma treatment, it is very important to have a precise and stable intraocular pressure (IOP) measurement because any fine tuning in corneal elasticity could lead to a dramatic shift in IOP.

## 1.2. MOTIVATION AND OBJECTIVES

Biomechanical properties changes in corneal tissues post refractive surgery may result in an unsatisfactory visual recovery and complications.

Noninvasive prediction of corneal biomechanics and structural strength is not quite accurate for either pre- or post-refractive surgery despite the great advance in assessment methods. This is due to the alteration happening to the corneal structure post-refractive surgery. Patients who undergo radial keratotomy (RK), photorefractive keratectomy (PRK), laser-assisted in situ keratomileusis (LASIK), or small incision lenticule extraction (SMILE), should have their stress concentration maps, potential creak zones, and potential errors in intraocular pressure (IOP) measurements determined for successful post refractive surgery assessment. Developing a noninvasive and accurate method for estimating corneal biomechanics post-refractive surgery will help in identifying many corneal pre- and postoperative pathogenesis.

Estimating biomechanical properties for normal and pathogenic corneal tissue pre- and postoperative is the ultimate objective of this research. Accounting for its high localization; ultrasonic estimation of soft tissue biomechanics will be used as non-invasive method for corneal tissue biomechanical properties estimation. Studying the efficacy of ultrasound probe that is capable of generating a highly localized acoustic force inside thin corneal tissue, as well as picking up the minimal displacement profile generated in response to this acoustic force is another objective of this research.

Objectives of this research can be stated as:

a. Developing an accurate, noninvasive and in-vivo measurement technique for corneal. viscoelasticity to facilitate the understanding of the pathogenesis and early diagnosis of ectatic corneal disorders.
b. Calculating the biomechanical properties for cornea pre- and post-refractive surgeries to assess the quality of the procedure and to early identify the pathologic changes early.
c. Proposing an alternative method to measure corneal biomechanical properties that uses single imaging modality, which is ultrasound imaging, other than using the existing technique that incorporates two modalities, the optical coherence tomography and the air-buff tonometry.

## 1.3. THESIS OVERVIEW

In this research, corneal biomechanics assessment pre- and post-refractive surgeries is studied from many directions, the mathematical modelling, the finite element modelling and the ultrasound transducer modelling. In general, two methods are proposed to assess the corneal biomechanics based on the excitation force and estimation of the resulting deformation; the focal peak axial deformation (FPAD) and radial shear wave speed (rSWS) methods. The finite element model is an Agar-Gelatin model in order to mimic the soft-tissue of the human cornea.

Firstly, a mathematical model is derived to predict the corneal biomechanical behavior in response to both the constant and transient force. Secondly, a FEM is implemented to simulate the deformation behavior for the whole corneal tissue in the two cases; pre- and post-surgery. Also, with the aid of the FEM, the two corneal biomechanics estimation methods are tested and the winner candidate; the radial shear wave speed method; is used in the next step. Thirdly, a complete simulation for the corneal biomechanics is implemented and carried out by means of three sub-models interacting with each other in order to simulate the corneal behavior in response to Acoustic Radiation Force Impulse (ARFI) and B-mode ultrasound imaging. These three sub-models are, the FEM, the scatterrers model and the ultrasound transducer model.

For each method of the two corneal biomechanics estimation methods, an ARFI is applied to the corneal tissue. The resulting deformation is utilized in both methods, where in FPAD method, the axial deformation is estimated and correlated to the corneal biomechanics by two mathematical equations. These two equations are derived by curve fitting values obtained from simulation experiments for different corneal elastic moduli. In the rSWS estimation method, the resulting deformation wave speed is estimated and is correlated to corneal tissue biomechanics by the shear wave speed inside soft-tissue equation.

Simulations are carried out for two general cases, the pre-refractive surgery case and the post-refractive surgery case. Case studies are simulated in both general cases, and their biomechanical properties are obtained from previous published experimental studies.

Finally, corneal biomechanics assessment parameters are estimated and their results and corresponding physical meaning are discussed for both pre- and post-refractive cases.

The proposed research is constituted of three major stages to reach its final output. These three stages are FEM, scatterrers model and Acoustic Radiation Force Impulse (ARFI) imaging stages.

A. <u>Finite Element Model (FEM) stage:</u>

In this stage, a full 3D FEM to build the general geometric shape of the cornea. Then the FEM is fed by the biomechanical properties of corneal tissue so that it exhibits the same behavior of living eye. ARFI map is imported to simulate the applied acoustic impulse on the cornea. Output displacements are exported from the Finite Element Analysis (FEA) software to Matlab software for subsequent imaging and calculation procedure.

B. <u>Scatterrers Model stage:</u>

This stage is dedicated to generate scatterrers phantom; that is coincident with the cornea model; in order to be used by the ultrasound probe to simulate the normal ultrasound (US) B-mode imaging procedure. This stage has two outcomes; the first is to optimize the US probe design to obtain optimum B-mode images of proper axial, lateral and temporal resolution; and the second outcome is to have a reference frame for the ARFI imaging procedure in order to estimate the biomechanical properties of the corneal tissue.

C. <u>Acoustic Radiation Force Impulse (ARFI) imaging stage:</u>

In this stage the acoustic force 2D map is generated and optimized for being applied to the FEM. The ARFI map is generated by using FOCUS tool; a tool used by Matlab software to simulate the acoustic pressure from US probe. This ARFI is generated according to customized US probe, so optimization of the probe is essential in the previous stage in order to have both proper pressure field and optimum US image.

The proposed research methodology can be summarized by the following steps:

1. Generating a Finite Element Model (FEM) corresponding to corneal tissue based on the dimensions obtained from medical literature.
2. Assigning the corneal tissue biomechanical properties obtained from relevant literature survey.
3. Ultrasound probe design and ARFI map generation.
4. Scatterrers model generation using anatomical corneal phantom image obtained from corresponding survey.
5. Reference frame imaging using the designed US probe parameters and the corneal phantom image.
6. Applying ARFI map to the FEM and obtaining relevant output of displacement profiles for the whole model.
7. Exporting the 2D displacement map to the scatterrers model and re-imaging the scatterrers model using ARFI imaging technique.
8. Estimating the displacement profiles for the corneal structure in order to calculate the corneal biomechanical properties.

## 1.4. THESIS ORGANIZATION

This thesis consists of six chapters as follows:

**Chapter 1 | Introduction:** introduces the work done and describes the objectives, motivation and organization of the thesis.

**Chapter 2 | Literature Review:** introduces the corneal anatomy and physiology medical background. Moreover, it presents the corneal biomechanics and related refractive disorders as well. It presents a review about the impact of post refractive surgeries on corneal biomechanics and the possible existing methods that are used in assessing corneal biomechanics. It concludes the optimum method amongst possible methods based on a literature-based comparison.

**Chapter 3 | Mathematical Modelling for Corneal Viscoelastic Behavior:** introduces the first step in assessing the corneal biomechanics by presenting a proposed mathematical model to predict the corneal behavior to applied forces both constantly and transiently.

**Chapter 4 | Finite Element Modelling of Corneal Biomechanics:** introduces FEM for assessing corneal biomechanics with FEA simulations which is the second step to predict the corneal biomechanics pre- and post-refractive surgeries. The corneal behavior is simulated by both pre-refractive and post-refractive FEMs. It introduces also two candidate methods to assess the corneal biomechanics, these two methods are rSWS method; which is used to assess the corneal biomechanics in case of high frequency ultrasound probe, and the FPAD; which is an alternative to rSWS method in case of low frequency ordinary ultrasound probe.

**Chapter 5 | Corneal Biomechanics Assessment Using High Frequency Ultrasound B-Mode Imaging:** introduces the three proposed models that are used in conjunction with each other in order to simulate the real corneal behavior as

captured by high frequency ultrasound probe. These models are; FEM, scatterrers' model and the ultrasound transducer model. It discusses also the parameters that can be used to assess the corneal biomechanics after refractive surgeries, and hence assessing the outcome of the refractive surgery itself.

**Chapter 6 | Conclusion and Future Work:** concludes the work and highlights the important points. It introduces the proposed future work that can be performed to enhance the results obtained in this research.

# Chapter 2 | Medical Background and Literature Review

## 2.1. INTRODUCTION

The cornea is the first surface of the human eye that covers the anterior chamber, pupil and iris. Cornea is a transparent and highly refractive surface in human eye's optical system that accounts for the most of its focusing power. Essentially, two thirds to 70% of the complete human eye's optical power are accounted for the cornea and the rest of the optical power is accounted for the anterior chamber and the lens [3], [6]. Approximately 43 dioptres; representing the refractive power of the human cornea; are quite enough for generating a precise image above the retinal receptor cells [7], [8]. The human cornea consists of five different layers; ordered from anterior to posterior; which are corneal epithelium, Bowman's layer, corneal stroma, Descemet's membrane and corneal endothelium; respectively [9]. Corneal tissue exhibits viscoelastic properties, which are related to the presence of fluids moving slowly across the collagen fibrils. Changes in corneal structure will arise in corneal biomechanics changes and alteration; where these changes affect the corneal shape and visual function of the human eye [10]. Changes may arise due to many refractive surgeries in the human eye. Such surgeries are developed empirically without having a full understanding of their effect on corneal biomechanics.

In this chapter, the corneal anatomy, physiology and different refractive surgeries that are performed on human eye and their locations inside the human cornea are outlined. Then, the effect on human cornea is presented based on the technical findings published by different research papers. Different assessment methods of corneal biomechanics are reported in this chapter as well.

The chapter is organized as follows; the first section presents a brief introduction about the problem definition and the impact of the research. The second section gives an introduction to cornea anatomy and physiology. Section three provides the necessary information about corneal biomechanics for the whole corneal tissue and corneal sublayers. Methods developed to estimate corneal biomechanics are also discussed in section three. The fourth section presents refractive disorders that occur in the human cornea and the common refractive surgeries that are necessary for correcting them. Section five discusses post refractive surgery effect on corneal biomechanics. Section six gives a closer investigation for ultrasonic methods that are used to estimate corneal biomechanics. The interaction between the technical findings from surgeries and their impact on the medical field is explained in the discussion. Finally, we summarize the problem and different approaches of solution in the conclusion.

## 2.2. CORNEAL ANATOMY AND PHYSIOLOGY

As the refractive power depends on the curvature of the transparent lens or mirror, the corneal refractive power will depend on its curvature. In fact, the cornea is not smoothly (uniformly) curved either at its anterior or posterior surface. It has an elliptical shape resulting in a difference in thickness between anterior and posterior surfaces at the apex of the cornea and its periphery. This elliptical shape results in a difference in radii of both anterior and posterior surfaces starting from the apex and ending at the proximal end of the cornea. The radii values are 7.8 mm and 6.7 mm for the anterior and posterior surfaces respectively [11]. This results in 551μ to 565μ corneal thickness at the apex and 612μ to 640μ corneal thickness at the periphery [11]. Hydration of cornea influences its thickness as well. Aging affects the cornea curvature as well, whereas the cornea is more spherical at childhood and experiences senility and more astigmatism during youth. The refractive power of the cornea is about 70% of the overall human eye refractive power, only about 20% to 30% of the refractive power is accounted for the rest of the eye optical system.

The cornea is divided into five layers which are corneal epithelium, Bowman's layer, corneal stroma, Descemet's membrane and corneal endothelium as shown in Fig. 2.1 [12]. In the following sub-sections, these layers will be discussed in detail according to their anatomy and their functions as well as their contribution to corneal viscoelasticity [13].

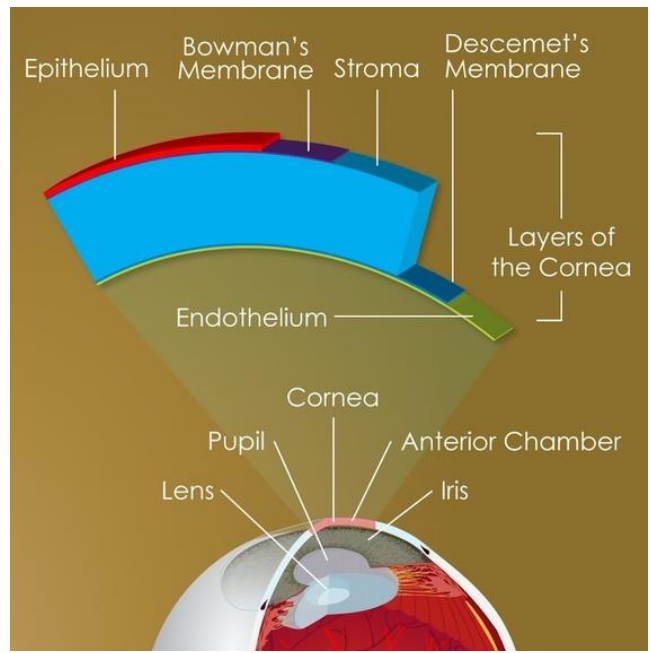

Fig. 2.1 Different Corneal Layers[12].

### 2.2.1. Corneal Epithelium

This layer is about 50 µm of the total thickness of the cornea and it consists of 6 sub-layers. The most outer sub-layer resembles skin in its flattening and is not keratinized. As it gets deeper, the sub-layer gets more tightly connected. A 50 nm to 60 nm thin basal membrane is found between this layer and the subsequent Bowman's layer. This membrane is responsible for regenerating the losses in these sub-layers. This layer is responsible for maintaining the tear film stability [14].

### 2.2.2. Bowman's Layer

It is a translucent membrane of 12 µm to 14 µm thickness consisting of collagen fibrils and lacks cells in its structure [15]. It is believed to support the corneal curvature as its fibrils are woven continuously with those found in the stroma [16].

### 2.2.3. Corneal Stroma

It accounts for approximately 90% of the corneal thickness. This layer consists of about 200 sub-layers consisting of collagen fibrils where each of which is about 1.5 µm to 2.5 µm thick. These sub-layers are almost parallel to the corneal surface, yet they are interconnected with each other. Collagen fibrils have an interconnecting fluid of water mucopolysaccharides that is responsible for corneal hydration. Corneal transparency is maintained by this layer and explained by two theories. The first one states that collagen fibrils arrangements scatter light rays passing through them, and due to destructive interference resulting from the whole mesh framework the scattered light rays are cancelling each other out [17]. The second one states that the distance between collagen fibrils has to be less than 200 nm; typically, in order of 60 nm to 70 nm to acquire uniform transparency [13].

### 2.2.4. Descemet's Membrane

It is a highly elastic acellular membrane and provides an obstacle for perforations and cuts. This layer resembles stroma in structure and collagen fibrils uniformity. It is about 10 nm thick.

### 2.2.5. Corneal Endothelium

It is a thin layer lining Descemet's membrane with a thickness of approximately 5 µm. it consists of a simple squamous monolayer that adjusts the fluid flow between aqueous and corneal stroma [18]. Lacking the property of regenerating dead cells, the endothelium in this layer flattens to compensate for the loss in cells due to death in contrast to the corneal epithelium. Flattening of these cells decreases the number of cells per unit volume; i.e. cell density, affecting the regulation of fluids between the two compartments, the aqueous and corneal stroma. Corneal edema is reached when the remaining cell density cannot compensate for the fluid balance and swelling happens due to excess of fluids. Such corneal edema causes visual impairing for the subject [18].

If counted as a separate layer; Dua's layer forms the anterior part of Descemet's layer with about 15 µm in thickness [19].

## 2.3. CORNEAL BIOMECHANICS

In this section, cornea biomechanics assessment methods are presented for different parts of cornea structure, and their corresponding disorders are explained as well. The cornea is considered as an elastic part of the eye where its resistance to elastic deformation could be measured.

Four common methods are reported in the literature to measure the elastic modulus of the cornea; ultrasonic, atomic force microscopy, strip testing, and globe inflation methods. Globe inflation and strip testing methods are valid for homogenous cornea structure, where they do not account for many elastic moduli corresponding to each part of the cornea and contributing for the overall elastic modulus of the whole cornea as one unit. However, few globe inflation approaches are reported to identify anisotropic material properties [20]–[26]. The experimental data [27], [28] obtained from these methods are summarized in Table 2.1.

Atomic force microscopy (AFM) enables the assessment of corneal layers' elasticity due to its reasonable resolution [29], [30]. As a result of providing very small forces and high resolution, each layer's contribution to overall corneal biomechanics has been assessed properly.

In the following sub-sections, the contribution of each corneal layer in estimating corneal overall biomechanical properties will be discussed.

Table 2.1 Experimental data obtained from different methods for measuring the elastic modulus of the cornea.

| Study | Method Implemented | Organ Under Study | Elastic Modulus (MPa) | Intra Ocular Pressure (IOP) (mmHg) | Number of Samples | Age of Donors |
|---|---|---|---|---|---|---|
| **Woo et. al.** [21] | Bi-Axial Shear | Human | 0.54 | 10 | 3 eyes post-mortem | Not mentioned in the experiment |
| **Hoeltzel et. al.** [25] | Uni-Axial Tensile | Human | 0.34 | 10 | 4 eyes post-mortem | 58, 81 |
| **Sit et. al.** [28] | Ultrasound Surface Wave Elastography | Human | 0.696±0.113 | 12.8 mmHg | 20 eyes healthy subjects | Mean age 51.4±7.2 Range 43-64 years |
| **Edmund et. al.** [27] | Indentation (In-Vivo) | Human | 9 | N/A | Not mentioned in the experiment | Not mentioned in the experiment |

### *2.3.1. Biomechanics of Corneal Epithelium*

The corneal epithelium has no or negligible effect on corneal biomechanics as it forms 10% of the overall corneal thickness [14]. Using AFM [31], the elasticity of epithelium is found to be much lower than other corneal layers.

### *2.3.2. Biomechanics of Bowman's Layer*

This layer contributes by about 20% of the total elasticity of cornea as obtained by [32] using AFM. An elastic modulus of about $109.8 \pm 13.2$ KPa is obtained using this method.

Transplantation of Bowman layer for severe keratoconus and progressive ectasia is possible by using a novel surgical technique proposed by [33]. This technique has shown steadiness for corneal ectasia.

### *2.3.3. Biomechanics of Corneal Stroma*

Since it represents about 90% of the total thickness of the cornea, this layer contributes to most of the corneal biomechanics [34]. Collagen fibrils organization in this layer has a great influence on its elastic modulus, where the anterior part; which is considered denser than the posterior part; has a larger value of elastic modulus of about 3 times higher than that of the posterior part [31], [35]. Using

Acoustic Radiation Force elasticity imaging, Mikula et al. [36] obtained an elasticity map for the cornea and found that peripheral anterior elasticity is less than the posterior one by two-thirds factor. This is due to the steepness in the inclination of the posterior part which means more interwoven collagen fibrils at this region in addition to their orthogonality at the center than being oblique at the periphery [37]–[39].

### *2.3.4. Biomechanics of Corneal Endothelium*

It is considered to affect the elasticity of the corneal indirectly as it regulates the hydration process inside the cornea. Losses in corneal endothelium cells result in an increase in corneal hydration where more water in absorbed by stroma [40], [41].

Experimental results for estimating the elasticity of different corneal layers are reported in [31, 32, 42, 43] are listed in Table 2.2.

Table 2.2 Estimated elasticity for different corneal layers using AFM.

| Corneal Layer | Elasticity |
|---|---|
| Anterior Epithelium Membrane [42] | 7.5±4.2 KPa |
| Bowman's Layer [32] | 109.8±13.2 KPa |
| Anterior Stroma [32] | 33.1±6.1 KPa |
| Posterior Stroma [43] | 89.5±46.1 KPa |
| Endothelium [31] | 4.1±1.7 KPa |

Keratoconus is considered the most common corneal disorder accompanied by corneal biomechanics changes [4]. Keratoconus causes two complications, breaks in Bowman's layer and decreasing in the stromal thickness as well as a decrease in content and elasticity of collagen fibrils anterior of the Descemet's membrane [10].

## 2.4. REFRACTIVE DISORDERS

Refractive disorders are classified into two major classes; which are low order aberrations and high order aberrations. Low order aberrations are divided into three major classes; namely; axially symmetrical, astigmatic and mixed disorders and are measured in diopters [44]. High order aberrations; which are deviations from the ideal wave front; are divided into two major disorders; spherical aberration and coma [21], [45], [46]. Figure 2.2 shows the hierarchal diagram for these disorders.

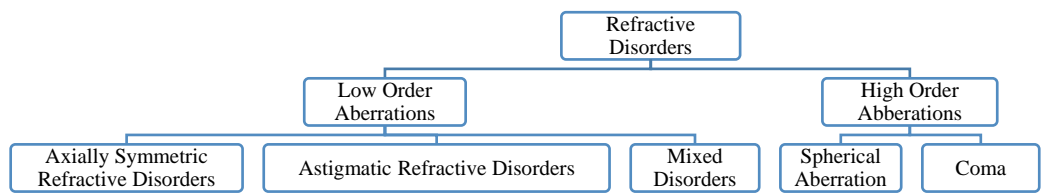

Fig. 2.2 Hierarchical diagram for refractive disorders.

Myopia and hyperopia are classified as axially symmetric refractive disorders. Myopia occurs when the image is formed blurred due to the focusing of the parallel light rays in front of the retina. Hyperopia (hypermetropia) happens when parallel light rays are focused; theoretically; to a point that is located behind the retina resulting in a blurred image on the retina.

The difference in curvature between both refractive components of the human eye optical system; cornea and lens; or one of them, gives rise to astigmatism. This difference in curvature gives rise to a difference in refractive powers between these two refractive media. This type of refractive disorders results in non-uniform light rays' refraction and distortion in image formation.

Mixed disorders are likely to occur in the human eye, where both axially symmetric and astigmatism are happening at the same time in the same subject.

In terms of aging, presbyopia happens where the lens is not flexible enough to accommodate its refractive power for near vision.

Compensating refractive disorders is achieved either by corneal curvature remodeling (corneal refractive surgery) or lens refractive surgeries and may result in changes in corneal biomechanics. Corneal refractive surgeries are achieved by two main methods; the LASER assisted method and the incision assisted method. Lens refractive surgeries are achieved by two main methods; Phakic Intra-Ocular Lens (PIOL) and Refractive Lens Exchange (RLE).

Based on viscoelastic properties of the cornea, another technique to compensate for refractive disorders by changing cornea shape is orthokeratology. This is achieved by flattening the cornea using an applied pressure temporarily using orthokeratology lenses. The osmotic pressure gradient is induced due to the applied pressure giving rise to liquid removal of the central epithelium. Liquid removal leads to changes in biomechanical properties of the cornea, however no further studies have introduced a complete assessment of orthokeratology efficacy and corneal biomechanics [47]–[49].

## 2.5. REFRACTIVE SURGERIES AND THEIR EFFECT ON CORNEAL BIOMECHANICS

Corneal refractive surgeries are divided into three main classes, namely flap, surface and incision procedures. Surface procedures use excimer LASER to ablate the corneal tissue just below the epithelium.

In LASER-assisted surgeries, the cornea is reshaped by the LASER beam where the cornea is ablated so as to make the incident light rays focus on the retina at the point of maximum vision (macula).

Surface procedures involve corneal epithelium to be removed away by any of the three means; mechanically, chemically or by LASER; in Photo-Refractive Keratectomy (PRK), while in LASER Assisted Sub-Epithelium Keratomileusis (LASEK) it is removed by alcohol. The microkeratome is used to remove corneal epithelium in epi-LASIK. In such procedures, the LASER beam is used to ablate epithelial tissue or sub-epithelial tissue only.

Flap procedures are achieved successfully by pleating the corneal flap aside using either microkeratome or femtosecond LASER to have an entrance to the stromal layer in LASER in-situ keratomileusis (LASIK) procedure. Flap procedures are characterized by utilizing the LASER beam deeper to ablate the inferior portion to the epithelial layer, i.e., anterior stromal layer.

Excimer LASER is used to correct disorders of refraction by removing a spherical lenticule from the cornea center in the myopia disorder. However, it is used to ablate the corneal periphery for the hyperopia disorder increasing in central cornea curvature.

Corneal incision procedures involve the usage of either a femtosecond LASER or a diamond knife to cut as few sufficient microscopic incisions as possible in the perpendicular plane to the steep corneal meridian in order to cancel the astigmatism of the eye by changing surface corneal curvature. Astigmatic Keratectomy (AK) is accomplished by perforating one or two central incisions. Radial Keratectomy (RK) is achieved by having these two perforations on both the distal limbal sides of the cornea [45], [50]–[52].

Corneal biomechanical properties are affected by refractive surgeries where it depends on both the depth of ablation of the LASER spot and the corneal flap thickness [53], [54]. One complication that may happen due to insufficient residual thickness of stromal bed post-refractive surgery is corneal ectasia. Corneal ectasia also happens due to subclinical keratoconus that is not diagnosed pre-refractive corneal surgery. Residual stromal bed thickness margin of not less than 250 µm is applied to limit the probability of occurrence of corneal ectasia, yet it does not prevent the occurrence of this case completely [55], [56]. Corneal ectasia happens also due to alterations in corneal biomechanics pre- and post-refractive surgery. The difference in corneal biomechanical properties of pre- and post-refractive surgeries has been reported for both LASEK and SMILE surgeries [57]. However, it is reported in [55] that SMILE has less effect on corneal biomechanics than LASEK accounting for preserving anterior stroma.

The cornea is claimed to have stiffer behavior after SMILE than after LASIK or PRK as SMILE maintains the stiffer layers in cornea. Even with parameters variation in the same surgery it is found to have different outcomes for the surgery.

Spiru et al. [58] reported lower elasticity for ex-vivo porcine cornea treated by femtosecond lenticule extraction (FLEX) than cornea treated with SMILE. They claimed that the cap-based technique is better for biomechanical stability [59]. Knox et al., reported that increasing LASIK flap depth from 90 µm to 160 µm alters the mechanical properties of the cornea in the horizontal plane than in the vertical plane [60]. A thin flap of depth 90 µm to 110 µm is reported to have no significant improvement on the cornea of rabbits as reported by the study of Wang et al. [61]. As in He et al. [62], higher Young's modulus is reported in rabbits' cornea after having SMILE with thickness of 160 µm compared to SMILE with a thickness of 100 µm. A significant factor leading to corneal ectasia is the percentage of altered tissue than a residual stromal bed and central corneal thickness as reported by Santhiago et al.[63]. Also, no significant difference in corneal biomechanics is reported by Kling et al. for corneas treated by SMILE and PRK[58]. Yet, it is expected for SMILE to give better results biomechanical properties than PRK.

Although epithelium thickness is reported to increase after laser refractive surgery, SMILE leads to an increase in the center of about 2.5 µm as reported in [22] and 15 µm as reported in [64] accompanied by a peripheral decrease in the thickness as reported in [65]. Moreover, hyperplasia for the central epithelium of about 1 µm is reported to occur after SMILE than after LASIK [66]. Residual refractive error reveals the difference between the pre-planned tissue removal and the achieved reduction in stromal tissue by about 8 µm and 11.9 µm thicker for SMILE, while it is 0.4 µm for LASIK 3 months post-refractive surgery in both surgeries [64], [66]. An increase in the refractive index of the stroma is observed after LASIK than in SMILE surgeries as reported in [67].

Table 2.3. summarizes results obtained from relevant studies that measured corneal biomechanics post-refractive surgery with LASIK and SMILE. The table is extended to include the results for PRK and LASEK as reported by[68].

Corneal crosslinking (CXL) is a procedure performed to overcome the biomechanical properties alterations that happens due to refractive surgeries leading to corneal ectasia. Such procedure is aimed to prevent the progression of resulting ectasia [62], [79]. Prevention of such progression is achieved by induction of new molecular bonds between lamellae and corneal collagen fibrils using riboflavin and UV-A light [22], [59], [80].

Newly generated cross links push the fibrils apart leading to an increase in intermolecular spacing and fibrils diameter [81]. This approach describes the increase of the corneal stiffness by about 320% in an ex-vivo study after the crosslinking procedure [80]. Crosslinking gives rise to an increase of about 22% and 16% for both anterior and posterior collagen fibrils diameter respectively due to the localization of corneal crosslinking effect in the anterior stroma [17], [82]. Figure 2.3 shows a scheme for different corneal refractive surgeries.

Table 2.3 Related studies comparing corneal biomechanics post-refractive surgery.

| Study | Refractive surgery | Conclusion |
|---|---|---|
| Shen et. al. [69] | SMILE, femtosecond LASIK | No significant difference between both surgeries. |
| Shen et. al. [69] | SMILE, LASEK | No significant difference between both surgeries. |
| Pederson et. al. [70] | SMILE, femtosecond LASIK | No significant difference between both surgeries. |
| Agca et. al. [71] | SMILE, femtosecond LASIK | No significant difference between both surgeries. |
| Sefat et. al. [72] | SMILE, femtosecond LASIK | No significant difference between both surgeries. |
| Zhang et. al. [73] | SMILE, femtosecond LASIK | No significant difference between both surgeries. |
| Al-Nashar et. al. [74] | SMILE, PRK | No significant difference between both surgeries. |
| Wang et. al. [75] | SMILE, femtosecond LASIK | Differences for myopia > 6D in favor of SMILE. |
| Wu et. al. [76] | SMILE, femtosecond LASIK | More effect after femtosecond LASIK. |
| Osman et. al. [77] | SMILE, femtosecond LASIK | More effect after femtosecond LASIK. |
| Kling et. al. [58] | SMILE, LASEK | More effect after LASEK. |
| El-Massry et. al. [78] | SMILE | Less biomechanical affectation with 160 μm than with 100 μm. |

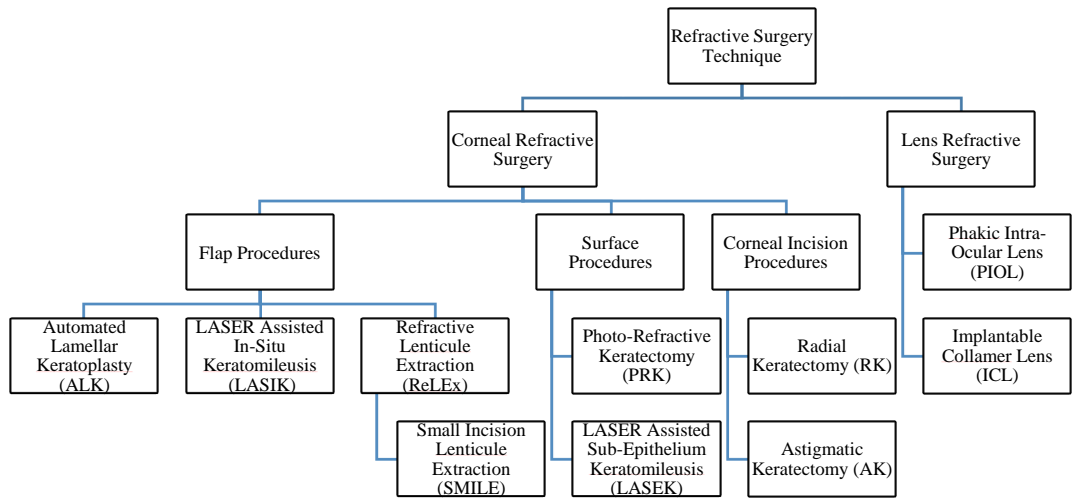

Fig. 2.3 Different corneal refractive surgeries map.

## 2.6. ULTRASONIC ESTIMATION OF CORNEAL BIOMECHANICS

This section aims to highlight many of the implemented techniques using ultrasound technology to estimate corneal biomechanics. Three trending methods are implemented in the field of corneal biomechanics assessment; namely, Harmonic Motion Imaging (HMI), Supersonic Shear Imaging (SSI) and Ocular Pulse Imaging (OPI).

For the major algorithm of supersonic ultrasound estimation for biomechanical properties of soft tissues, a high frame rate of B-mode scanning is adopted, obtaining a stack of deformed frames along with a reference frame [83], [84]. A transient ultrasound push is applied either using an external force inductor or internally by focusing the scanner's acoustic force to a single point inside the tissue [85]. A deformation wave inside the tissue is generated corresponding to the applied force. The speed of this wave is estimated as it is a characteristic for tissue's mechanical properties; such as shear and Young's modulus. The wave speed is calculated using Eqn. 2.1. Time-Of-Flight (TOF) [86], [87], Lateral Time-To-Peak (TTP) [71], [88] or cross-correlation [4], [22], [89], [90] algorithms are implemented to estimate the speed of the resulting tissue deformation wave; i.e. Shear Wave Speed (SWS).

$$C = \sqrt{\frac{\mu}{\rho}} \qquad (2.1)$$

where C is the shear wave speed, μ is the shear modulus of the tissue, and ρ is the density of the tissue material. It is feasible now to estimate tissue stiffness using the estimated wave speed. The whole-time sequence of the algorithm is presented in Fig. 2.4.

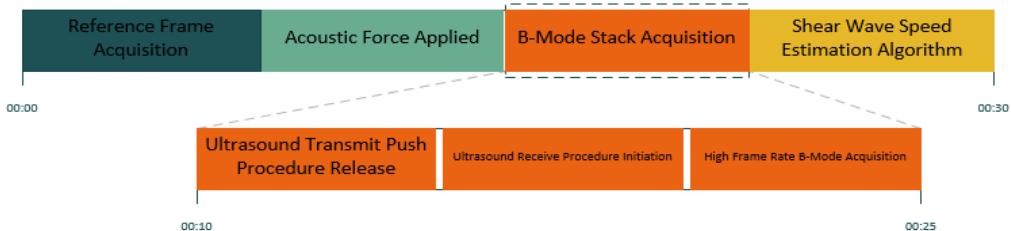

Fig. 2.4 Time sequence ultrasonic estimation algorithm.

The advantage of SSI is that it ensures a linear behavior of cornea tissue as it induces micrometer and sub-micrometers displacements which is much lower than the cornea thickness. This advantage is not present in other non-ultrasonic estimation methods, where the induced displacements are of larger values greater than corneal tissue thickness leading to non-linearity behavior.

Tanter et al. [87], managed to achieve a high-resolution 2D elasticity maps of 150μm resolutions for corneal biomechanics using SSI technique, where they used a 15 MHz linear array ultrasound probe made up of 128 elements and 0.2 mm pitch spacing and 12 mm elevation focus to perform the imaging. Fresh porcine eyes placed in a water cap are used as an ex-vivo tissue under study. The ultrasound probe is positioned at a height of 5 mm above the porcine eyes and inside the water cap for acquiring the imaging stack. The frame rate is limited by the Time-Of-Flight (TOF) of the ultrasonic wave inside the tissue, and for the corneal tissue it is about 40 kHz. This is because of the thickness limitation of ophthalmic tissues which is 20 mm in most of the deeper cases. Tanter et al. [87] used a frame rate of 20 KHz. Average elasticity of $190\pm32$kPa is achieved for four different specimens.

Zohu et al. [91] managed to obtain a relationship between the Intra-Ocular Pressure (IOP) and the SWS of the cornea. Ex-vivo porcine eyes are used in their experiment. These eyes are immersed in gelatin phantom to fix them in position. An ultrasound probe is used to measure the resulting waves of a localized harmonic excitation source at one side of the cornea. The experiment is performed for IOP range of 5 mm to 30 mm with 5 mm step.

Chen et al. [92] measured the corneal biomechanics by means of a high-resolution ultrasound probe. A dual-element transducer of 8 MHz and 32 MHz is used, where the 8 MHz element is used for excitation of the tissue deformation and the 32 MHz is used for the wave propagation monitoring. Their designed probe achieved about 72 µm resolution in the lateral direction and 140 µm in the axial direction. Thus, achieving higher resolution than that obtained by [87]. Two agar-gelatin phantoms are used in their experiment, which are 3% and 7% agar respectively, with resulting shear wave speeds of 0.64 m/sec. and 1.27 m/sec. for each phantom. For donor human corneas, the results obtained are $2.45\pm0.48$ m/sec. which corresponds to $11.52\pm7.81$ KPa.

Touboul et al. [93] studied the biomechanical changes that are induced by the Iontophoresis-assisted trans-epithelial corneal collagen cross-linking (I-CXL) using the SSI technique. The treated corneas are found to have a sharp stiffness in comparison to untreated corneas.

Zhang et al. [94] used an ultrasound probe and external excitation source to generate shear waves inside both ex-vivo bovine untreated corneas and treated corneas with CXL. Results obtained confirm that treated corneas with CXL have higher elasticity and lower viscosity than those of untreated corneas. In another research, Zhang et al. [95] have attempted to correlate the age-related stiffness with group shear wave speed of the crystalline rabbit lens proposing that these group velocities are a promising biomarker for crystalline lens stiffness. The advantage of this experiment is that it measures the group velocities in vivo for the rabbits.

Ocular Pulse Imaging (OPI) and Ocular Pulse Elastography (OPE) [96] techniques have the advantage of ocular pulse occurring naturally in human eyes.

OPE measures the corneal strains during each IOP cycle, thus allowing the measurement to take place along any meridian of the eye with high spatial resolution and through the whole thickness of the cornea.

Harmonic Motion Imaging (HMI) is developed by [97], in order to distinguish different elastic moduli objects at a very small scale. An ultrasound probe is designed to have two concentric disc elements, 4 MHz and 40 MHz. The 4 MHz element aims to generate a periodic Acoustic Radiation Force (ARF) to induce periodic deformation of the sample under test. The 40 MHz element is used to track the displacements in the sample. This design is capable of having 154μm axial resolution and 314 μm. This makes the design of such a probe worse than the methods proposed in [29] and [92] where obtained results from these three methods are listed in Table 2.4.

Table 2.4 Obtained results for corneal biomechanics from applying different ultrasonic estimation methods.

| Author | Method Implemented | Organ Under Study | Results Obtained |
|---|---|---|---|
| Bercoff et. al. [83] | Supersonic Shear Imaging | Porcine ex-vivo eyes (Corneal tissue) | 190±32 KPa (Elastic Modulus) |
| Chen et. al. [92] | Dual element transducer + conventional ultrasound imaging technique | Donor human corneas | 11.52±7.81 KPa. (Elastic Modulus) |
| Touboul et. al. [93] | Supersonic Shear Imaging | Rabbit Cornea | 27.1 KPa (Elastic Modulus) |
| Zhang et. al. [94] | ARFI | Bovine ex-vivo cornea | 1145 ± 267 kPa (Elastic Modulus) |
| Palvatos et. al. [96] | Ocular Pulse Imaging | Animal (Porcine) and human donor eyes | 0.000925±0.000003 (Strain Values) |

## 2.7. DISCUSSION

Different methods are used for the assessment of corneal biomechanics such as strip testing, globe inflation, atomic force microscopy or ultrasonic methods. These methods have different levels of feasibility, accuracy, and nature of the test.

Even though strip testing is an accurate method for estimating corneal biomechanics, it is not used for clinical application. As a result of using ex-vivo tissue and its sample destruction nature, this method is non-convenience for clinical practice.

Globe inflation, which is the second method; is widely used in the clinical estimation of corneal biomechanics, but it lacks the accuracy of the estimated stiffness. It is a feasible non-destructive test and suitable for patients who have not undergone refractive procedures. However, it is not suitable for post-refractive procedures patients, as the applied air pressure may distort the healing cornea. For the last reason it is not recommended for patients who have recently got refractive surgery.

The third method of estimation is Atomic Force Microscopy (AFM), which is an accurate method, but lacks the feasibility of preparation and measuring setup for specimen under study.

Finally, the ultrasonic estimation method has the advantage of feasibility to use, the accuracy of its results, and is non-destructive. For these three reasons it is considered to be a promising method for clinical practice. Table 2.5 presents a comparison between the four mentioned methods.

Table 2.5 Comparison between different corneal biomechanics estimation methods.

| Methods of Assessment / Points of Comparison | Strip Testing | Globe Inflation | Atomic Force Microscopy | Ultrasonic Estimation |
|---|---|---|---|---|
| Feasibility | Not feasible | Feasible | Not feasible | Feasible |
| Accuracy | Accurate | Not Accurate | Accurate | Accurate |
| Nature of test (Destructive/Non-destructive) | Destructive | Non-destructive | Non-destructive | Non-destructive |

Many points should be discussed when dealing with ultrasonic estimation methods. The first one is the resolution of the scanning probe typically, the lateral resolution. Lateral resolution is of great impact with wave propagation, as the wave propagates laterally off the excitation point. High lateral resolution leads to a more accurate estimation of wave propagation. This is because the generated wave is of micrometer order requiring a high-frequency probe.

Peak displacement is the magnitude of the displacement of a specific point in the three directions x-, y- and z-direction. Clearly, the peak displacement of the focal node; and any node; have a decreasing behavior with increasing the stiffness at the same excitation frequency.

This decreasing behavior is due to the dashpot element effect. The decrease in the peak displacement occurs because the point under investigation cannot keep up with the frequency of excitation; which is not altered or changed; at higher stiffness values. Thus, smaller displacement is achieved at higher stiffness values. Larger displacement values are observed at smaller stiffness values, where points under investigation can reach their peak displacement (deformation) easily after excitation as the dashpot element responds well enough for the same excitation frequency at smaller values of stiffness.

Also; peak axial displacement generated at the focal point (the point of excitation) is related to the frequency of the scanning probe. Higher frequency means higher resolution and higher accuracy of axial peak tracking.

A trade-off arises between the peak displacement and the resolution of the transducer used in tracking the propagating wave. As the peak displacement of the points becomes smaller and smaller, their tracking process becomes more difficult. This is due to the resolution of the tracking transducer is fixed at some axial and lateral resolution values, but the peak displacements tracked are variable, and their variation is related to the stiffness of the soft tissue, as shown in Fig. 2.5. Hence, there is an optimum resolution; axially and laterally; for tracking the propagating shear wave at each stiffness value. The limitation here is in increasing order.

The insufficient tracking results in an underestimation of the tracked wave, leading to estimating a value of the wave-less than its real value.

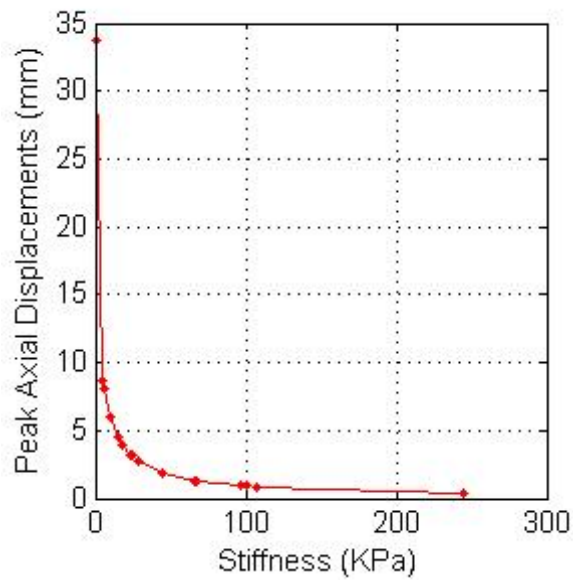

Fig. 2.5 Relationship between peak axial displacement and soft tissue stiffness.

## 2.8. SUMMARY

In this chapter, corneal disorders and their relevant refractive surgeries are presented. In addition to the effect of these refractive surgeries on the biomechanical properties of the cornea. Different assessment methods for cornea biomechanics are overviewed in conjunction with their limitations. These methods are strip testing, globe inflation, atomic force microscopy and ultrasonic estimation. In terms of feasibility globe inflation and ultrasonic estimation are better than strip testing and atomic force microscopy. While in terms of accuracy the globe inflation shows insufficient accuracy to assess cornea biomechanics. On the other hand, the nature of the test reveals that strip testing is a destructive method. Comparison between these methods shows that ultrasonic methods are better than other methods in terms of feasibility, accuracy and it is non-destructive. Eventually, it is recommended to use ultrasonic-based methods in the assessment of biomechanics of the cornea after refractive surgery.

# Chapter 3 | Mathematical Modelling for Corneal Viscoelastic Behavior

## 3.1. INTRODUCTION

Deformation behavior of soft tissues due to stress is investigated by measuring their biomechanics properties. Recently, researches are giving attention to biomechanical properties of ocular tissues so as to understand the progression of ocular diseases such as: keratoconous, and post-refractive ectasia which are characterized by corneal tissue deformation. Also, elasticity changes before and after the refractive surgeries are of their interest as they affect the developing of post-refractive ectasia. Biomechanical properties are reported to be changing due to keratoconous corneas and refractive surgeries [27], [98].

Understanding of cornea biomechanics leads to better understanding of these corneal alterations and how to treat them. Many researches have focused on investigation of the corneal biomechanics ex-vivo such as the work reported in [87]. However, ex-vivo cornea experience swelling and loss of its tear film leading to changes in its biomechanical properties [27]. Hence, there is a great need for in-vivo cornea studies that allows for normal and diseased cornea biomechanics. Mathematical modelling of corneal biomechanics is the first step in understanding its behavior in normal and diseased states. A suitable model should describe the elastic and viscous components of the cornea as its tissue exhibits viscoelastic behavior when loaded under certain forces transiently [98].

Human cornea experiences a viscoelastic behavior when subjected to a transient stress[99]. This behavior can be represented by two components; elastic component and viscous component. The elastic component gives an instantaneous deformation while the viscous component gives a damping deformation. Elastic and viscous components can be modelled by spring; with elasticity of (E); and dashpot; with viscosity of ($\eta$); system components respectively. The spring represents the pure elastic behavior to an applied load, while the dashpot represents the time-dependent viscous resistance to that load.

There are multiple models that are used to describe the viscoelasticity behavior, amongst them Maxwell model, Kelvin-Voigt model and the standard linear solid model [100]. The three models are presented in Fig. 3.1. Maxwell's model consists of a spring and a dashpot connected in series to each other, while Kelvin-Voigt model consists of the same two components but connected in parallel to each other.

Standard linear solid model in Kelvin-Voigt representation has an extra spring element connected in series to the parallel spring-dashpot connection.

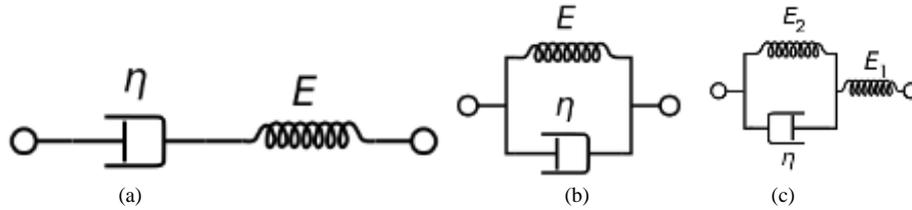

Fig. 3.1 (a) Maxwell Model, (b) Kelvin-Voigt and (c) Standard linear solid model in Kelvin-Voigt representation.

Several researches proposed mathematical models for studying corneal biomechanics. One of them is the model reported in [99], where two Kelvin-Voigt models in series with a spring is proposed to compare the effect of fast versus slow viscoelastic behavior. In [101], a Kelvin-Voigt model is proposed to study the effect of elasticity and viscosity separately on corneal hysteresis. In [102], a mathematical model based simulation is proposed to investigate induced corneal vibrations due to air-puff and its relationship to intra-ocular pressure and viscoelasticity of the cornea. Other mathematical models are proposed to study the heating effect of different wavelengths of UV lasers used in corneal ablation procedures as well [103]–[105].

This chapter proposes a mathematical model to investigate the corneal tissue viscoelastic behavior when subjected to a transient force under the effect of intra-ocular pressure and tear film pressure.

## 3.2. Methodology

In this chapter, both the Kelvin-Voigt and the standard linear solid models are used to simulate the human cornea viscoelastic behavior. Maxwell's model is discarded due to its indefinite creep behavior and because it does not fully recover from deformation due to the pure viscous component.

The complete vertical cross-section of human eye with cornea represented as standard linear solid model is shown in Fig. 3.2. The corneal tissue is modelled using Kelvin-Voigt model and standard linear solid model. Its behavior using both models is simulated under two scenarios; while applying a constant load and a transient load respectively. The constant load is simulated as a unit step function, and the transient load is simulated as a sinusoidal burst of 3 cycles applied for 40% of the simulation time. The transient load is chosen to be a sinusoidal burst as it simulates the acoustic radiation force burst as an internal actuator generated from an ultrasound transducer during elastography procedure.

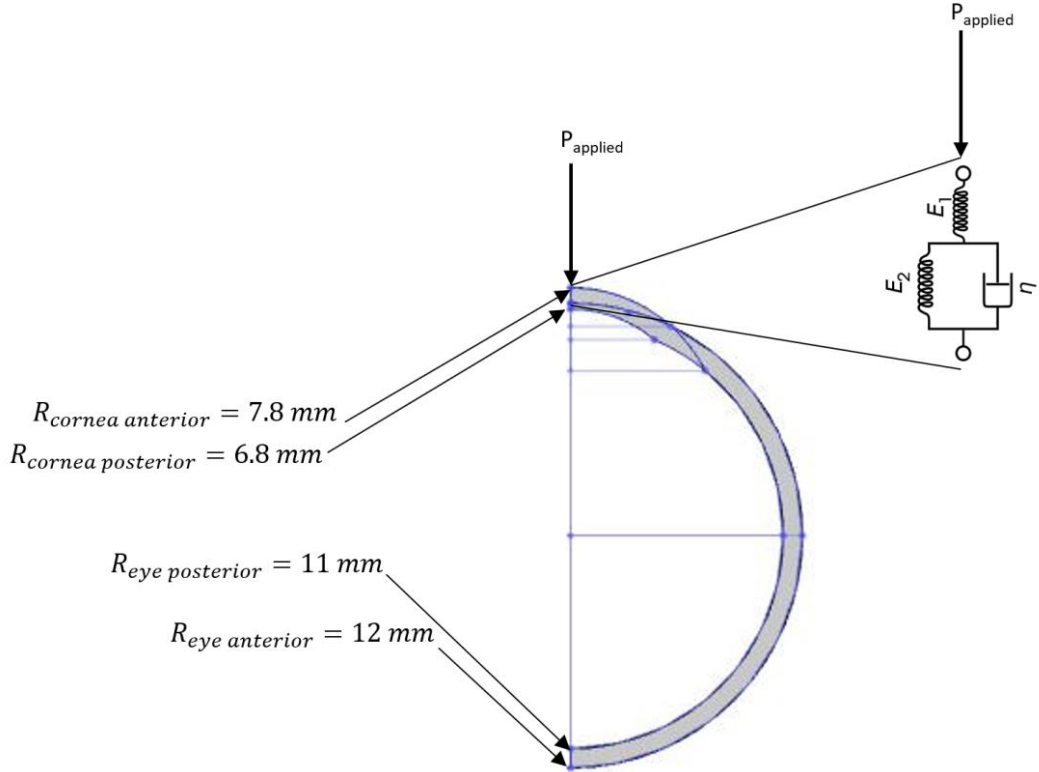

Fig. 3.2 Complete vertical cross-section of human eye, and corneal tissue represented with standard linear solid model.

The corneal tissue elasticity is set to 150 KPa to simulate a normal corneal tissue [106]. Either the constant load or the transient load are applied at the corneal apex. The simulation time is 5 msec for both Kelvin-Voigt and standard linear solid model to study their strain behavior. Constant load is applied for the full 5 msec, while the transient load is applied for only 2 msec. The complete simulation parameters are listed in Table 3.1.

For a Kelvin-Voigt material; the stress-strain equation as a function of time is given by Eqn. 3.1.:

$$\sigma(t) = E\varepsilon(t) + \eta \frac{d\varepsilon(t)}{dt} \qquad (3.1)$$

where $\sigma$ is the applied load in ($Pa$), $E$ is the modulus of elasticity, $\varepsilon$ is the strain in ($mm$) and $\eta$ is the viscosity in ($Pa.sec.$).

The strain as a function of time due to Kelvin-Voigt model will follow Eqn. 3.2. when subjected to a sudden stress:

$$\varepsilon(t) = \frac{\sigma_o}{E} (1 - e^{\frac{-t}{\tau_R}}) \qquad (3.2)$$

where $\tau_R = \frac{\eta}{E}$ is the retardation time and t is the simulation time in seconds.

Strain as a function of time due to standard linear solid model is given in Eqn. 3.3.:

$$\varepsilon(t) = \frac{\sigma_o}{E}\left(1 - e^{\frac{-t}{\tau_R}}\right)\left(e^{\frac{-t}{\tau_R}}\right) \qquad (3.3)$$

When applying either a constant or a transient load *Papplied*; the total load experienced by the cornea is given by the following Eqn. 3.4.:

$$P_{resultant} = P_{applied} + P_{tear\ film} - IOP \qquad (3.4)$$

where $P_{tear\ film}$ is the tear film pressure working in the same direction of the applied load and equals 4.15 mmHg [107], and IOP is the intra-ocular pressure. Tear film pressure is responsible of maintaining corneal tissue constituents in place together.

Table 3.1 Simulation parameters for Kelvin-Voigt and Standard Linear Solid (SLS) models.

| Simulation Parameters | Simulation Type | |
|---|---|---|
| | *Constant Load* | *Transient Load* |
| **Excitation Frequency** | 0 Hz | 1 KHz |
| **Number of cycles per burst** | ---- | 3 |
| **Excitation window per simulation time** | 100% | 40% |
| **Applied Load** | 800 Pa | 27 KPa |
| **Simulation Duration** | 5 msec | |
| **Young's Modulus** | 150 KPa | |
| **Viscosity** | 100 Pa.sec. / 250 Pa.sec. | |
| **Tear Film Load** | 4.15 mmHg ~ 550 Pa | |
| **Intraocular Pressure (IOP)** | 15 mmHg ~ 2050 Pa | |

## 3.3. RESULTS AND DISCUSSION

Simulation is performed using MATLAB program (R2020b) and using a machine of 8 GB RAM and core i5 processor with a best value of execution time of 3.3 msec for transient load scenario and 4 msec for constant load scenario.

### 3.3.1. *Constant Load*

For the Kelvin-Voigt model, when a constant load is applied, it starts to deform instantaneously till it reaches an asymptote. The model describes the instantaneous (elastic) deformation due to sudden load but it fails in describing the viscous damping deformation effect of the viscoelastic material.

The applied constant load is shown in Fig. 3.3 (a). Kelvin-Voigt model time dependent strain (deformation) behavior due to applying constant load is presented in Fig. 3.3 (b). The material does not recover from its deformation but it continues to creep constantly. The deformation amplitude due to the applied constant load is about 4.4 mm. The material reaches its peak deformation amplitude at 3 msec approximately.

The standard linear solid model is able to describe both the instantaneous (elastic) deformation and the viscous damping deformation of the material. The time dependent strain behavior due to constant load is shown in Fig. 3.3 (c). The deformation amplitude due to this constant load is about 1.1 mm. The material reaches its peak deformation amplitude at 0.45 msec approximately. The material returns to its original state recovering from its deformation.

From the constant load simulation, it is clear that the standard linear solid model is able to describe the corneal tissue viscoelasticity more effectively. This is because it models the material recovery from deformation and because it predicts more reasonable deformation amplitude value compared to Kelvin-Voigt model.

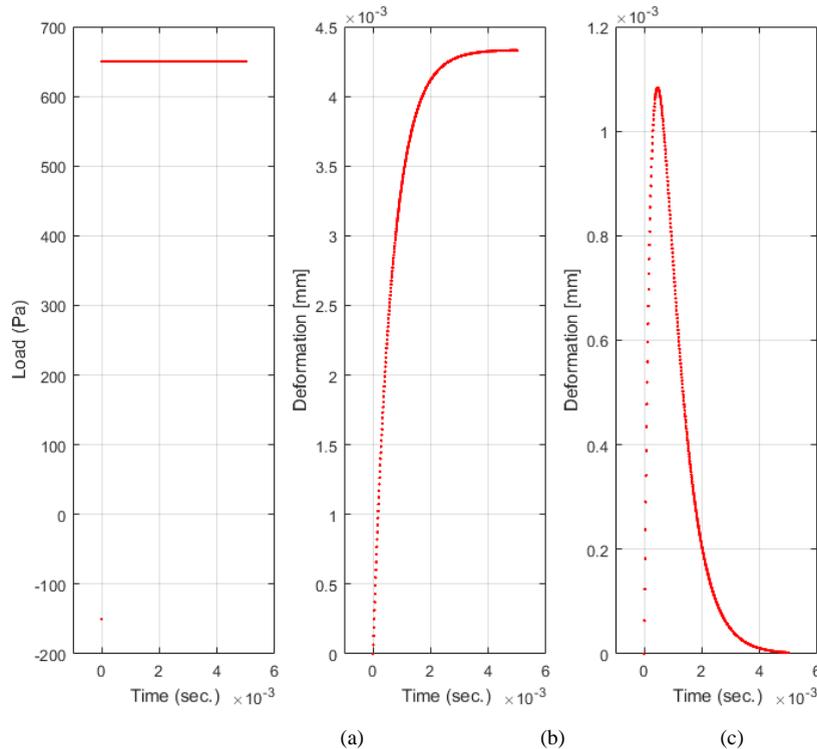

Fig. 3.3 (a) Constant load, (b) Kelvin-Voigt model time dependent strain behavior and (c) Standard linear solid model time dependent strain behavior.

### 3.3.2. *Transient Load*

The applied transient sinusoidal load burst is presented in Fig. 3.4 (a). Kelvin-Voigt model yields a sinusoidal strain behavior that increases with time. The model describes again the instantaneous elastic deformation effectively. However, it fails in describing the viscous damping behavior. The deformation amplitude reaches 0.007 m approximately and continues to increase with each new cycle of excitation which is not logical for corneal tissue. This behavior is presented in Fig. 3.4 (b).

Standard linear solid model behavior is presented in Fig. 3.4 (c). The model predicts the material to have an instantaneous elastic deformation combined with exponential damping for deformation behavior. The deformation amplitude is approximately 1.2 mm which decreases significantly with each cycle of load excitation till reaches original state. The deformation amplitude is reasonable for cornea tissue in response to the applied transient load compared to the deformation amplitude predicted by Kelvin-Voigt model.

In terms of temporal localization, where the deformation is supposed to be highly localized within short temporal duration, the predicted behavior due to Kelvin-Voigt model is not suitable as it does not fulfill the temporal localization for the corneal deformation. On the other hand, standard linear solid model predicts properly the temporal localization of the corneal deformation.

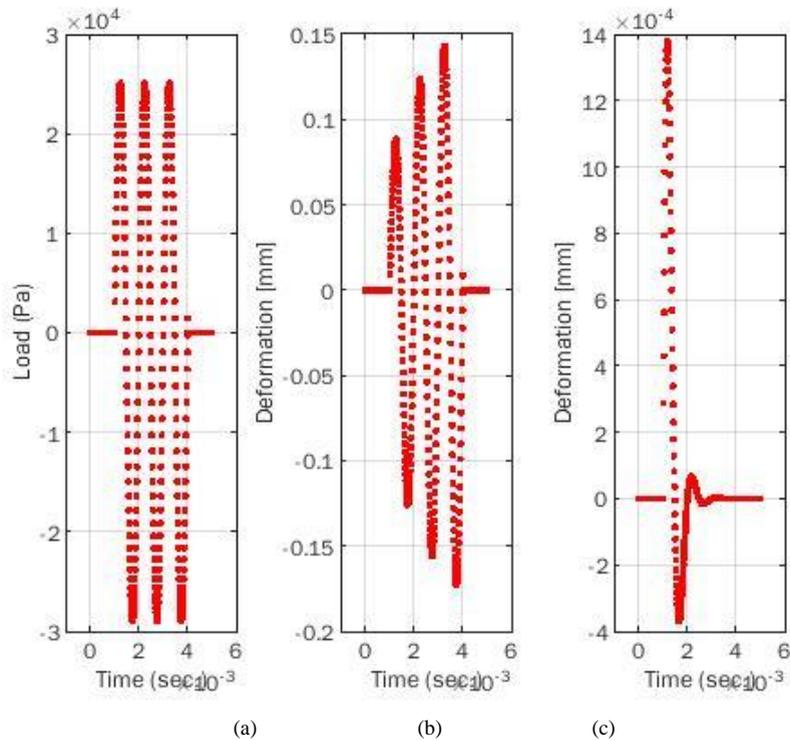

(a)                    (b)                    (c)

Fig. 3.4 Sinusoidal load burst, (b) Kelvin-Voigt time dependent strain behavior and (c) standard linear solid model in Kelvin-Voigt time dependent strain behavior.

The deformation behavior is calculated for different elastic moduli starting from 150KPa to 3MPa to investigate the mathematical model behavior for a wide range of elastic moduli using SLS model. This is done in order to understand the behavior in both pre-refractive and post-refractive states of the cornea. The complete behavior of the mathematical model for transient load is presented in Fig. 3.5. It is observed that the behavior is a non-linear behavior.

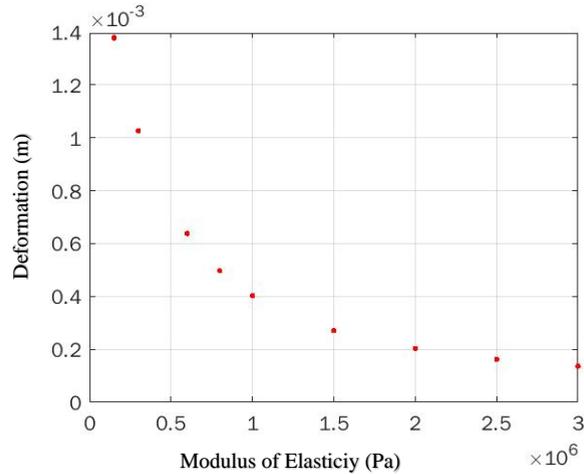

Fig. 3.5 Generalized mathematical model behavior for transient load.

## 3.4. SUMMARY

In this chapter, two mathematical models are proposed to simulate corneal viscoelastic behavior. Their performance is compared when applying a constant and transient load respectively. The simulation time is 5 msec for both models. The constant load is applied for the whole simulation time, while the transient load is applied for about 2 msec.

From the obtained results of the two proposed models and with comparison of these two models with each other, it is concluded that standard linear solid model describes the human corneal behavior in response to constant and transient load more efficiently. This is because of the lack of Kelvin-Voigt model in describing the viscous damping behavior of corneal biomechanics. Also, because the deformation amplitude in Kelvin-Voigt is greater than the reasonable values for the corneal tissue for the same applied load either constantly or transiently.

On the other hand, standard linear solid model predicts a reasonable deformation amplitude for corneal tissue for the same applied load either constantly or transiently. This model also describes the viscous damping behavior of the corneal tissue efficiently.

# Chapter 4 | Finite Element Modelling of Corneal Biomechanics

## 4.1 INTRODUCTION

Cornea is the first translucent layer in the human eye that provides most of its refractive power. Corneal tissue protects human eye against microbial infection and trauma as well [108], [109]. Two thirds of human eye refractive power are accounted for cornea [6], [110] which covers anterior chamber along with pupil and iris. Generating proper image just above the retina is obtained by approximately 43 dioptres of corneal refractive power [7], [111]. Cornea has elliptical shape, where it is thin at the center with gradual thickening as moving to either of its distal peripheries. Normal human eye has central thickness range of about 440 µm to 650 µm with an average value of 540±30 µm [108], [112]. Cornea exhibits stiffness and viscoelasticity against internal and external stress forces that distorts its structure and shape. Corneal stiffness is minimum in radial direction perpendicular to collagen fibrils while its highest stiffness is found to be in the direction of collagen fibrils [9]. Most of corneal strength is due to stroma layer, while weak contribution if obtained from other corneal layers such as epithelium, Descemet's membrane and endothelium layers [9]. Tensile strength tests performed on human cornea shows that anterior stroma part is stronger than posterior part of stroma [108]. In terms of viscoelasticity; tissue response to applied stress is dependent on strain rate. Corneal tissue's strain increases gradually when applying continuous force where collagen fibrils and lamellae exhibits viscous sliding in proteoglycan matrix is responsible for energy loss [109].

Predicting efficiency of refractive laser surgeries and identifying patients with high probability to develop corneal iatrogenic ectasia is achieved by biomechanical investigation for corneal tissue before and after refractive surgery [113]–[115]. Researches have confirmed that diagnosis and monitoring the progression for ectatic corneal disease is significantly related to corneal biomechanics assessment while topographic assessment is still insignificant for diagnosis of such disease [116], [117].

Corneal tissue shape imaging techniques were considered as an alternative for biomechanics imaging techniques. Corneal shape reflects its biomechanical properties, however these properties are appropriately assessed by applying force to corneal tissue and tracking its behavior to this force [118]. More than one corneal tomographic image is needed to properly estimate corneal biomechanics leading to the concept of corneal multimodal imaging [117]. Detection of mild corneal ectasia anomalies is improved by placido disk-based corneal topography [116], [119], [120]. This led to the need for 3D tomographic imaging for anterior chamber generally, giving rise for more detailed structure of corneal shape from 3D images [112], [117], [121], [122]. Improving detection of subclinical ectasia by means of corneal tomography imaging is studied in many papers that include patients with identified subclinical ectasia [123]–[128]. Tendency to develop post-LASIK ectasia is observed have high relevance to corneal tomography parameters as presented by relevant researches involving patients with such complications [129]. As a result of the advance in tomographic imaging techniques the sub-corneal imaging can be performed by two methods, very-high frequency ultrasound [130]–[133], and spectral-domain optical coherence tomography (SD-OCT) [134]–[136]. This led to reaching the order of epithelium layer thickness imaging.

Investigating corneal ectasia hazard in terms of corneal biomechanics is endorsed by later studies that are independent on evaluation of such complication based only on corneal shape [137], [138].

Introduction of non-contact tonometry techniques used by Ocular Response Analyzer (ORA; Reichert Ophthalmic Instruments, Buffalo, NY) in 2005 allowed for estimating corneal in vivo biomechanical properties measurement [139]. Reflection from incident infrared light beam over corneal surface allows tracking of corneal deformation in response to a collimated air puff focused to push about 3-6 mm of cornea central apex [139]–[141]. ORA pressure-derived parameters are reported to have a low significance in diagnosing keratoconus due to their distributions' overlap leading to poor sensitivity value of about 75% [142]–[144]. Although, pressure-derived parameters are of low value for keratoconus compared to healthy eyes [144].

Combining ultrafast corneal imaging along with corneal deformation analysis approaches are proposed and investigated. Supersonic shear-wave imaging as a high speed imaging technique has been investigated in [145].

Feasibility to use ultrafast high resolution ultrasound imaging to evaluate corneal tissue viscoelasticity by means of quantitative maps for ex-vivo porcine cornea using supersonic shear imaging is investigated in [87]. Supersonic shear imaging is achieved by a transducer sound push along with high frame rate B-mode ultrasound imaging procedure of several kilohertz to capture corneal tissue displacement in response to this push. Corneal displacement gives rise to shear wave propagation along the lateral direction of corneal tissue. This results in a shear wave that has an implicit relationship to the investigated tissue's elasticity. This relationship is given by Eqn. 4.1. & 4.2.:

$$C = \sqrt{\frac{\mu}{\rho}} \qquad (4.1)$$

$$\mu = \frac{E}{2(1+\vartheta)} \qquad (4.2)$$

where C is the shear wave speed, $\mu$ is the shear modulus, $\rho$ is the density in Kg/m3, E is the Young's modulus and $\vartheta$ is the Poisson ratio for tissue under investigation respectively.

Supersonic shear imaging yields a non-invasive real-time quantitative elasticity maps for tissues under study.

Another ultrasonic technique that is used to estimate focal corneal biomechanical properties is surface wave elastometry. It is used to assess corneal biomechanics relevant to glaucoma, ectatic disease and refractive surgery. This technique involves measurement of propagation time of ultrasound surface wave between two fixed distance transducers along the radial direction of corneal tissue [146].

In this chapter, an ultrasound based approach to assess corneal biomechanics is proposed. This approach is based on two independent techniques; both relies on transient ultrasonic acoustic force. The first one relies on estimating radial SWS (rSWS) of the corneal tissue. This wave is due to acoustic force applied transiently, leading to quantitative assessment of corneal biomechanics. The second technique estimates the focal peak axial displacement (FPAD) for corneal tissue giving rise to qualitative assessment of corneal tissue biomechanics. This approach has the advantage of using one imaging modality to assess corneal biomechanics with both quantitative and qualitative points of view.

This chapter is organized as follows, introduction about cornea, its biomechanics behavior and different methods for assessment of cornea biomechanics with related brief literature survey is presented in section 1. the proposed methodology demonstrating the corneal FEM generation, Acoustic Radiation Force Impulse generation, acquisition sequence and shear wave speed estimation are introduced in sections 2 and 3. Simulation results are reported in section 4. Finally, discussion and summary sections are presented in section 5 and 6.

## 4.2 METHODOLOGY

In this research, a 3D FEM is implemented to study the effect of post-refractive surgery on corneal biomechanics by COMSOL multi-physics v5.4 software. FEM used in this study is chosen to resemble agar-gelatin phantom for the same tissue under study, where agar is used to simulate scatterers for ultrasound waves and gelatin is used to maintain phantom stiffness [147]. Corneal biomechanics properties are assigned to this model for different elastic moduli [87], [148]–[150]. The proposed technique is adopted from work presented in [88], [151]. Corneal biomechanics are then estimated by speed of the resulting propagating shear wave. A point force generated by a focused ultrasound beam is used to initiate corneal tissue deformation resulting in a shear wave propagation laterally off-axis. The generated force is applied transiently for about 1 msec. seconds. Generated shear wave due to corneal tissue deformation in response to the applied transient force is then tracked using two fixed probing nodes. One node at the focal point of the probe and the second one is laterally distal along the corneal tissue. The resulting deformation is tracked at these two nodes for consecutive time frames. High frame rate is deployed to accurately estimate the resulting wave speed. Wave speed is estimated using MATLAB software version R2019a. Frame rates of several kilohertz to hundred kilohertz are used in order to reduce estimation error of resulting shear wave speed. In this research, two frame rates are used, typically 10Khz and 100Khz. Quantitative elasticity maps for tissue under study are generated using shear wave speed estimation between the focal probing node and any certain node of choice located spatially on the corneal tissue. Focal probing node is dependent on the focal point of the ultrasound transducer during conventional B-mode imaging process. The block diagram of the proposed approach is presented in Fig. 4.1.

### 4.3.1. Corneal FEM Generation

3D FEM is generated for a vertical cross section of complete human eye as shown by Fig. 4.2. Model's geometry and dimensions are set to those from medical literature for average human eye [108], [112]. The FEM material mechanical properties are set to simulate human eye in both pre- and post-refractive surgeries. These properties are listed in Table 4.1.

#### 4.3.1.1. FEM Drawing

FEM is implemented to study the effect of post refractive surgery on cornea biomechanics for the human eye. For feasibility and time saving; the model is implemented to simulate one half of the human eye as a vertical cross section. FEM is drawn in 3D as two intersecting spheres with two shifted centers along the z-axis. One sphere is drawn to represent the complete human eye, while the second sphere represents the human cornea. As human cornea is located at the top of the eye's structure, its part in the model is shifted along the z-axis. The complete (revolved) 3D FEM in different planes is shown in Fig. 4.3. The complete dimensions are listed in Table 4.2. These spheres are drawn as circles' sectors in plane geometry and are revolved along the z-axis for drawing feasibility.

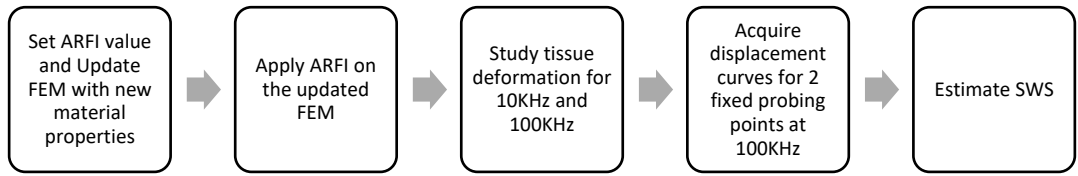

Fig. 4.1 Block diagram of the proposed approach.

Table 4.1 FEM Mechanical Properties.

| Property | | Value |
|---|---|---|
| Young's modulus (range of values) | Pre-refractive (High E in MPa) | 1, 1.5, 2, 2.5 and 3 MPa [149] |
| | Post-refractive (Low E in KPa) | 3, 30, 140, 300, 600 and 800 KPa [148] |
| Poisson Ratio | | 0.499 |
| Heat Capacity at constant pressure | | 2348 |
| Thermal Conductivity | | 0.21 |
| Density | | 911 |

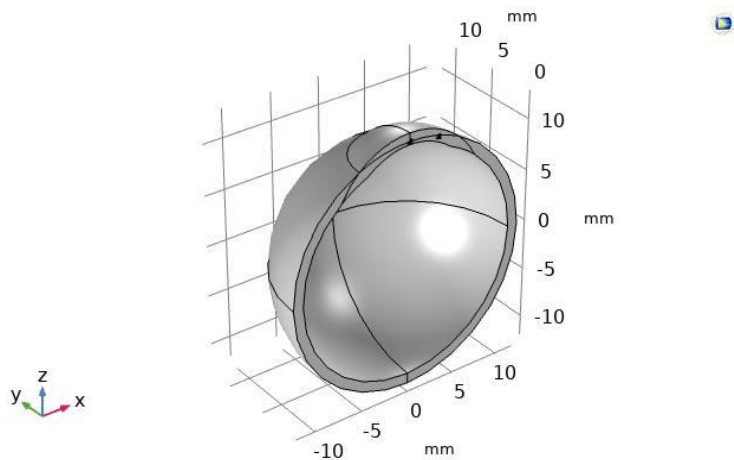

Fig. 4.2 Isometric vertical cross section geometry of average complete human eye.

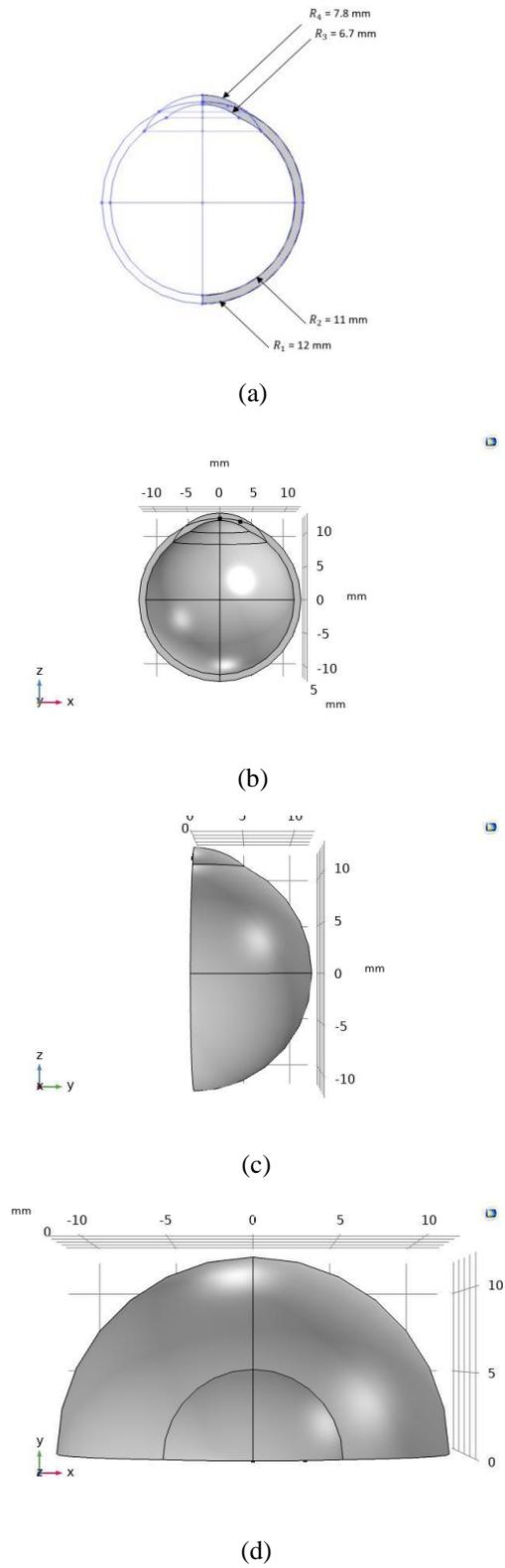

(a)

(b)

(c)

(d)

Fig. 4.3 (a) 2D view, (b), (c) and (d) 3D FEM of human eye in three different plan views.

Table 4.2 Model geometrical dimensions[108], [112].

|  | Circle 1 | Circle 2 | Circle 3 | Circle 4 |
|---|---|---|---|---|
| **Radius (mm)** | 12 | 11 | 6.7 | 7.8 |
| **Sector Angle (degrees)** | 180 | 180 | 90 | 90 |
| **Center X (mm)** | 0 | 0 | 0 | 0 |
| **Center Y (mm)** | 0 | 0 | 5 | 5 |

### 4.3.1.2. FEM Solid Mechanics

FEM is implemented to act as an isotropic linear visco-elastic model. The model is simulating Standard Linear Solid mathematical model. Fixed constraints at the lower boundaries of cornea are set to simulate cornea iris ciliary muscles. A point load is set at the center of corneal tissue to simulate ARFI from ultrasound scanner, this is shown in Fig. 4.4.

### 4.3.1.3. FEM Meshing

FEM is meshed as a triangular physics-controlled mesh by using COMSOL software. The mesh is chosen to be narrower at the volume of interest; i.e. the cornea volume, for accurate measurements as shown in Fig. 4.5.

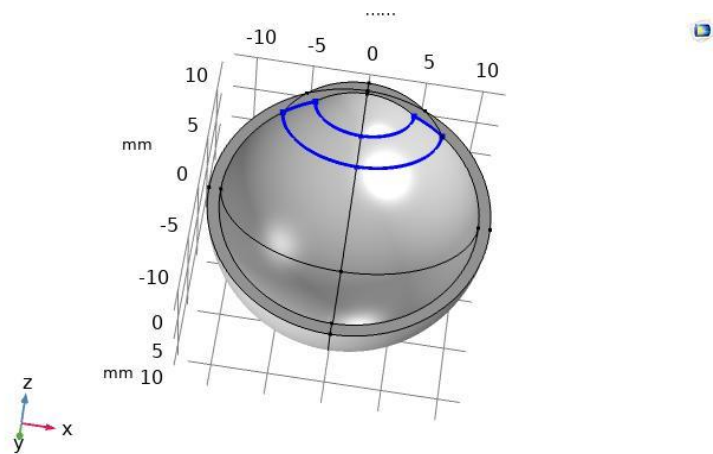

(a)

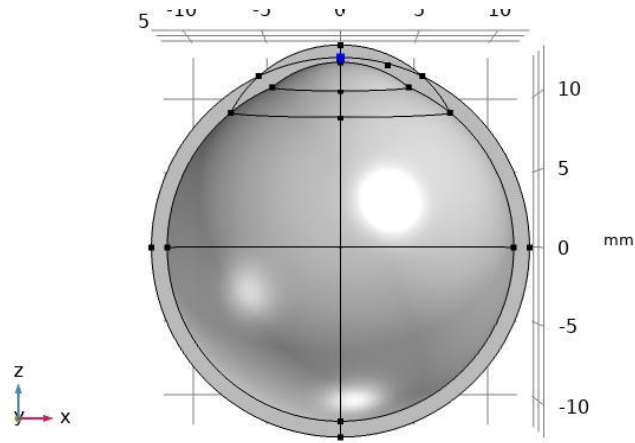

(b)

Fig. 4.4 a) Fixed boundaries as iris ciliary muscle, b) Point load as ultrasound transient force.

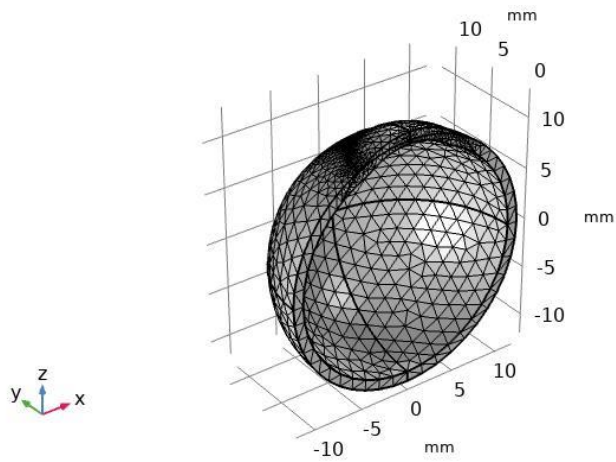

(a)

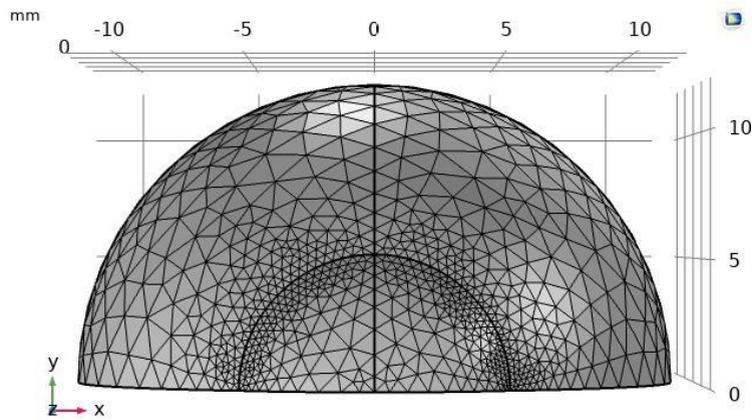

(b)

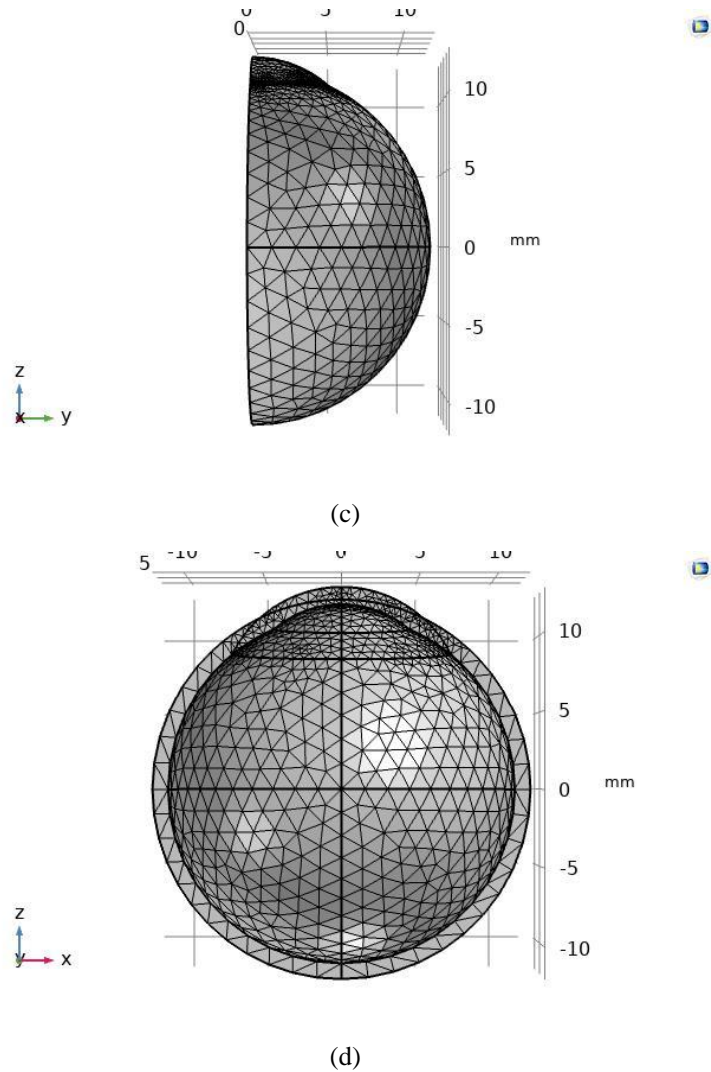

(c)

(d)

Fig. 4.5 Corneal FEM Mesh (a) Isometric view, (b), (c) and (d) are plan, side and elevation views respectively.

### 4.3.2. Acoustic Radiation Force Impulse Generation

Acoustic Radiation Force (ARF) is considered as an internal tissue actuator that leads to quantitative elasticity measurement. The resulting push from ARF impulse is highly localized where the push power damps exponentially around the push location giving rise to accurate mechanical properties estimation. Acoustic Radiation Force (ARF) from ultrasound transducer generates local tissue displacements and shear waves. It is the resulting pressure from ARF that induces tissue displacement. Pressure field is simulated for a linear array ultrasound transducer that is modelled by FOCUS toolbox in MATLAB (Focus is a toolbox in MATLAB for simulating ultrasound transducer pressure, intensity and force field and spatial distribution). Transducer parameters are listed in Table 4.3. The resulting force field is then applied on the FEM to simulate its effect on tissue biomechanics. Induced tissue displacements; forming shear wave; are then tracked by B-mode ultrasound imaging procedure after applying force field for about 1-10 microseconds. Tracking these displacements give rise to an estimation for shear wave speed which is a characteristic for each specific tissue and is related to tissue biomechanics.

Transducer's imaging beam wave apodization is used to reduce grating and side lobes of the resulting sound pressure wave. 2D Pressure field is computed using Fast Near Field Method (FNM) across the plane of excitation inside the soft tissue. Figure 4.6. shows the ultrasound wave pressure distribution along the focal axial and lateral direction for the ultrasound transducer. The FEM behavior to this pressure field at the focal plane is investigated, and the resulting shear wave is tracked. The focal point is set to 28.5 mm in the axial direction. The highest sound pressure value is observed around the preset focal point along the axial line passing through the pressure field as shown in Fig. 4.6. The axial beam intensity and force are calculated as well and presented in Fig. 4.7. and Fig. 4.8. respectively. The ARFI is obtained by Eqn. 4.3.:

$$F = \frac{2\alpha I}{C} \tag{4.3}$$

where $\alpha$ is the acoustic absorption coefficient in $(\frac{dB}{cm.MHz})$, $I$ is the temporal average intensity of the beam in $(\frac{Watts}{m^2})$, and $C$ is the speed of sound in tissue in $(\frac{m}{sec.})$.

Table 4.3 Ultrasound Transducer Parameters.

| Parameter | Value |
|---|---|
| Number of elements | 192 |
| Elements' width | 170 µm |
| Kerf | 30 µm |
| Focal depth | 28.5 mm |
| Center frequency (fo) | 12 MHz |

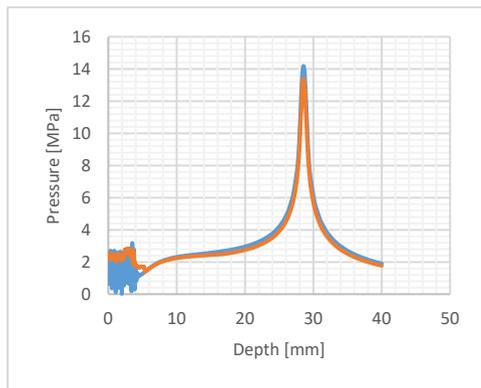 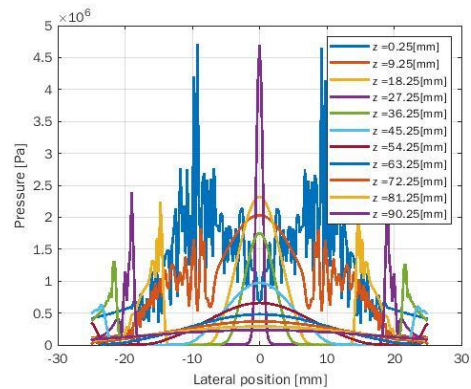

(a)

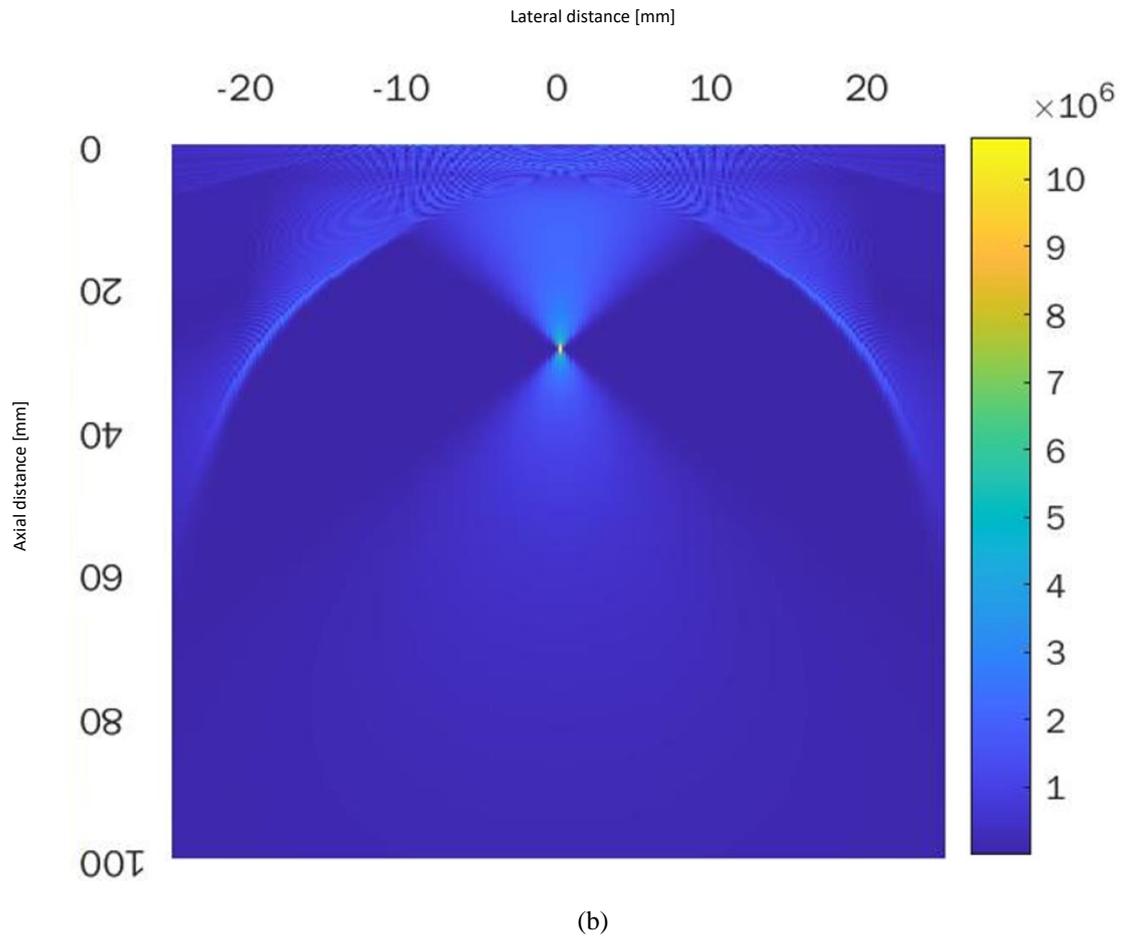

(b)

Fig. 4.6 Pressure distribution along (a) focal axial (centerline: blue, averaged: red) and lateral direction of the transducer, (b) 2D pressure field map.

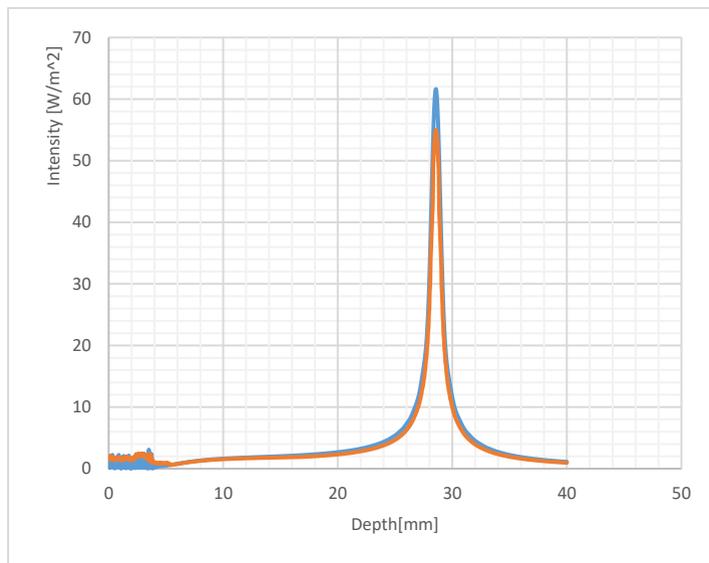

Fig. 4.7 Intensity distribution along focal axial direction of the transducer (centerline: blue, averaged: red).

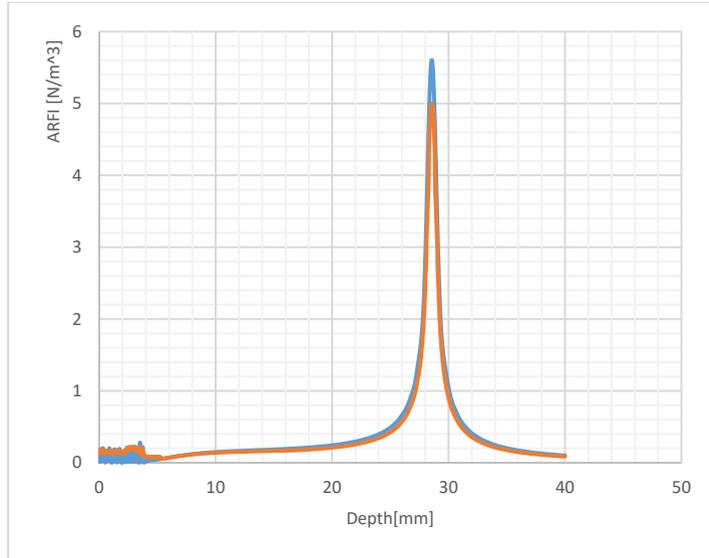

Fig. 4.8 ARFI distribution along focal axial direction of the transducer (centerline: blue, averaged: red).

### *4.3.3.  Acquisition Sequence*

A plane wave is sent to the medium under investigation so as to acquire a reference frame for tissue under study before deformation takes place. B-mode imaging procedure is used for obtaining this reference frame, and the subsequent deformed frames as well. Deformed frames are acquired by the same B-mode imaging procedure but with a much higher frame rate in order to accurately track the induced deformation without aliasing. Figure 4.9. shows the complete timing diagram of the whole acquisition sequence for two acquisition periods.

The ARFI is applied just after the reference frame is acquired. It is applied for about 1-10 microseconds. at the focal zone of the ultrasound transducer. In this study, two frame rates are used to compare between estimated results. This comparison leads to the optimum frame rate for the tissue under investigation. These two frame rates are typically 10 KHz and 100 KHz. Quantitative elasticity map can be constructed by continuous adaptation of focal zone and the probing node radially across the spatial domain for the tissue under study.

### *4.3.4.  Shear Wave Speed Estimation*

Time-To-Peak (TTP) displacement technique is utilized to estimate the shear wave speed inside corneal tissue. The estimation process is defined as the difference between the TTP at the focal node and the TTP at the distal probing node along the radial path of the corneal tissue. Each node yields a displacement magnitude profile with respect to time. TTP is characteristic for every tissue, where the arrival time of the peak is related to tissue biomechanics. There is no dependence on ARFI, where ARFI changes only the amplitude of the deformation for tissue under study with correspondence to exponential damping effect happening accompanying localized excitation methods. Simply, the shear wave speed is estimated as the distance difference between two nodes divided by the time difference at which the TTPs are occurring for these two nodes. However, for a good estimation of the shear wave speed, more nodes have to be involved in probing where the average is calculated.

SWS is calculated using Eqn. 4.4.:

$$C_{avg} = \frac{\Delta x}{\Delta t} \tag{4.4}$$

where $C_{avg}$ is the average velocity across the lateral position in ($\frac{m}{sec.}$), $\Delta x$ and $\Delta t$ are the difference in distance between probing nodes and difference in times of peak arrival at these nodes respectively.

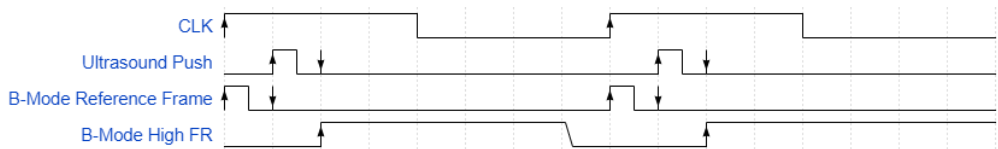

Fig. 4.9 Timing diagram for two periods of the acquisition sequence.

## 4.4   RESULTS

3D displacement maps with stress overlapped as color coded for 3MPa at 100KHz frame rate for cornea model at successive time stamps are presented in Fig. 4.10. These maps show the progress of shear wave propagation across cornea agar-gelatin phantom starting from the reference frame through deformed frames within simulation time. Similar 3D displacement maps are obtained for cornea agar-gelatin FEMs having different elasticity moduli, typically 3KPa, 30KPa, 140KPa, 300KPa, 600KPa, 800KPa, 1MPa, 1.5MPa, 2MPa and 2.5MPa.

1D displacement magnitude profiles for each FEM are extracted from cornea 3D displacement maps as focal and radial probing nodes' magnitude displacement profiles as shown in Fig. 4.11 for post-refractive FEMs and Fig. 4.12 for pre-refractive FEMs. Their corresponding focal peak axial displacements, power fitting, logarithmic fitting, squared error for both fittings and their corresponding Mean Square Errors (MSE) are listed in Table 4.4. Focal peak axial displacements along with curve fitting for these discrete points with the power and logarithmic equations are presented in Fig. 13. Curve fitting with power equation yields Eqn. 4.5. that can be used for estimating points that are not included in the experiment; i.e. Young's moduli that are not included in the study:

$$Y = 51.388 \, (X)^{(-0.322)} \qquad (4.5)$$

Curve fitting with logarithmic formula yields Eqn. 4.6.:

$$Y = 6.1256 - 0.395 \ln(X) \qquad (4.6)$$

where Y is the value of the peak axial displacement occurring at the focal probing node and X is the value of Young's modulus for this particular peak axial displacement.

The focal and radial probing nodes magnitude displacement profiles for each agar-gelatin FEM of the corneal tissue involved in shear wave speed estimation are obtained as well. These profiles are obtained at 100 KHz frame rate as shown in Fig. 4.11 and Fig. 4.12.

Average estimated shear wave speeds along with theoretical values from mathematical formula, percentage error and accuracy for both frame rates; 10KHz and 100KHz respectively are listed in Table 4.5 and are presented in Fig. 4.14. and Fig. 4.15.

A bar chart representing average estimated shear wave speeds along with theoretical wave speed values for the two frame rates for each FEM is presented by Fig. 4.14. Another bar chart for percentage errors between estimated wave speeds and theoretical values for each FEM at both frame rates is presented in Fig. 4.15.

The 2D spatial map for the temperature rise due to the pushing is calculated using bio-heat transfer equation [152] and is presented in Fig. 4.16.

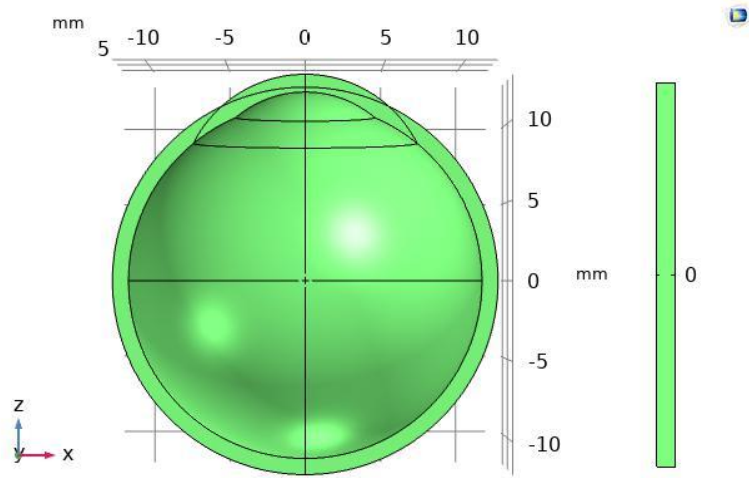

a. Reference frame at $t_0 = 0$ sec.

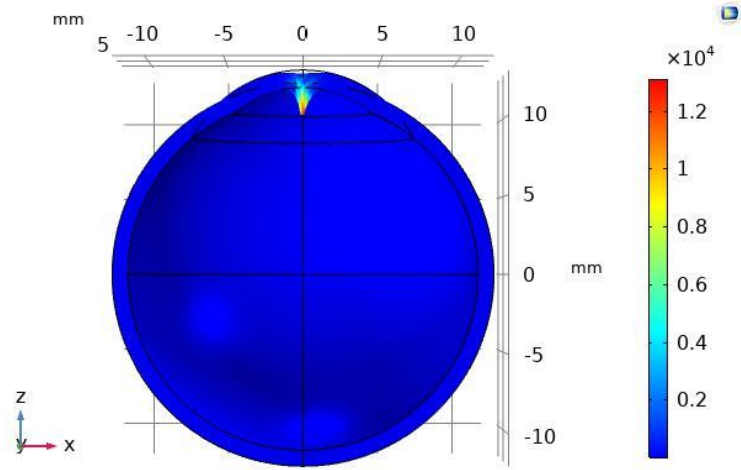

b. ARFI at $t_1 = 0.002$ sec.

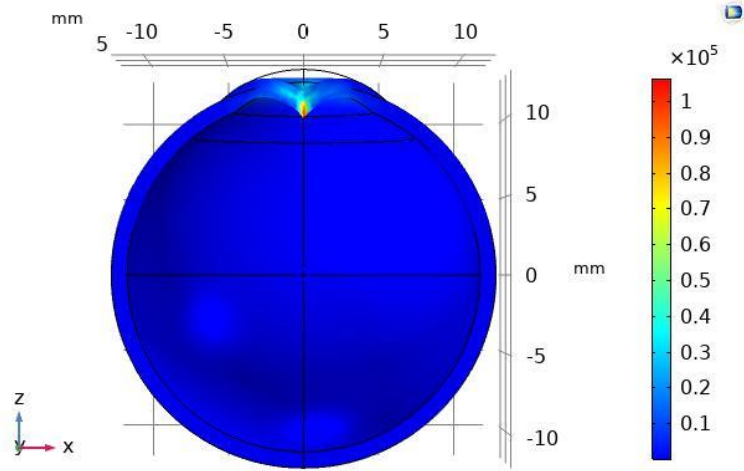

c. at t₂ = 0.005 sec.

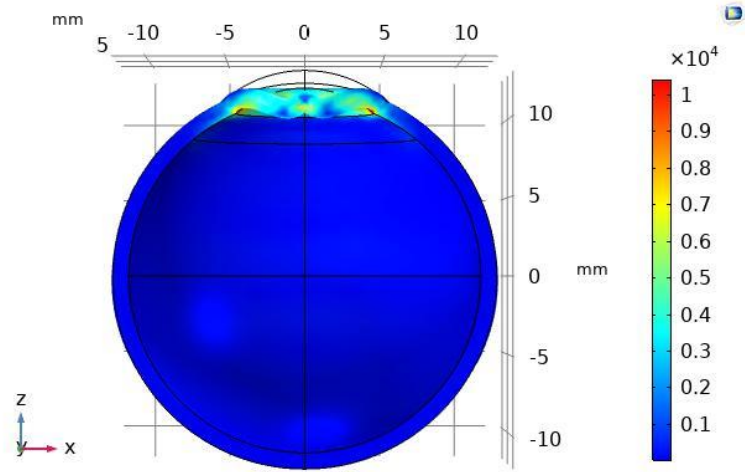

d. at t₃ = 0.008 sec.

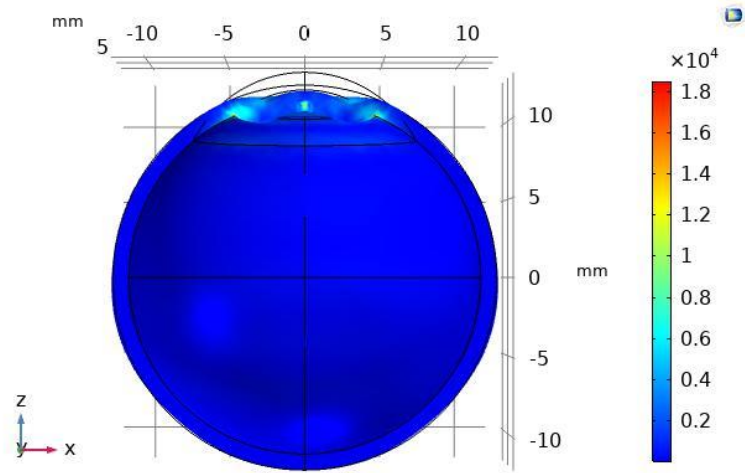

e. at t₄ = 0.01 sec.

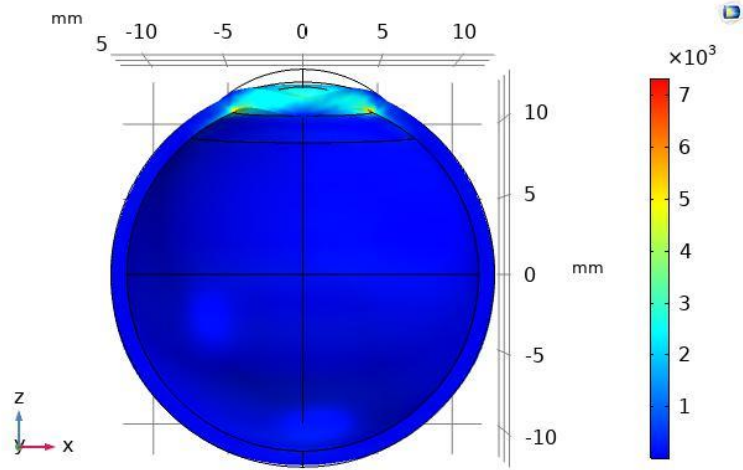

f. at $t_5 = 0.013$ sec.

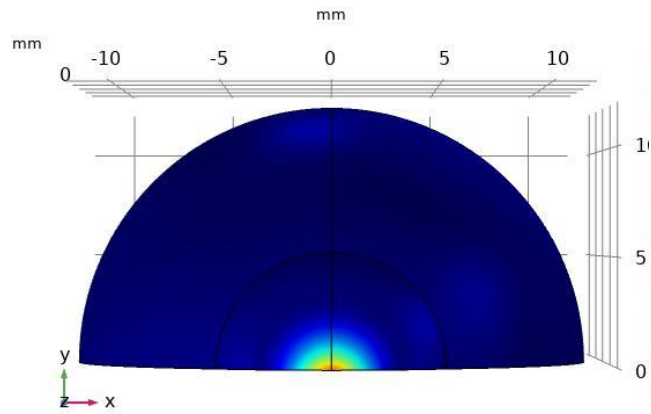

g. at $t_1 = 0.002$ sec.

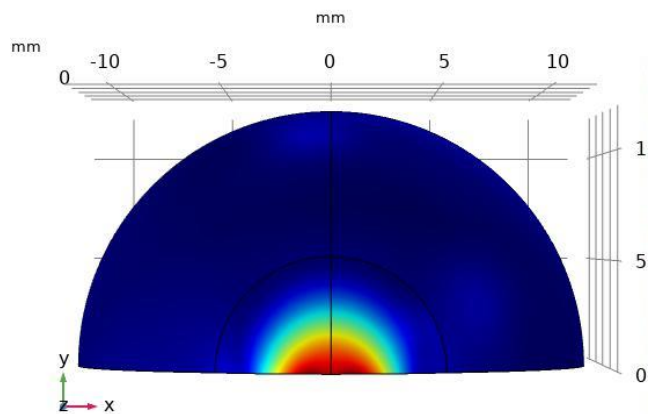

h. at $t_2 = 0.005$ sec.

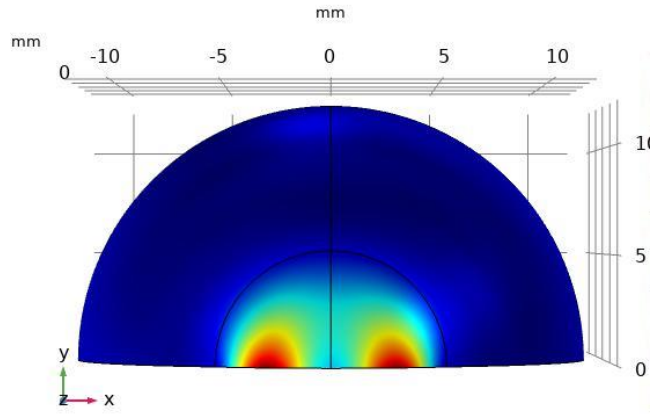

i. at $t_3$ = 0.008 sec.

Fig. 4.10 3D Displacement maps for 3MPa agar-gelatin FEM at 100 KHz frame rate at different simulation time stamps, a-f: vertical view, g-i: plan view.

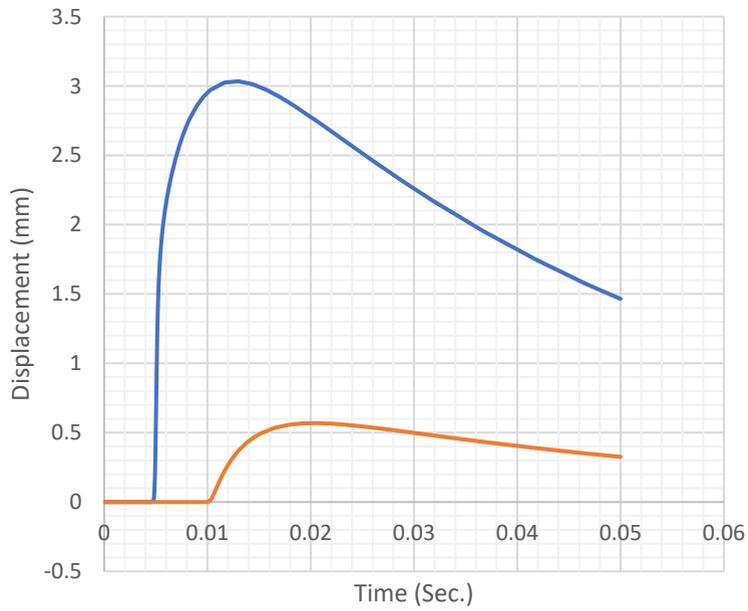

a. 3KPa FEM.

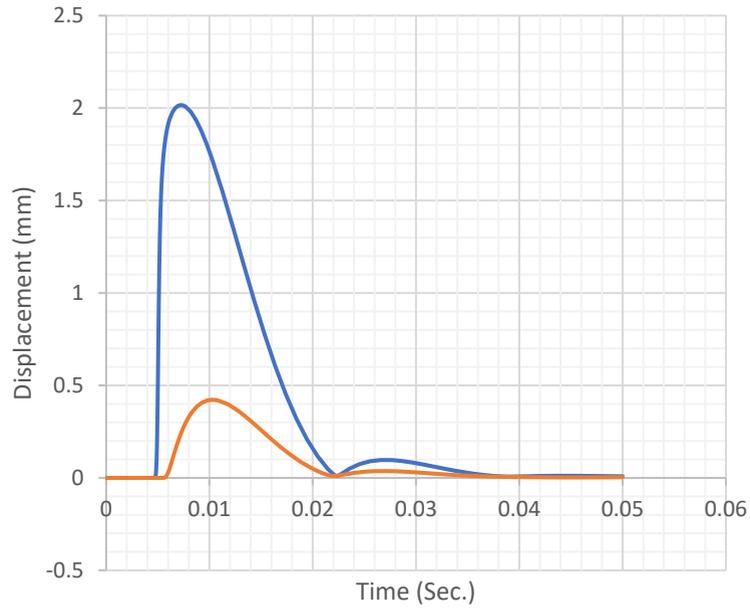

b. 30KPa FEM.

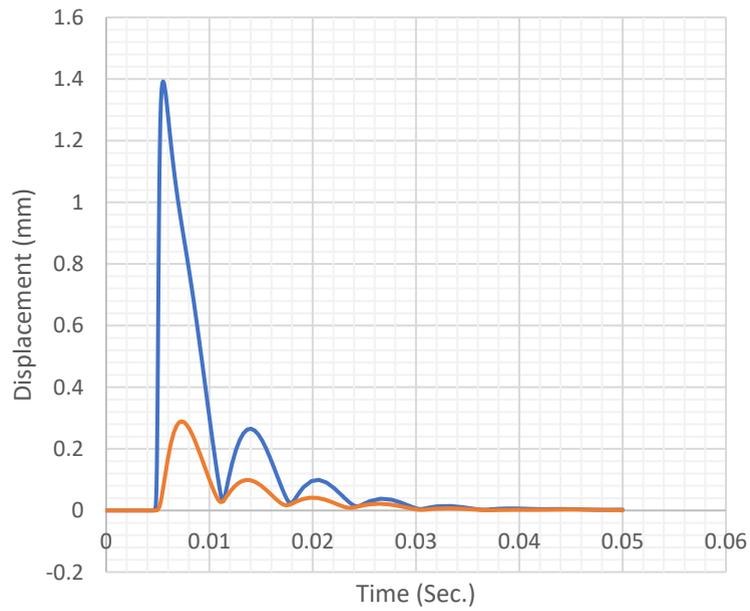

c. 140KPa FEM.

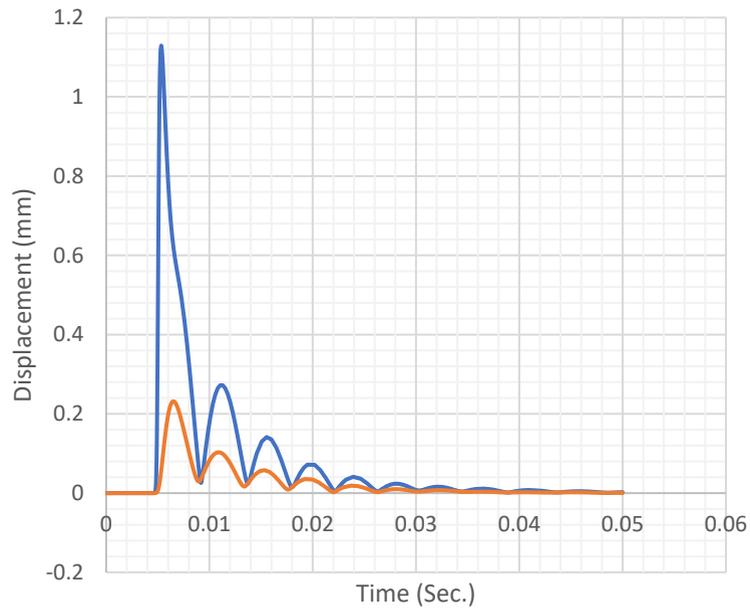

d. 300KPa FEM.

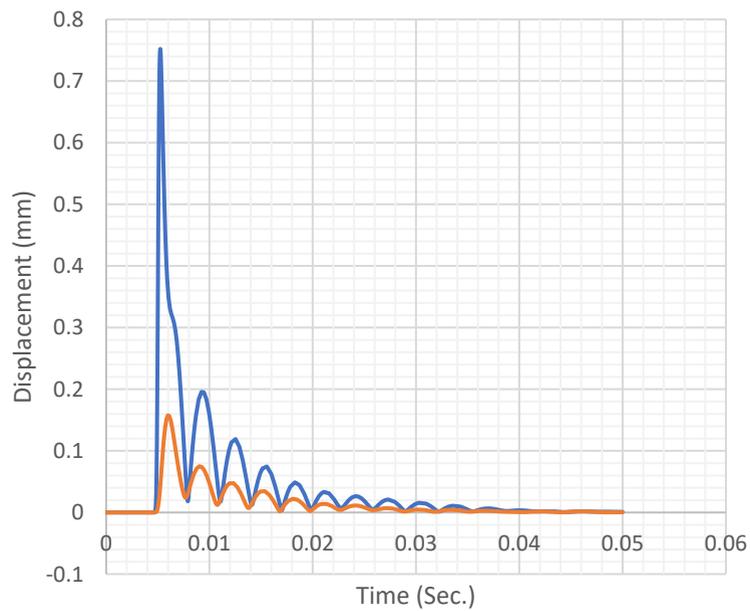

e. 600KPa FEM.

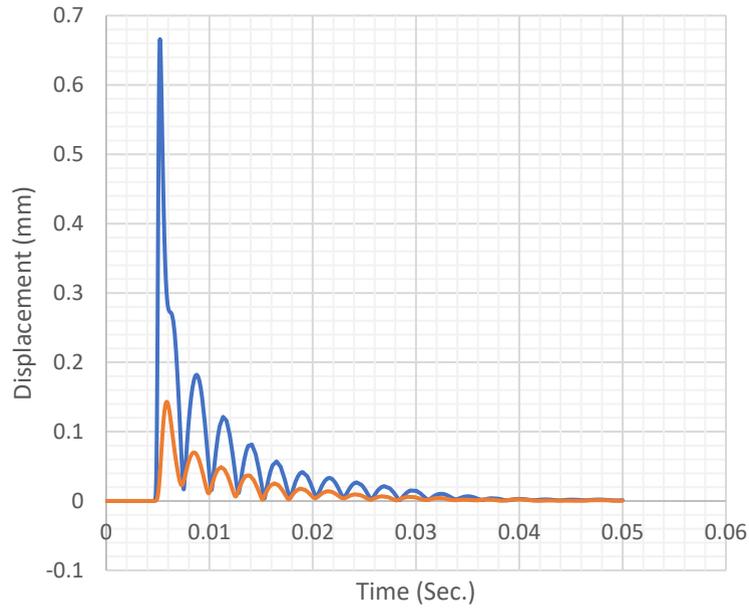

f. 800KPa FEM.

Fig. 4.11 Displacement magnitude profiles for focal and Radial probed nodes for each cornea agar-gelatin post-refractive FEM (Blue line = focal node displacement, Red line = radial node displacement).

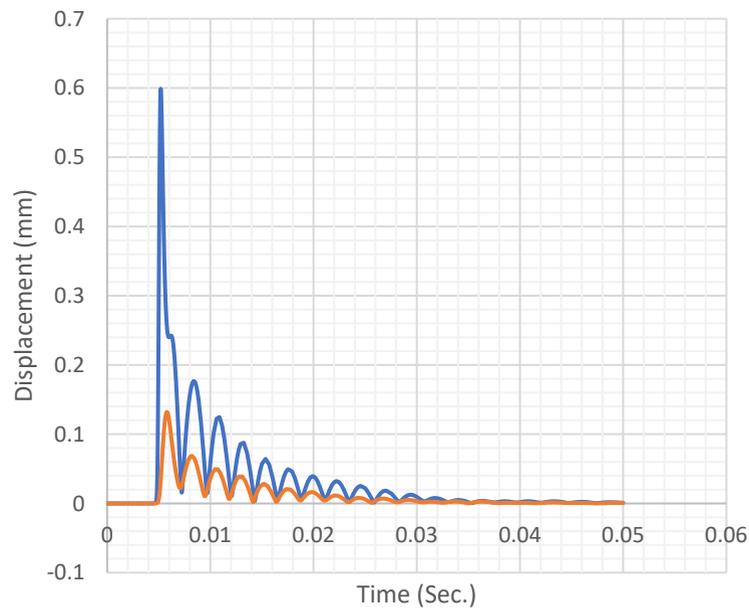

a. 1MPa FEM.

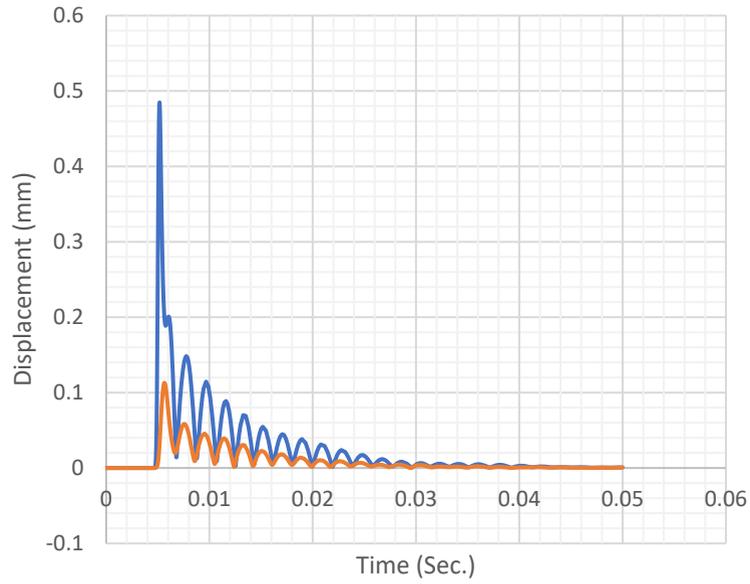

b. 1.5MPa FEM.

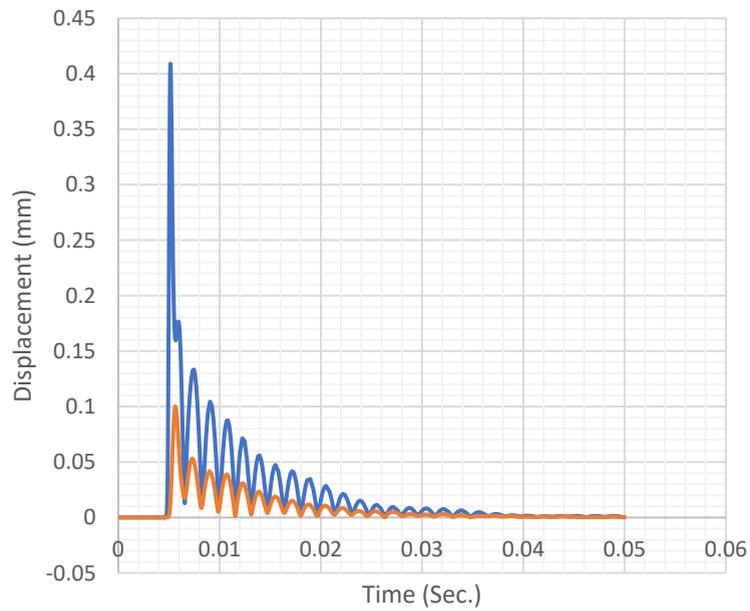

c. 2MPa FEM.

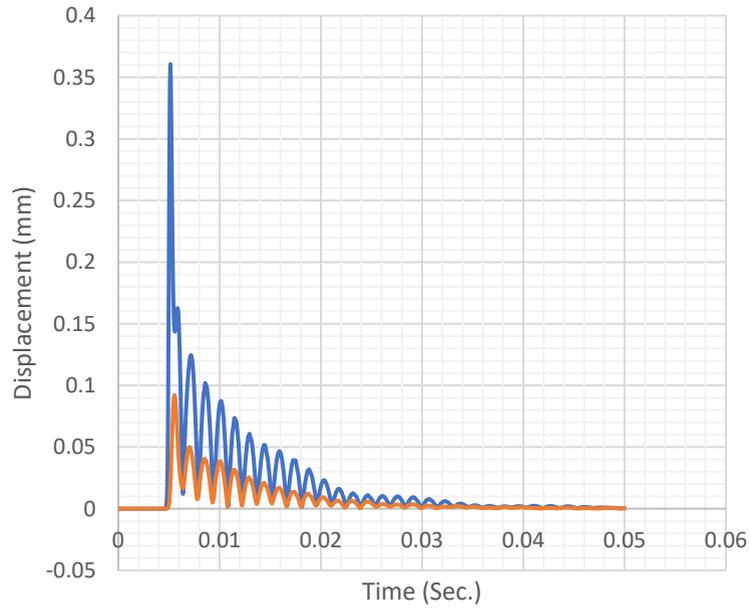

d. 2.5MPa FEM.

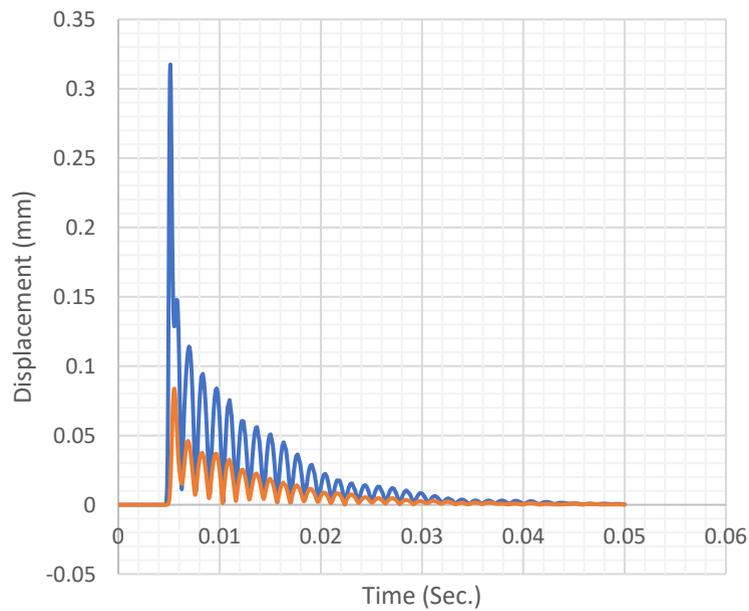

e. 3MPa FEM.

Fig. 4.12 Displacement magnitude profiles for focal and Radial probed nodes for each cornea agar-gelatin pre-refractive FEM (Blue line = focal node displacement, Red line = radial node displacement).

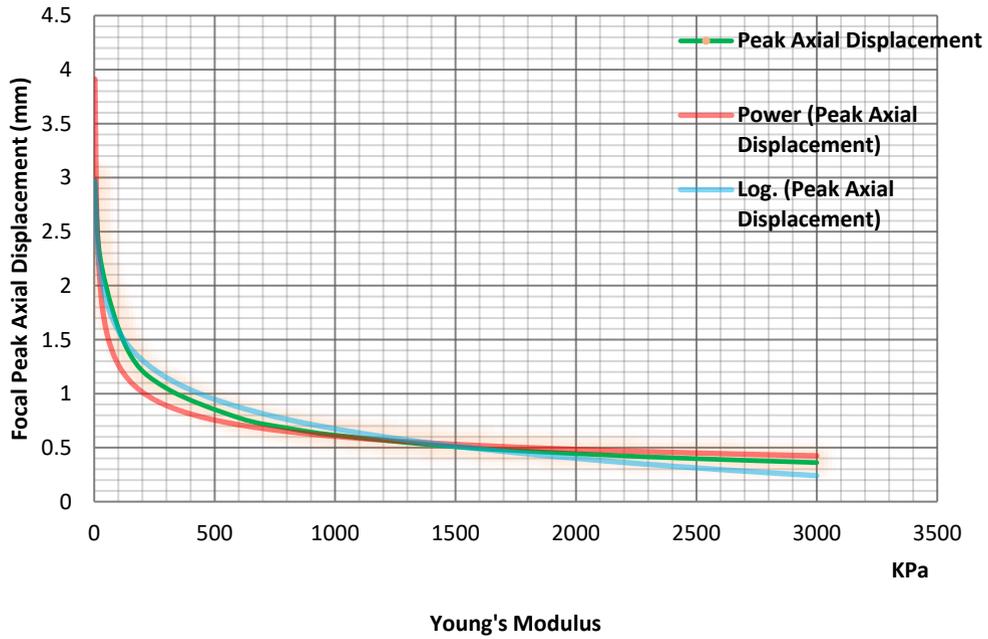

Fig. 4.13 Focal peak axial displacements, power and logarithmic fitted curves for each of the agar-gelatin FEMs involved in the study.

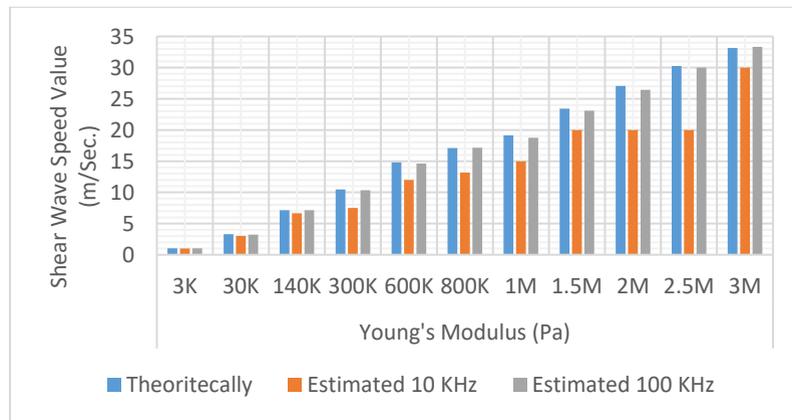

Fig. 4.14 Shear Wave Speed value obtained theoretically and by estimation at both 10 KHz and 100 KHz.

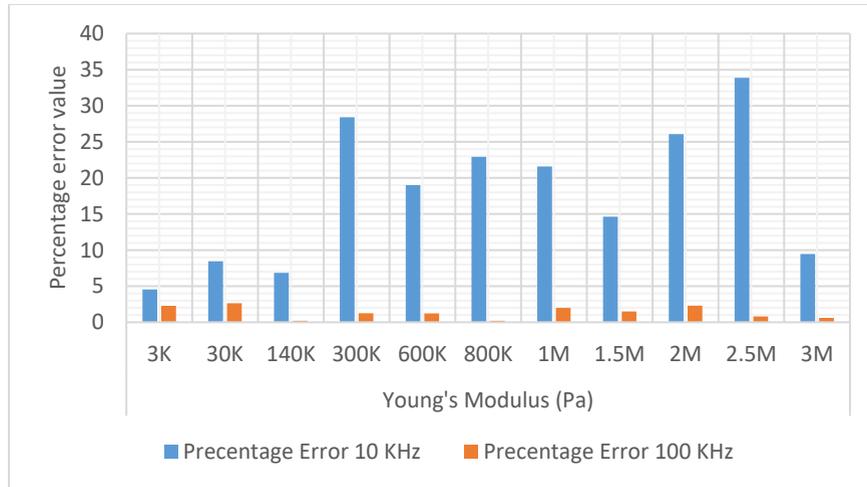

Fig. 4.15 Percentage error of estimation with respect to theoretical value of SWS at both 10 KHz and 100KHz frame rates.

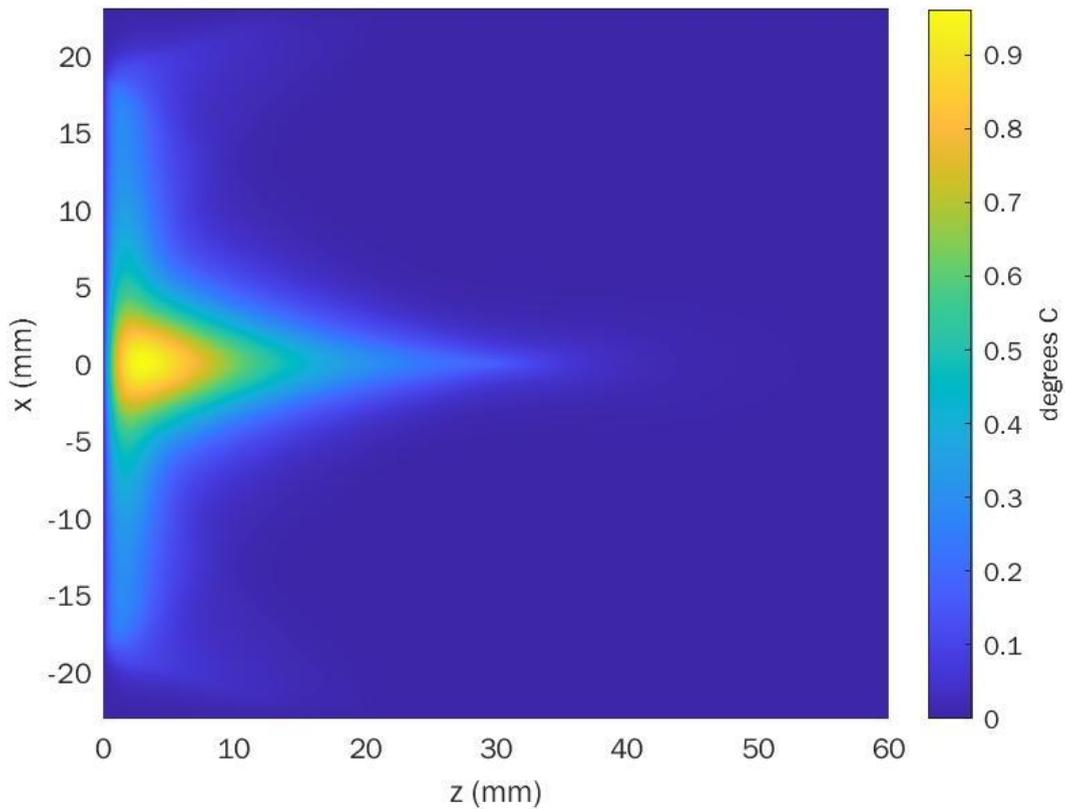

Fig. 4.16 Temperature rise due to pushing beam.

Table 4.4 Peak axial displacements and their corresponding power fitting, logarithmic fitting, square error and mean square error values for each of the agar-gelatin FEMs involved in the study.

| Young's Modulus (Pa) | Peak Axial Displacement (mm) | Power Fitting | Logarithm ic Fitting | Square Error(Expone ntial) | Square Error(Logarit hmic) |
|---|---|---|---|---|---|
| 3KPa | 2.9772 | 3.8976 | 2.9630 | 0.8472 | 0.0002 |
| 30KPa | 2.1895 | 1.8569 | 2.0535 | 0.1106 | 0.01850 |
| 140KPa | 1.3996 | 1.1307 | 1.4450 | 0.0722 | 0.0020 |
| 300KPa | 1.0494 | 0.8847 | 1.1440 | 0.0271 | 0.0089 |
| 600KPa | 0.7762 | 0.7077 | 0.8702 | 0.0046 | 0.0088 |
| 800KPa | 0.6816 | 0.6451 | 0.7566 | 0.0013 | 0.0056 |
| 1M | 0.6155 | 0.6003 | 0.6684 | 0.0002 | 0.0028 |
| 1.5M | 0.5077 | 0.5269 | 0.50831 | 0.0003 | 3.48353E-07 |
| 2M | 0.4444 | 0.4802 | 0.3946 | 0.0012 | 0.0024 |
| 2.5M | 0.3980 | 0.4469 | 0.3065 | 0.0023 | 0.0083 |
| 3M | 0.3615 | 0.4215 | 0.2345 | 0.0035 | 0.0161 |
| | | | | **MSE** = 0.0973 | **MSE** = 0.0067 |

Table 4.5 Theoretical, estimated velocities at 10hkz, 100khz, percentage error and accuracy for each FEM at each frame rate.

| From Mechanical Model | | | | | | | | |
|---|---|---|---|---|---|---|---|---|
| | | Theoretically | Estimated | | Precentage Error(%) | | Accuracy | |
| | | | 10 KHz | 100 KHz | 10 KHz | 100 KHz | 10 KHz | 100 KHz |
| **Young's Modulus (Pa)** | 3K | 1.0477 | 1 | 1.0238 | 4.5536 | 2.2733 | 95.446 | 97.726 |
| | 30K | 3.3131 | 3.0333 | 3.2258 | 8.4455 | 2.6362 | 91.554 | 97.363 |
| | 140K | 7.1572 | 6.6666 | 7.1428 | 6.8539 | 0.2006 | 93.146 | 99.799 |
| | 300K | 10.4770 | 7.5 | 10.3448 | 28.415 | 1.2624 | 71.584 | 98.737 |
| | 600K | 14.8168 | 12 | 14.6341 | 19.011 | 1.2330 | 80.988 | 98.766 |
| | 800K | 17.109 | 13.18 | 17.1428 | 22.924 | 0.1977 | 77.075 | 99.802 |
| | 1M | 19.128 | 15 | 18.75 | 21.582 | 1.97855 | 78.4171 | 98.0214 |
| | 1.5M | 23.4274 | 20 | 23.0769 | 14.630 | 1.49640 | 85.3697 | 98.5036 |
| | 2M | 27.0517 | 20 | 26.4285 | 26.067 | 2.30361 | 73.9324 | 97.6963 |
| | 2.5M | 30.2447 | 20 | 30 | 33.872 | 0.80927 | 66.1271 | 99.1907 |
| | 3M | 33.1314 | 30 | 33.3333 | 9.4516 | 0.6092 | 90.5483 | 99.3907 |

## 4.5 DISCUSSION

Due to the small dimensions of corneal tissue in reality, it is difficult to deploy more than two probing points along the radial direction inside the cornea to estimate the shear wave speed accurately with ordinary lateral resolution of ultrasound transducers. Deploying more probing points inside corneal tissue implies reduced spatial distance between these points. Reduction in spatial distances leads to better estimation of shear wave speed. Yet, this reduction is subjected to dependence on the lateral resolution of the ultrasound transducer. A trade-off between better estimation and the manufacturing limitation takes place, where due common existent lateral resolution of transducers, two probing points are considered to be of appropriate choice.

For 3MPa FEM, which is presented in Fig. 4.10 an elevated transient stress is shown at the focal point where the ARFI is applied. Tissue deformation is at its peak value at the focal point as well with taking into consideration the temporal domain. This elevated stress along with the deformation (from figures) are dampened rapidly both in spatial domain; as shown by 3D displacement maps and in temporal domain as shown from the 1D displacement profiles; as the shear wave propagates away. This is confirmed by observation of 1D displacement profiles for both focal and radial probing nodes for each FEM.

The force is rapidly dampened with time which indicates that the tissue absorbs the force efficiently reducing the resonance effect on cornea. Hence the time period during which the stress resulting from ARFI is reduced eliminating the possibility to induce corneal rupture due to ARFI. Also, rapid damping in time and in spatial domain confirms high localization of ARFI, thus reducing the surrounding tissue damage and increasing the accuracy of estimation. The same applies for all other stated FEMs in this study.

From Table 4.5 along with bar chart presented in Fig. 4.14. We can see that estimation of SWS at 100KHz is more accurate than estimation found at 10KHz using the proposed method depending on the tracking of SWS utilizing 1D displacement profiles at both focal and radial probing node. Actually, estimation using any of 10KHz or 100KHz is nearly the same till Young's modulus is greater than 140KPa, it is at this elasticity value where estimation using 100KHz gives more optimum results than using 10KHz. It is also clear that percentage error from estimation is nearly constant and is minimum at 100KHz as well. While estimation at 10KHz yields high fluctuating percentage error. Nearly constant percentage error or even a fitted percentage error by mathematical formula, is useful when there is a need for precise error cancellation. This can be achieved by estimation at 100KHz on contrary to estimation at 10 KHz.

From Table 4.5 the accuracy of estimation using 10KHz is at its highest value of approximately 95.5% at 3KPa FEM. However, accuracy of estimation for the 100KHz is observed to be maximum at 800KPa with approximately 99.8% of the theoretical calculated value. Also we can see that the accuracy range for the 100KHz is about 2.4% whereas for the 10KHz is approximately 29% with maximum accuracy of about 95.5% and minimum accuracy of 66.1%.

Focal peak axial displacements for corresponding Young's moduli FEMs that are estimated using 100KHz frame rate are listed in Table 4.4. Two curve fitting formulas are obtained from these discrete data points, namely; power and logarithmic fitting. Their corresponding squared errors and MSEs are calculated and listed in Table 4.4. Mean Square Error values for both fitting formulas predicts the winner formula for predicting elasticity moduli that are not involved in this study between 3KPa and 3MPa based on the estimated data points collected from simulation. According to MSE values, logarithmic fitting can describe the data more accurately and predict missing non-involved data points with minimal error with MSE of approximately 0.006. Yet, power fitting loses the competition against logarithmic fitting with MSE of approximately 0.09.

Focal peak axial displacements along with both fittings; power and logarithmic typically; are plotted in Fig. 13. It is clear from the figure that logarithmic fitting is better for the overall prediction with minimum error. Yet, with a closer look at the plot it is observed that for models with elasticity lower than 1.5MPa, logarithmic fitting is optimum, and for models with elasticity higher than 1.5MPa power fitting is optimum. At 1.5 MPa both fitting approximately gives the same value of 0.51 mm compared to 0.5 mm obtained from the FEM simulation.

Fitting can be used in a reversed process to estimate the most optimum elasticity modulus for corneal tissue under investigation, with only prior knowledge of picked up peak axial displacement value at the focal node. This focal peak axial displacement value is fed to the proper formula from both formulas to estimate corresponding Young's modulus.

For tissues pre-refractive surgery, as reported in literature, where tissue's Young's modulus is considered to be of order of MPa, it is optimum to use power fitting to obtain an accurate estimation. On the other hand, post-refractive surgery tissue, as reported in literature, tends to have lower Young's modulus than in order of fractions of MPa or in order of KPa. Hence, it is optimum to use logarithmic fitting to obtain an accurate estimation.

This method of estimation is endorsed when using ultrasound transducers with low temporal resolution, typically lower than 100KHz. Involving this method eliminates the dependence on lateral and temporal resolutions of the transducer as they are required by SWS estimation method as well. It is now a matter of axial resolution dependency for the utilized transducer. It also eliminates the need for huge computation needed by SWS estimation method.

In terms of temperature rise due to pushing beam, the increase in temperature is concentrated at the proximal area of the transducer forming a delta shaped area of temperature of about 0.9 C. Hence, the temperature rise is concentrated in front of the transducer directly. From literature, there is a debate about the acceptable temperature rise due to ultrasound waves. As in [153], a temperature rise from 1 C to 2 C is acceptable for all body organs. However, tissue temperature may increase by as much as 10 C, raising safety concerns even though the acoustic output was still within the Food and Drug Administration's recommended maximum output exposure level for diagnostic ultrasound [154]. Moreover, The FDA regulates the temperature rise (TR) of tissue to be no more than 6 C as stated by [155]. In general, the anterior chamber contains a watery-like fluid called aqueous humor, and the posterior chamber contains a gel-like fluid called vitreous humor. fluids inside the human eye either in the anterior chamber or the posterior chamber act as a heat sink for any generated heat. By consequence, the thermal effect induced due to the acoustic radiation force impulse is reduced by these fluids.

In terms of the mechanical index, FDA regulates the acceptable mechanical index to be in order of 1.9 and to be defined as spatial-peak value of the peak rarefactional pressure, derated by 0.3 dB/cm-MHz at each point along the beam axis, divided by the square root of the center frequency. Our calculations show that the obtained mechanical index to be 1.2 which is lower than the regulated value. Our value is less than 1.4 that is concluded in [154].

## 4.6  SUMMARY

In this chapter, estimating shear wave speed for human corneas pre- and post-refractive surgery is investigated. Eleven different elastic moduli FEMs are constructed to simulate human cornea behavior using finite element analysis. Transient focal acoustic radiation force impulse is applied to each of the eleven FEMs to induce corneal tissue deformation giving rise to shear wave propagation. Generated shear wave is tracked using B-mode ultrasound imaging procedure with two high frame rates; typically, 10 KHz and 100 KHz. Two fixed probing nodes are used to track the induced shear wave, focal node and radial node. This lack in utilizing more probing nodes is due to small dimensions of the human cornea in addition to manufacturing limitations of ultrasound transducers leading to relatively low transducer lateral resolution.

Shear wave speed is estimated by obtaining wave arrival time difference between the probing nodes divided by the spatial distance between them. The estimated speeds are related to the elastic moduli of FEMs under study.

Moreover, the focal peak axial displacements for each FEM are calculated which are used to fit curves between peak axial displacements and corresponding elastic moduli inducing them.

Both methods can be used to estimate cornea elasticity pre- and post-refractive surgery. Yet, focal peak axial displacement method is recommended for ultrasound transducers with low frame rate capabilities. Shear wave speed estimation method is recommended for high frame rate capabilities.

Also, focal peak axial displacement method is recommended for qualitative measurement of tissue under study as it involves single node for elasticity estimation. While shear wave speed method is recommended for quantitative measurement for whole cornea elasticity as either the focal node or radial node locations' can be modified inside tissue under study. Modifying these nodes locations can be set to cover a spatial matrix covering tissue under study yielding corresponding quantitative elasticity map.

# Chapter 5 | Corneal Biomechanics Assessment using High Frequency Ultrasound B-Mode Imaging

## 5.1 INTRODUCTION

Cornea is the transparent component of human ocular system that acts as a protective part enveloping other human ocular components such as lens. It preserves shape of human eye and accounts for most of its refractive power [148], [156]. Shape, transparency and uniformity of human cornea are important factors when assessing its role to vision. However, human vision degradation is caused by cornea shape alteration due to age, corneal diseases and refractive surgery [148]. Developing post-surgery ectasia is common among patients undergoing refractive surgeries. Hence, assessing corneal biomechanics post-refractive surgery is of great importance to ophthalmologists [113], [114], [157]. Refractive surgery outcomes are subjective to corneal biomechanics [57], [109], [158], [159].

Studying corneal behavior has been extensively investigated by many models and techniques [160]. Age related factors are studied in [161]–[163]. X-Rays diffraction is used to investigate the inter-fibril spacing of stromal collagen fibrils and inter-molecular spacing of collagen molecules in [161]. The inter-fibril spacing of stromal collagen is found to be decreasing with age as reported in [162]. In the work presented in [163], diameter of collagen fibrils is observed to be age dependent, where there is a change of about 9% in subjects older than 65 years old compared to those under 65 years old. Cornea stiffness related to age is investigated in [164], [165], where the cornea is found to be stiffer when getting older.

In terms of refractive surgery, myopic astigmatism is achieved by small-incision lenticule extraction (SMILE) [165]. SMILE is claimed to have minimal effect on corneal biomechanics compared to other refractive surgeries such as LASIK (laser-assisted in-situ keratomileusis) and femto-LASIK (femto-second laser-assisted in-situ keratomileusis) [76], [166]. SMILE has shown less effect on corneal biomechanics than femto-LASIK [76]. Corneal safety and sensitivity are studied in multiple studies [167]–[169]. LASEK (laser-assisted sub-epithelial keratectomy) is reported to have less effect on corneal biomechanics with minimal risk of developing ectasia as well [170]. However, some cases are reported to have corneal biomechanics alteration after performing LASEK[171]. Moreover, both LASEK and SMILE are reported to induce changes in corneal biomechanics, where SMILE has the advantage of inducing less biomechanics changes than LASEK does due to the preservation of stiffer anterior stroma [57].

Ultrasound techniques are used to estimate corneal biomechanics. The main algorithm for estimating soft tissue biomechanics is by acquiring a stack of deformed frames for the tissue along with a reference frame utilizing high frame rate B-mode imaging [83], [84]. Transient ultrasound acoustic radiation force impulse (ARFI) is applied by means of ultrasound focusing techniques [85]. This ARFI generates deformation wave inside tissue under study that alters the characteristic of tissue's mechanical properties such as, Young's modulus (E) and shear modulus (μ). This induced shear wave speed is related to tissue shear modulus by Eqn. 5.1. & 5.2.:

$$C = \sqrt{\frac{\mu}{\rho}} \qquad\qquad (5.1)$$

$$\mu = \frac{E}{2(1+\vartheta)} \qquad\qquad (5.2)$$

where C is the shear wave speed, μ is the shear modulus, ρ is the density in $Kg/m3$, E is the Young's modulus and ϑ is the Poisson ratio for tissue under investigation respectively.

Supersonic Time-Of-Flight (TOF) [86], [87], Lateral Time-To-Peak (TTP) [89], [151] or cross correlation [89], [90], [172], [173] algorithms are implemented to estimate the speed of the resulting tissue deformation wave; i.e. Shear Wave Speed (SWS).

Till now, there is no non-invasive technique for clinical ultrasound in-vivo device that is used to assess the corneal biomechanics as a single modality combining the pushing element and the imaging element. All available approaches are still proof of concepts and utilize two different elements, external one for generating the push and another one for capturing the deformed frames [28]. Even, these approaches do not give spatial distribution for the corneal biomechanics, instead they give a mean estimate for the whole cornea tissue area.

In our research, we propose a new technique where the same ultrasound transducer is used to generate highly localized ARFI to act as internal actuator to induce tissue deformation and to perform B-mode imaging procedure to capture the propagation of the deformation wave. In our research also, we can continuously change the investigated zone of cornea to cover the whole cornea area and to obtain a spatial distribution of cornea biomechanics.

In this chapter, a 3D FEM in conjunction with 2D scatterer model for human cornea and an ultrasound transducer model are implemented to study the effect of post-refractive surgery on the biomechanics of corneal tissue using ultrasound B-mode imaging. Different elastic moduli are assigned to the 3D FEM to simulate different cornea tissue biomechanics [87], [148]–[150]. Cornea dimensions are adopted from medical literature for average human cornea [108], [112].

The chapter is organized as follows, section 1 gives an introduction about the cornea, their biomechanics effect and related refractive surgeries, and a brief survey on ultrasound techniques used for assessing corneal biomechanics. Section 2 shows a full description for the proposed approach. Section 3 describes the methodology of how the three proposed models interact with each other and the acquisition sequence and the shear wave speed estimation method. Experimental results are reported in section 4. Finally, discussion and summary are presented in sections 5 and 6 respectively.

## 5.2 THE PROPOSED APPROACH

Our proposed system is visualized in Fig. 5.1 (a) and (b), where (a) is a real life photo for the ultrasound probe used on top of human eye, and (b) is a drawing representing the ultrasound transducer is used on top of the human eye lid [174]. The same ultrasound transducer is used to generate the ARFI and to perform the B-mode imaging procedure. The ultrasound transducer will be fixed in the same position during the whole corneal biomechanics estimation process. At the beginning of the estimation process, the transducer is used to generate the ARFI at the corneal apex, then the ordinary B-mode imaging procedure is initiated to capture the departing deformation wave across the lateral direction.

The proposed algorithm for estimating corneal biomechanics for each elastic modulus is presented in the block diagram shown in Fig. 5.1 (c). It shows three main blocks: First block resembles the mechanical model of the cornea, the second resembles the ultrasound transducer model and the third resembles the scatterer model of the cornea. The three models are mutually interacting with each other to produce a deformed video stack of the cornea.

### 5.3.1. Cornea Mechanical FEM

A vertical cross section for the human cornea is modeled as 3D FEM and shown in Fig. 5.2 (a). The model dimensions are assigned to be equal to those of average human cornea 440 µm to 650 µm with an average of 540±30 µm [108], [112]. The mechanical properties are set to those of cornea pre- and post-refractive surgery and are listed in Table 5.1. Triangular mesh is used by COMSOL software to mesh the volume of the cornea model. The mesh is set to be narrower at the volume of interest for obtaining accurate results and wider at the remaining volume to speed up the processing time. The cornea mesh is shown in Fig. 5.2 (b).

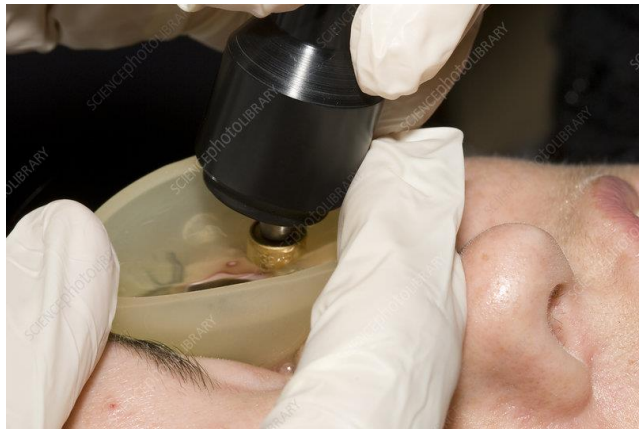

(a) Real-life photo of the ophthalmic ultrasound investigation procedure [175].

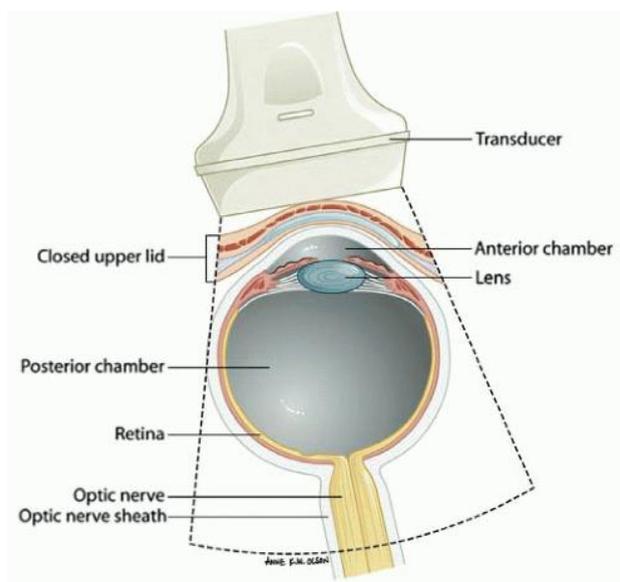

(b) drawing representing the location of the probe with respect to the human eye.

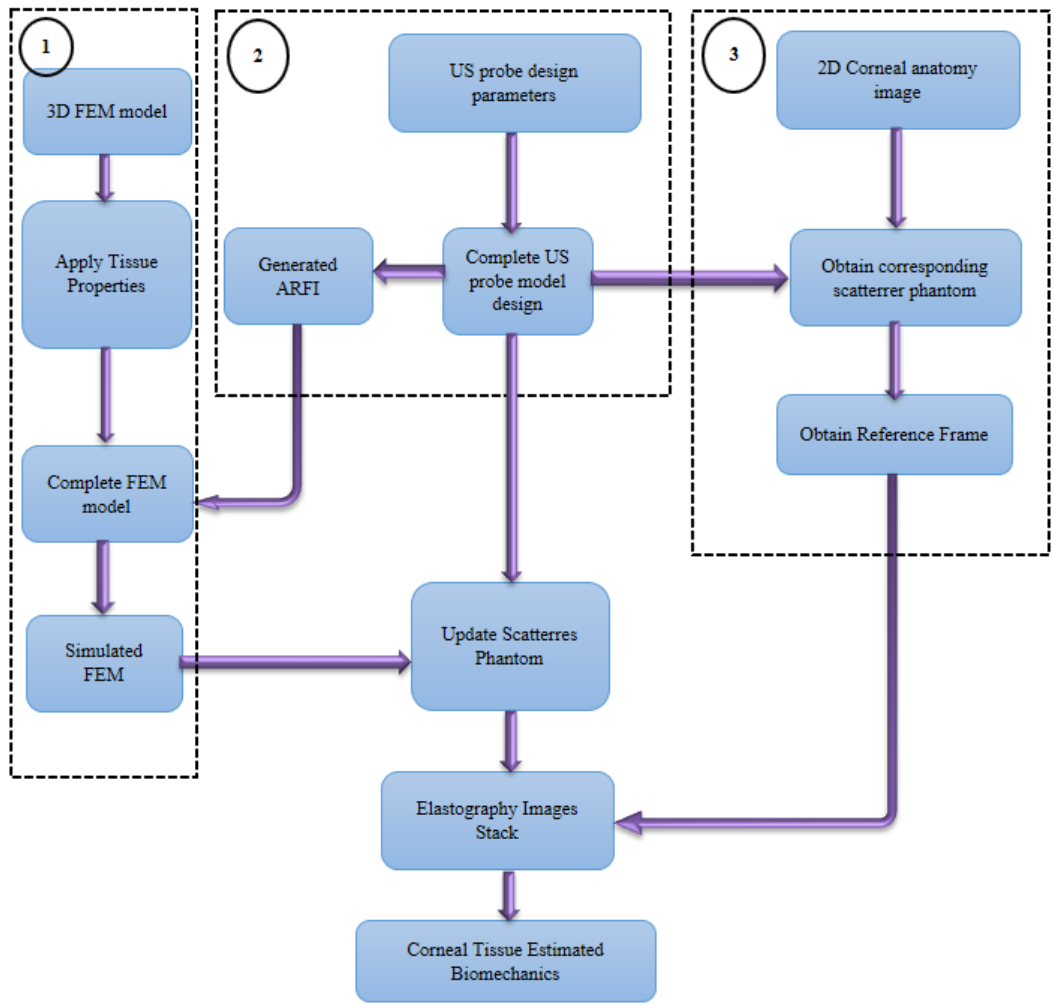

(c) block diagram of the proposed approach.

Fig. 5.1 (a) Real-life photo and (b) a drawing representing the location of the probe with respect to the human eye, (c)block diagram of the proposed approach.

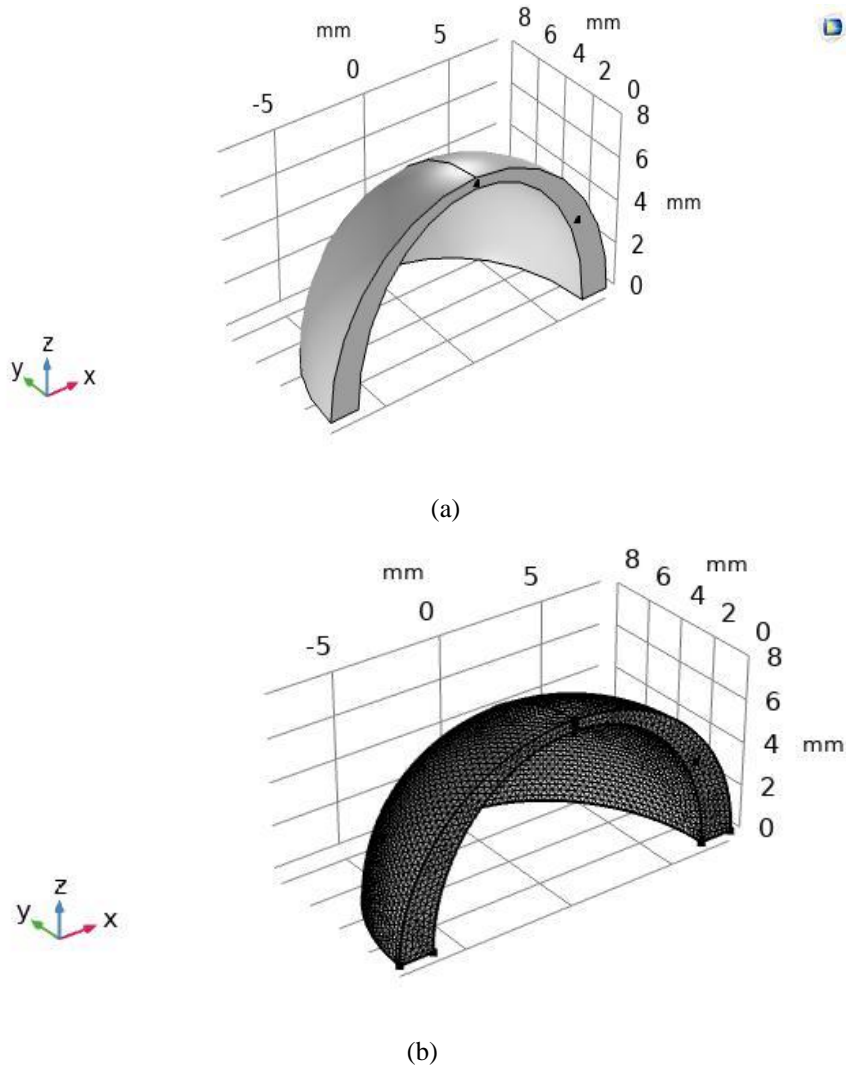

(a)

(b)

Fig. 5.2 (a) Vertical cross section of human cornea 3D FEM, (b) Its corresponding Mesh.

Table 5.1 Mechanical properties of human cornea.

| Material Property | Value | |
|---|---|---|
| Young's modulus (range of values) | Pre-refractive (High E in MPa) | Post-refractive (Low E in KPa) |
| | 1, 1.5, 2, 2.5 and 3MPa [174] | 140, 300, 600 and 800KPa [112] |
| Poisson Ratio | 0.499 | |
| Heat Capacity at constant pressure | 2348 | |
| Thermal Conductivity | 0.21 | |
| Density | 911 | |

### 5.3.1.1.FEM Drawing

Human cornea is modeled using FEM as a vertical cross section to simulate its behavior for pre- and post-refractive surgery. As a consequence, corneal biomechanics changes from pre-refractive to post-refractive surgery in terms of Young's modulus. This change affects cornea behavior to external applied forces. The FEM is constructed from two intersecting shapes, a circle and an ellipse in the 2D plan. Then the intersection is revolved around the z-axis to construct the volume of the model. Complete model in 2D plane drawing is presented in Fig. 5.3 and its dimensions are listed in Table 5.2.

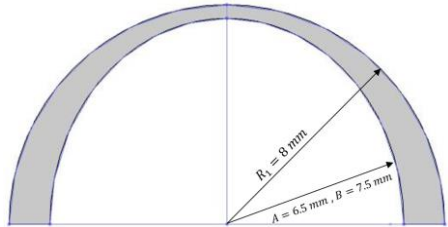

Fig. 5. 3.Cornea 2D plan view.

Table 5.2 Geometrical dimensions of human cornea.

|  | Circle 1 | Ellipse |
|---|---|---|
| **Radius (mm)** | 8 | A = 6.5, B = 7.5 |
| **Sector Angle (degrees)** | 180 | 180 |

### 5.3.1.2.FEM Solid Mechanics

Isotropic linear visco-elastic approach is adopted for this model as it simulates Standard Linear Solid mathematical model in Kelvin-Voigt form efficiently in Finite Element Analysis (FEA) software. Peak ARFI is applied at the apex of the cornea which is set to be at the focal point of the ultrasound beam. The fixed boundaries simulating the behavior of the ciliary muscles of cornea are set to be in the lower face of the cornea at 6 vertex points. This is shown in Fig. 5.4.

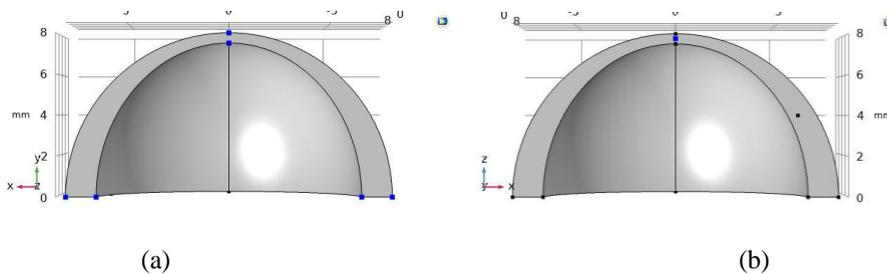

(a)                                      (b)

Fig. 5.4 a) Fixed boundaries as iris ciliary muscle, b) Focal point where ultrasound transient force is applied.

### 5.3.2.  Ultrasound Transducer Model

Ultrasound transducer is used to generate the internal push inside the cornea tissue that induces tissue deformation. As well it is used to perform B-mode imaging on the deformed tissue at subsequent time stamps to capture the propagation of the deformation wave. Therefore, the ultrasound transducer performs two functions in the transient elastography procedure, internal actuator for inducing tissue deformation and imaging to capture the induced displacement propagation.

In this model, two toolboxes in MATLAB are used to simulate the behavior of the ultrasound transducer in both cases of generating the ARFI and performing the B-mode linear imaging procedure. FOCUS toolbox is used for simulating the upltrasound transducer while generating the ARFI. However, Field II toolbox is used for simulating the ultrasound transducer imaging procedure. The transducer parameters used in both simulations are listed in Table 3.

Table 5.3 Ultrasound transducer parameters.

| Parameter | Value |
|---|---|
| Center frequency (fo) | 12 MHz |
| Number of elements | 192 |
| Number of active elements | 128 |
| Elements' width | 170 µm |
| Kerf | 30 µm |
| Focal depth | 28.5 mm |

The purpose of the ARFI is to generate a highly localized ultrasound push inside the tissue. This localization gives rise to quantitative estimation of tissue's mechanical properties. The imparted ultrasonic power due the focused push dampens highly around the focal push location in all direction exponentially. The generated ARFI induces cornea tissue to deform yielding a deformation (displacement) wave that propagates away from the focal point. The generated wave is considered to be the shear wave.

Fast Near Field Method (FNM) method from FOCUS toolbox is used to simulate the ultrasound linear transducer behavior while generating the ARFI. Pressure map is simulated for the ultrasound transducer and is presented in Fig. 5.5. The focal point is set to be around 28.5 mm in the axial direction (depth direction). The focal point is that point where the peak ultrasound pressure generated from the transducer is applied inside the tissue. It is the imparted ultrasonic pressure that is responsible for inducing the tissue deformation and hence the shear wave. We have simulated the transducer 2D pressure map, transverse cross-sectional pressure distribution across the 2D map at different axial depths, focal axial pressure at the center and the average axial pressure. In addition, focal axial intensity, average axial intensity, focal axial force and average axial force are simulated as well. The resulting force field is then applied to the 3D FEM to induce deformation and to simulate tissue behavior to such applied force for different elastic moduli. Tissue deformation at different consecutive time frames is then picked up by using the B-mode imaging with the ultrasound transducer model implemented by Field II toolbox. The force is applied for about 1 msec. and the B-mode imaging procedure starts just after applying of the ARFI. Captured B-mode images resulting from imaging the wave propagation gives rise to an estimate of the shear wave speed (SWS) that is related to the tissue biomechanics.

The 2D pressure distribution is shown in Fig. 5.5 (a) where the highest pressure value is observed around the preset focal point of 28.5 mm. Multiple transverse cross-sectional pressure distributions at different focal planes are shown in Fig. 5.5 (b) where again the highest smoothest pressure distribution is observed at 27 mm which is the closest focal plane to the focal point. Pressure 2D distribution for the imaging beam is shown in Fig. 5.6 (a) where the peak axial pressure value is observed at 28.5 mm which is coincident with the preset focal point. Imaging beam pressure axial distribution at the focal beam (blue line) along with averaged pressure axial distribution (red line) are shown in Fig. 5.6 (b) where peak pressure axial value is observed at 28.5 mm. Multiple lateral pressure distributions at different depths for the imaging beam are shown in Fig. 5.6 (c) where acoustic force is calculated according to Eqn. 5.3.:

$$F = \frac{2\alpha I}{C} \qquad (5.3)$$

where $\alpha$ is the acoustic absorption coefficient in ($\frac{dB}{cm.MHz}$), $I$ is the temporal average intensity of the beam in ($\frac{Watts}{m^2}$), and $C$ is the speed of sound in tissue in ($\frac{m}{sec.}$).

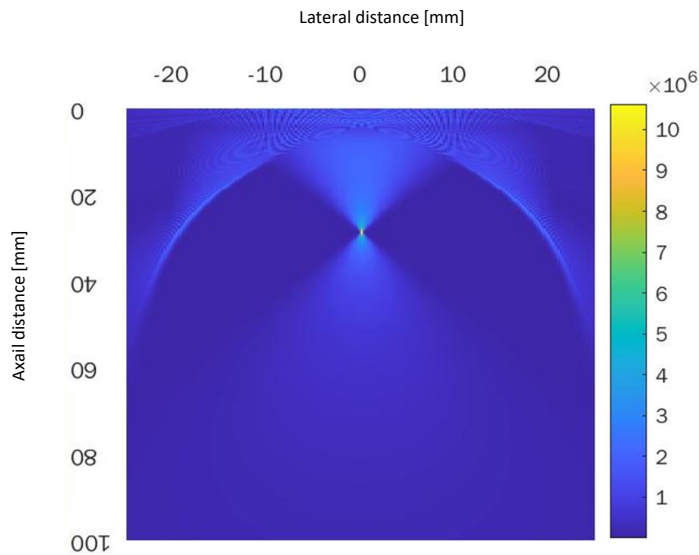

(a)

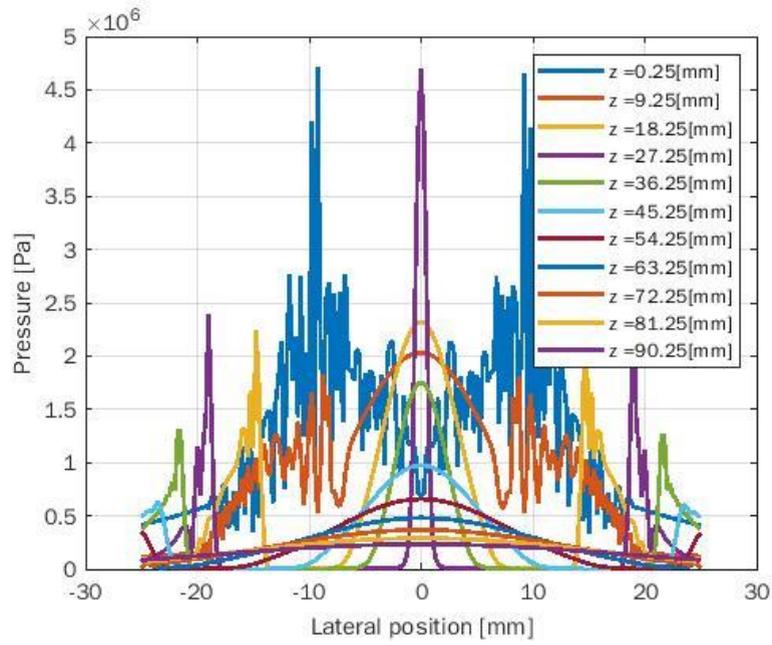

(b)

Fig. 5.5 (a) Simulated 2D pressure map for ultrasound transducer model, (b) multiple transverse cross-sectional pressure distributions.

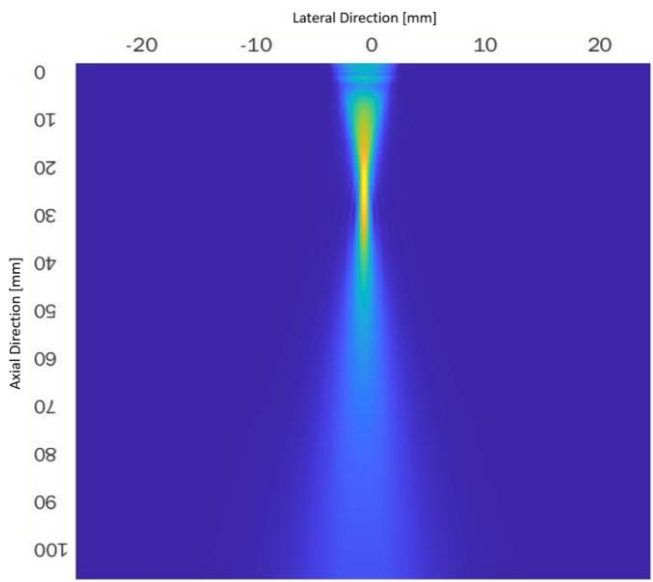

(a)

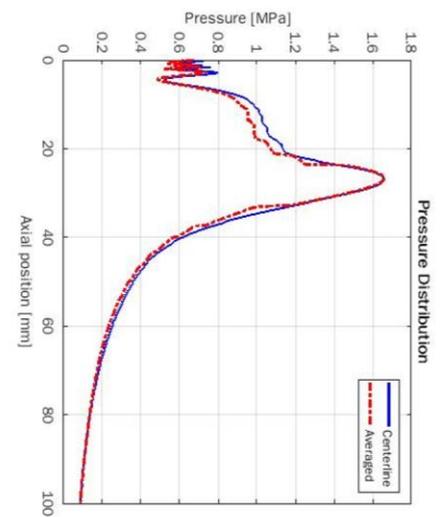

(b)

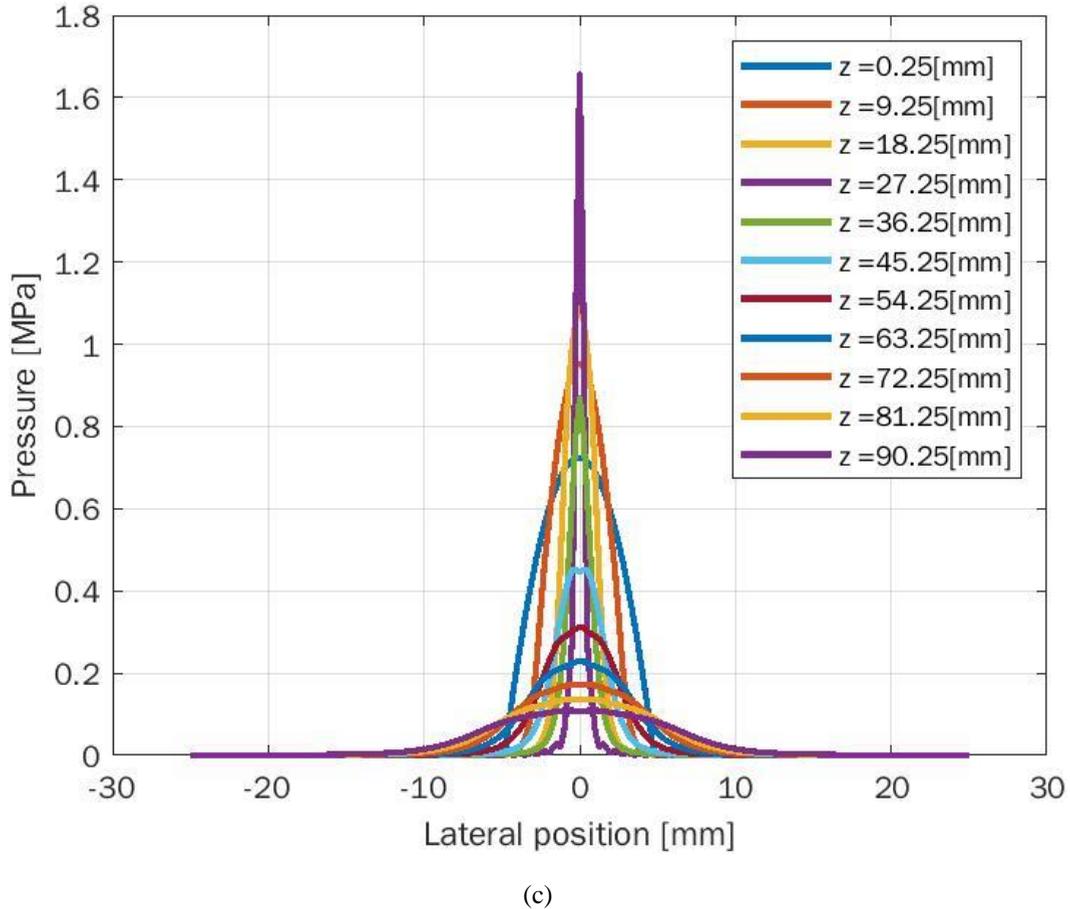

(c)

Fig. 5.6 (a) 2D pressure distribution map. (b) pressure axial distribution. (c) multiple transverse cross-sectional pressure distributions.

### 5.3.3. Cornea Scatterer Model

FIELD II toolbox is used to generate the equivalent scatterrer model of the 3D FEM of cornea. The FEM consists of nodes with random spatial locations in-between and specific spatial locations at the boundaries. It is at these nodes where the ultrasound echoes are calculated to reconstruct an ultrasound line in the B-mode image. Nodes' spatial locations are exported to a comma separated value (csv) file with a fixed file structure; where x, y and z positions for each node are listed in single row. Each row is starting from the nodes' stationary spatial locations at the reference frame at the first three columns and the subsequent columns holds the nodes' new spatial locations at each time frame of simulation. Then we import this file to MATLAB for reading the scatterrers' spatial location at both the reference frame and each deformed frame in order to construct the B-mode image. A snapshot of corresponding file structure is presented in Table 5.4 where the presented spatial locations are x and z only as they are responsible for deformation in our FEM. The complete scatterrers model as seen by the ultrasound transducer is shown in Fig. 5.7.

Table 5.4 Scatterrer file snapshot.

| X | Z | u (mm) @ t=0 | v (mm) @ t=0 | u (mm) @ t=0.0174 | v (mm) @ t=0.0174 | : | : | : |
|---|---|---|---|---|---|---|---|---|
| -7.19115 | 0.647299 | 0 | 0 | -0.024530482 | -0.001285963 | : | : | : |
| -7.25 | 1.33E-15 | 0 | 0 | -0.006800931 | -0.003734764 | : | : | : |
| -8 | 1.47E-15 | 0 | 0 | 0 | 0 | : | : | : |
| -6.5 | 1.32E-15 | 0 | 0 | 0 | 0 | : | : | : |

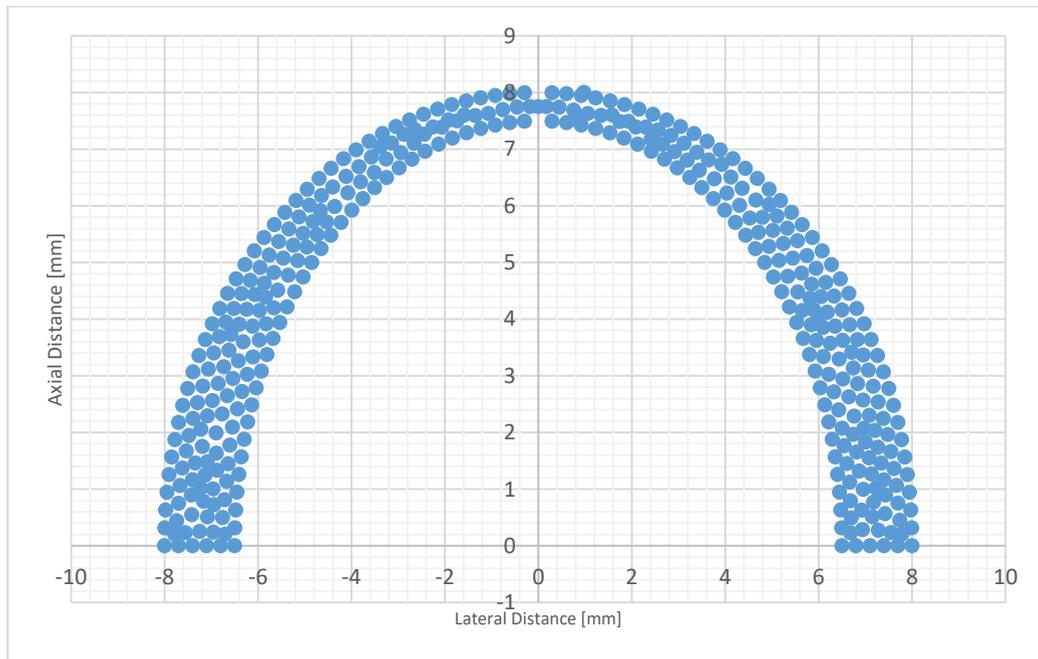

Fig. 5.7 Complete human cornea scatterrer mode.

## 5.4. METHODOLOGY

Corneal biomechanics are estimated from the speed of propagating shear wave in response to acoustic radiation force impulse (ARFI) generated by the ultrasound transducer. This force impulse is generated by a focused ultrasound push and applied to the corneal tissue transiently for about 1 msec. ARFI causes deformation of corneal tissue as a wave propagating away from the focal point of application. The resulting deformation is then tracked using ultrasound transducer via high frame rate B-mode imaging procedure. Monitoring sequence is then obtained for the cornea model for complete 10 msec. of B-mode imaging resulting in a video stack for deformation propagation. Deformation is tracked in either direction laterally around the cornea apex by means of two fixed points in the video stack. One point is the focal point and the other one is laterally distal from cornea apex in either lateral direction. This process is repeated for different elastic moduli cornea models introduced in this study.

The proposed methodology is explained as follows:

i. 3D FEM, 2D scatterrer and ultrasound transducer models are implemented in parallel to simulate the average human cornea dimensions and linear array ultrasound transducer respectively, as discussed in section 2.1.

ii. The ultrasound transducer model is used to:

    a. Obtain a reference frame for the cornea before applying the ARFI (unloaded cornea frame).

    b. Generate ARFI to act as internal stimulus for cornea tissue to deform.

iii. The generated ARFI is then applied to the 3D FEM to simulate the mechanical behavior of the cornea (loaded cornea) for different time frames corresponding to the frame rate of the ultrasound B-mode imaging. We use a frame rate of 100 KHz for high frame rate imaging procedures to reduce the error in the estimated values from ultrasound imaging, as discussed in section 2.3.

iv. The simulated time frames from the mechanical FEM model are then used to update the spatial locations of the scatterrers inside the scatterrer model.

v. The ultrasound transducer is used to image the updated scatterrer model to obtain different time frames.

vi. The corneal tissue biomechanics are estimated from the reference frame and the stack of deformed frames.

vii. This process is repeated for every cornea with elastic moduli included for this study.

### 5.4.1. Acquisition Sequence

Acquisition sequence starts by acquiring a reference frame for corneal tissue under study. This reference frame is acquired by transmitting a plane wave to the medium of interest (corneal tissue). Then an ARFI is sent to the corneal tissue for a short duration of time of about 1 msec. to resemble a transient force. This transient force is used to induce corneal tissue deformation giving rise to shear wave propagation inside the tissue. In this study we apply ARFI at a focal zone of 28.5 mm inside the tissue.

Just after applying of the ARFI to the corneal tissue, high frame rate B-mode imaging starts to capture the migrating shear wave in both lateral directions away from the focal zone. The frame rate used in this study is 100 KHz enabling the accurate tracking of the wave peak through acquisition time of 10 msec. The complete timing diagram for the complete B-mode imaging procedure is shown in Fig. 5.8. Apodization of the ultrasonic wave is performed to help reduce both the grating and side lobes of ultrasound beam.

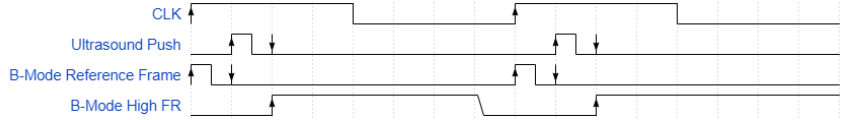

Fig. 5.8 Timing diagram for two periods of the B-Mode acquisition sequence.

### 5.4.2. *Shear Wave Speed Estimation*

Estimating shear wave speed is achieved by lateral Time-To-Peak displacement (lateral TTP) technique. The wave speed is estimated by dividing the lateral distance between the two probing nodes over the estimated times of wave peak arrival at these nodes. The two probing nodes are chosen to be the focal zones of two distal ultrasound beam, one node is at the focal zone of the central ultrasound beam and the other is at the focal zone of any of the lateral beams. The second probing point can be in either lateral direction as well, i.e. to the left or the right of the central beam. Time-To-Peak displacement is a property for every tissue biomechanics, where tissues with different elastic moduli yields different TTPs. It is worthy to be noted that, dependence on ARFI value only changes the amplitude of tissue deformation where two different elastic moduli tissues having same ARFI value will yield two different peak deformation values. Moreover, these two tissues will yield two different TTP values. TTP is estimated for tissue under study from displacement curve obtained for the probing node with proper tracking frequency considerations. Equation 5.4. is used to calculate the estimated SWS:

$$\mathbf{C_{avg}} = \frac{\Delta \mathbf{x}}{\Delta \mathbf{t}} \qquad\qquad (\mathbf{5.4})$$

where $C_{avg}$ is the average velocity across the lateral position, $\Delta \boldsymbol{x}$ and $\Delta \boldsymbol{t}$ are the difference in distance between probing nodes and difference in times of peak arrival at these nodes respectively.

## 5.5. RESULTS

Figure 5.9 shows the reference frame and peak axial deformation frame after applying ARFI for 140 KPa corneal model. The focal point is observed to reach its peak deformation along the axial direction in Fig. 5.9 (b). These two figures are chosen for illustration purposes, where all other different elastic moduli models included in this research have undergone the same procedure and yielded the relevant outcomes.

The reference frame shown in Fig. 5.9 (a) is used as a base for estimating the axial deformation of the cornea through subsequent time frames after applying ARFI, and estimating the peak axial deformation as well. It is considered the starting point in transient elastography imaging procedure. This reference frame is obtained for all cornea models with different elastic moduli namely; 140 KPa, 300 KPa, 600 KPa, 800 KPa, 1 MPa, 1.5 MPa, 2 MPa, 2.5 MPa and 3 MPa. The field of view (FOV) is limited to the lateral distance obtained by the ultrasound probe during B-mode imaging. The field of view (FOV) in this experiment is about 8 mm in the lateral direction and about 12 mm in the axial direction. Cornea tissue is observed to occupy axial location from 28 mm to 28.5 mm at the apex which is relevant to literature information about corneal thickness at the apex to be about 0.5 mm.

Peak corneal deformation after applying ARFI is shown in Fig. 5.9 (b) where the 8 mm wide FOV is considered to be optimum for monitoring the lateral deformation of the cornea. This is an empirical result obtained by try and error. The shear wave is observed at this frame clearly, where it propagates in either lateral direction along the corneal tissue.

A complete simulation for shear wave propagation inside corneal model of 140 KPa along the temporal domain is performed and shown in Fig. 5.10. Shear wave propagation is shown for only eight different time stamps along with the reference frame at the beginning for convenience in presentation. The monitoring sequence starts with the reference frame at 0 sec. before which ARFI is applied and end up at 10 msec.

A complete simulation time sequence with 100 KHz frame for each corneal model is obtained. The deformation along these frames is tracked at each time frame and compared with the reference frame. Complete 1D deformation curves for both the focal point and the lateral point are then constructed from the peak deformation tracking procedure and are shown in Fig. 5.11. The time axis starts from 3 msec. to 10 msec. and not from 0 sec. since there are no significant information before deformation to be shown along the curve.

Figure 5.12 shows the estimated times at which focal peak deformation takes place for each corneal model. These times enables the estimation for the frame at which the peak deformation happens, hence enabling the estimation of the focal peak deformation itself accurately (Deformation Amplitude). This is achieved by means of the reference frame and the estimated peak deformation frame. Reference frame along with peak deformation frame for each corneal model is shown in Fig. 5.13. These frames are used to estimate the peak deformation value for each cornea model and the Deformation Amplitude Ratio at 2 mm in either lateral direction distal from the apex as well.

1D deformation curves allow for estimating the SWS for each cornea model by calculating the time difference between TTPs for each of the two probing nodes divided by their spatial distance. Theoretical, estimated shear wave speed along with speed estimation accuracy for each cornea model are listed in Table 5.

Deformation amplitude is estimated accurately for each cornea model by comparing the ultrasound B-mode focal beam for the reference frame and corresponding deformation frame. The reference beam along with all deformed beams for each cornea model are presented in Fig. 5.14 The deformation amplitude is considered to be the difference between the axial location at which the cornea tissue appears and the axial location at which each other cornea tissue for different cornea models appear. The deformation amplitudes are shown in Fig. 5.15.

Figure 5.16 shows the deformation amplitude ratios for each elastic modulus cornea model. They are estimated from B-mode images where the focal peak axial deformation happens. It is the ratio between the apex deformation to 2 mm nasal or temporal peak axial deformation. Deformation amplitude ratio defines how the cornea apex deforms with respect to either 1 mm or 2 mm nasal or temporal deformation of corneal tissue.

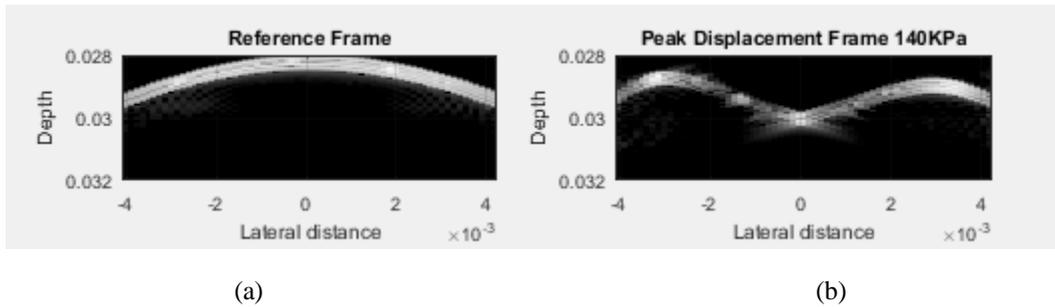

(a)                                                                 (b)

Fig. 5.9 (a) Reference frame, (b) Peak deformation frame for cornea model.

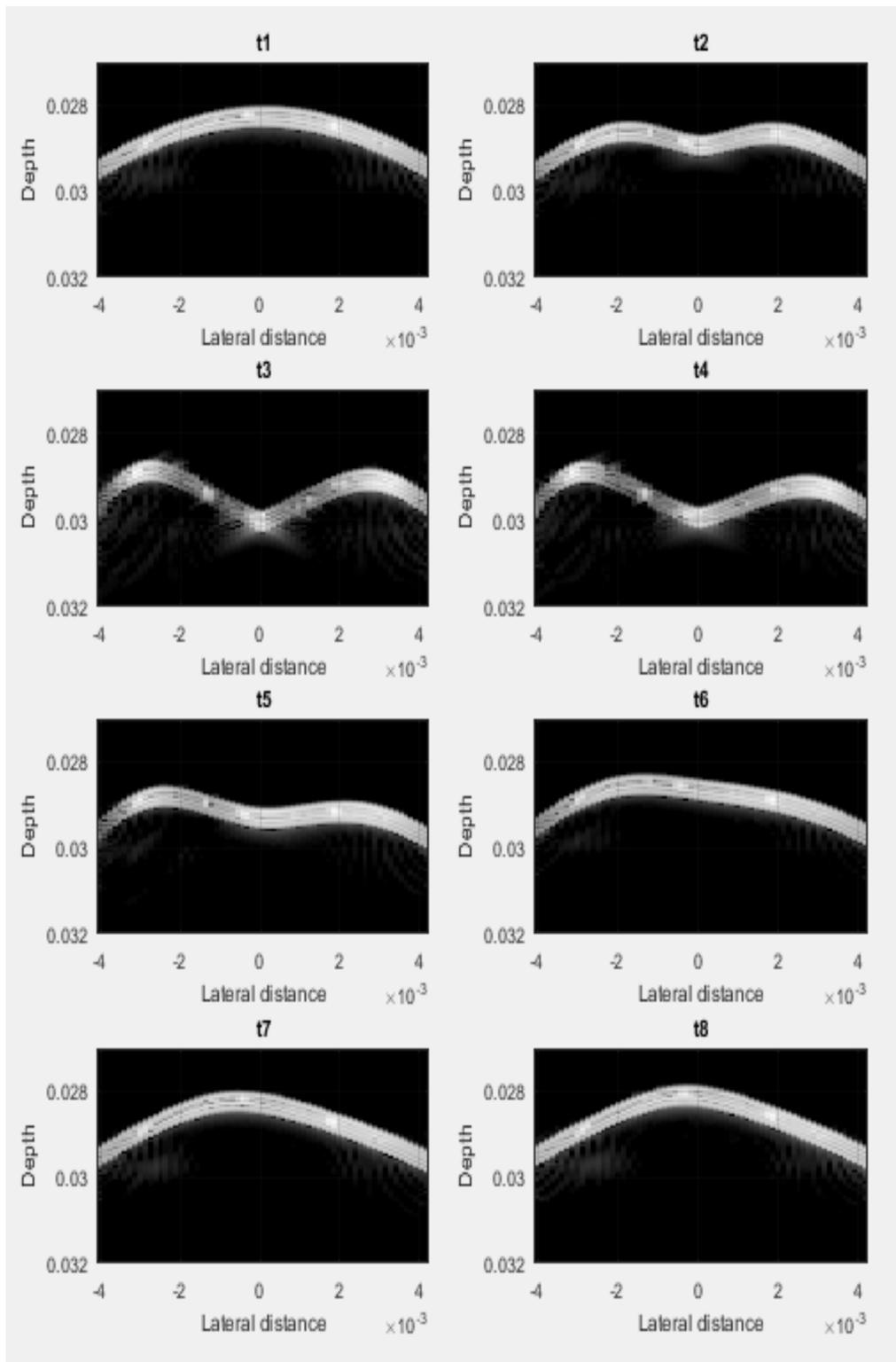

Fig. 5.10 Shear wave deformed frames stack for 140KPa.

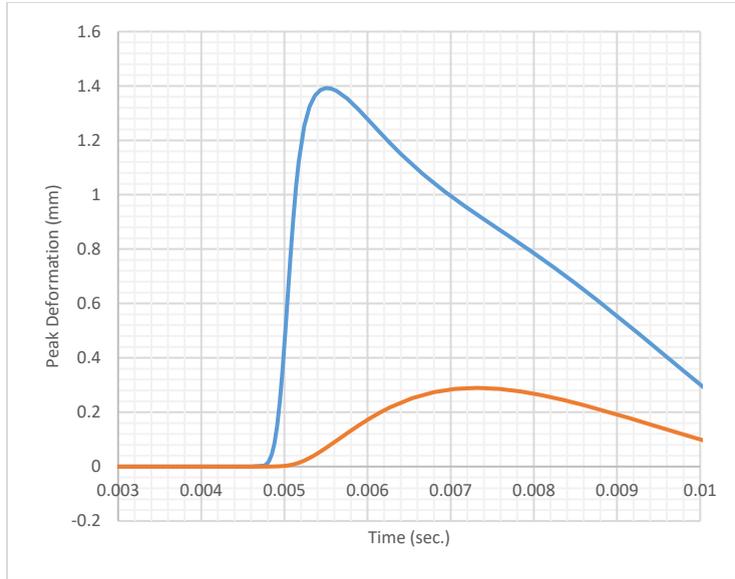

(a)   140KPa.

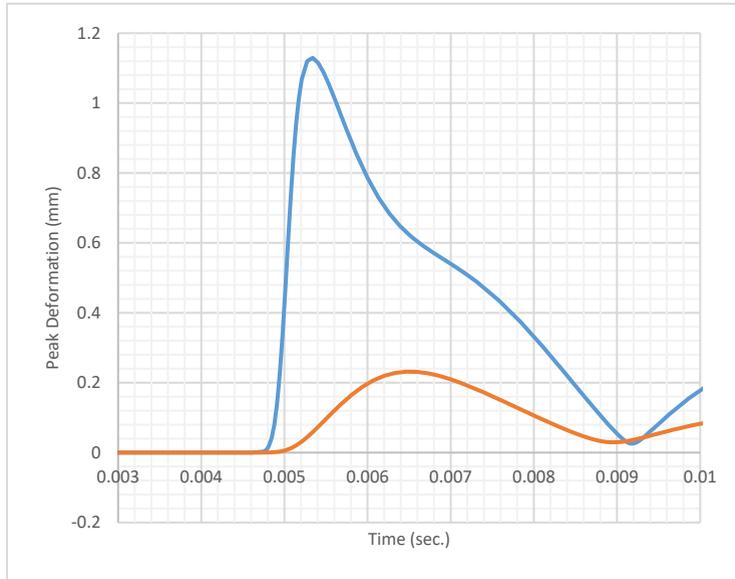

(b)   300KPa.

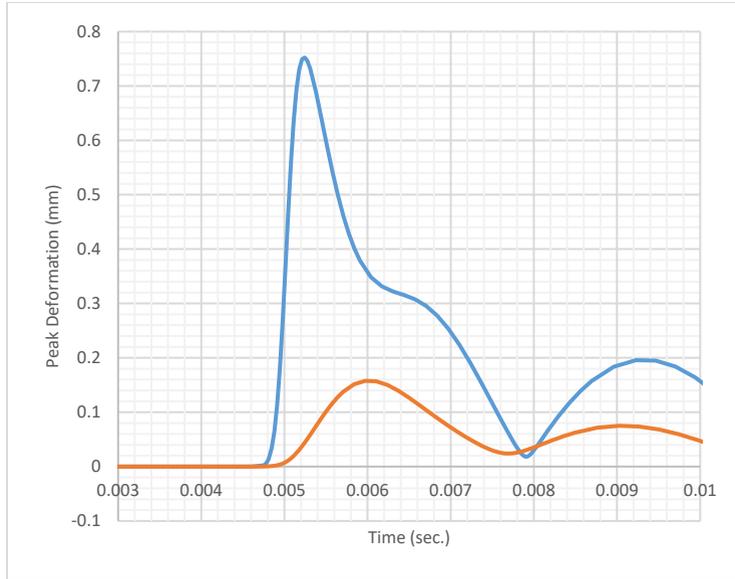

(c)    600KPa.

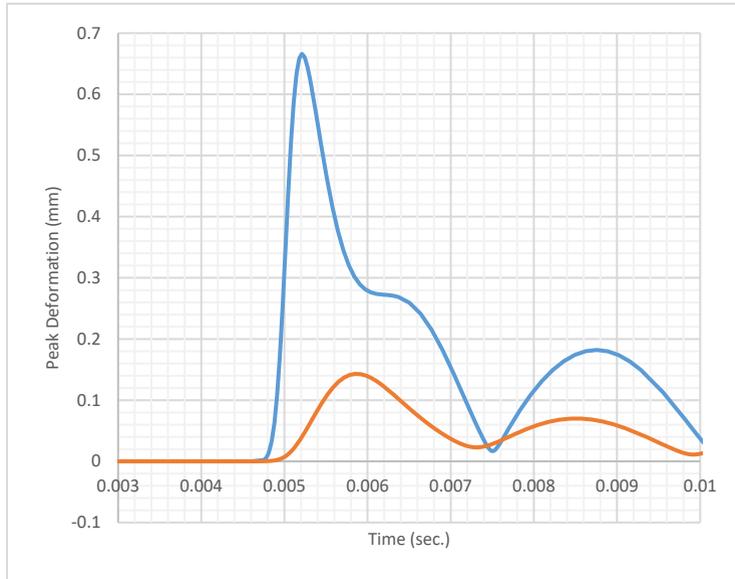

(d)    800KPa.

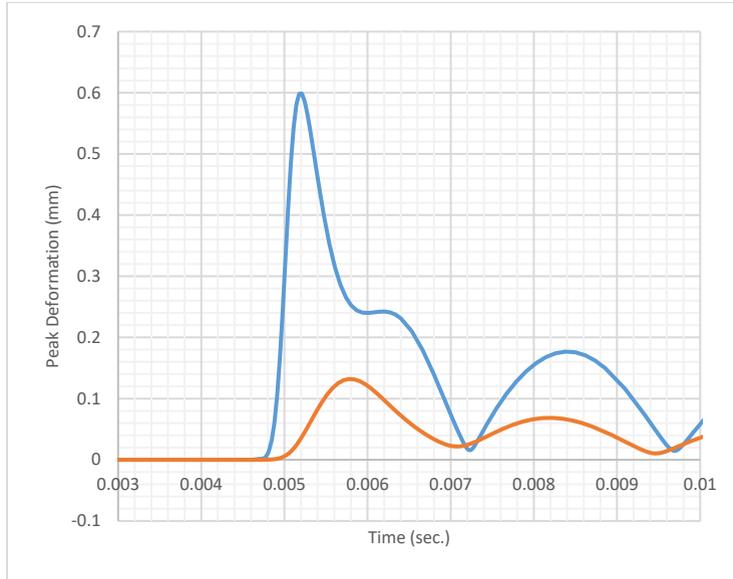

(e)   1MPa.

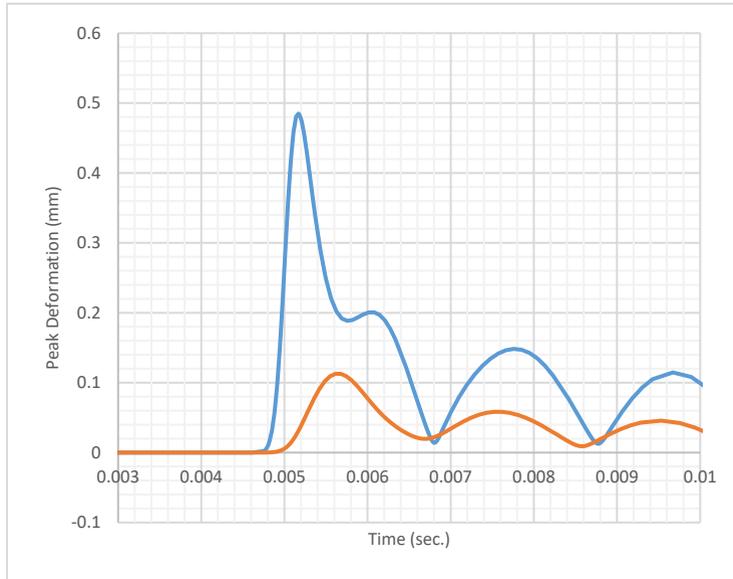

(f)   1.5MPa.

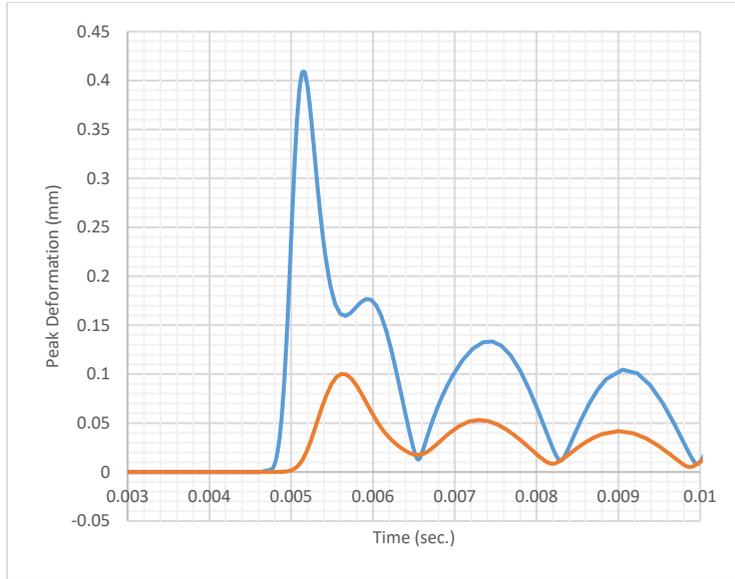

(g)   2MPa.

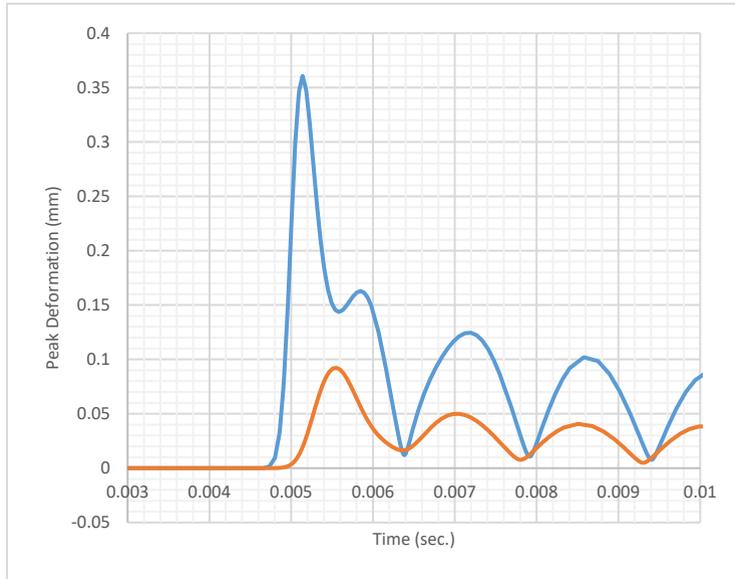

(h)   2.5MPa.

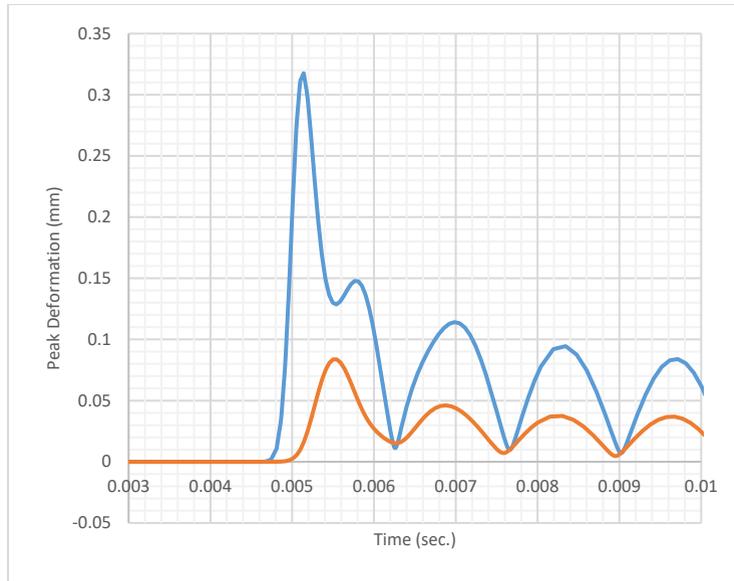

(i)    3MPa.

Fig. 5.11 1D Deformation profiles for focal and lateral probed nodes for each cornea model (Blue line = focal node displacement, Red line = radial node displacement).

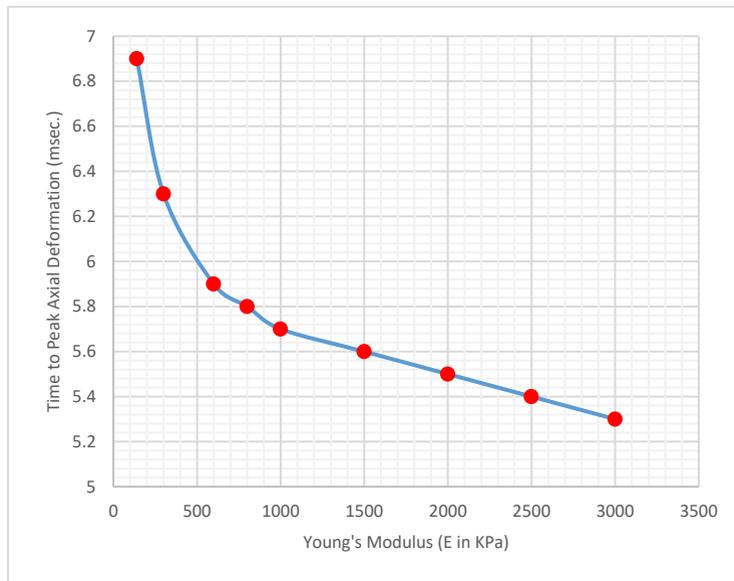

Fig. 5.12 Time-To-Peak (TTP) deformation for each cornea model.

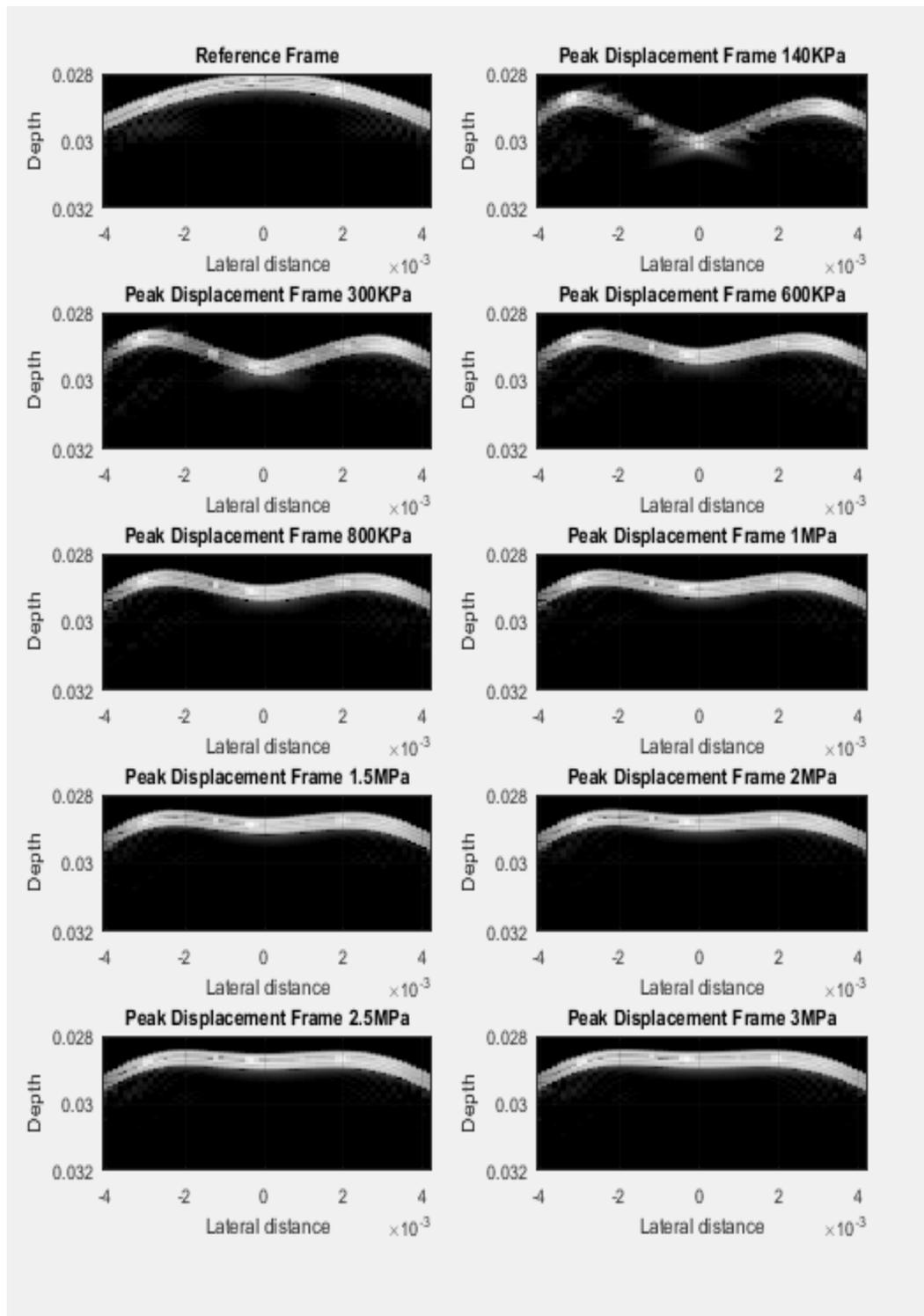

Fig. 5.13 Reference frame along with peak deformation frame for each cornea model.

Table 5.5 Theoretical, estimated SWS and accuracy for each cornea model.

| Young's Modulus (Pa) | Theoretically (m/sec.) | Estimated (m/sec.) @ 100 KHz | Accuracy (%) |
|---|---|---|---|
| 140K | 7.157217283 | 7.142857143 | 99.79936 |
| 300K | 10.4770933 | 10.34482759 | 98.73757 |
| 600K | 14.81684744 | 14.63414634 | 98.76694 |
| 800K | 17.10902172 | 17.14285714 | 99.80224 |
| 1M | 19.1284678 | 18.75 | 98.02144 |
| 1.5M | 23.42749283 | 23.07692308 | 98.5036 |
| 2M | 27.05173859 | 26.42857143 | 97.69639 |
| 2.5M | 30.24476319 | 30 | 99.19073 |
| 3M | 33.1314781 | 33.33333333 | 99.39074 |

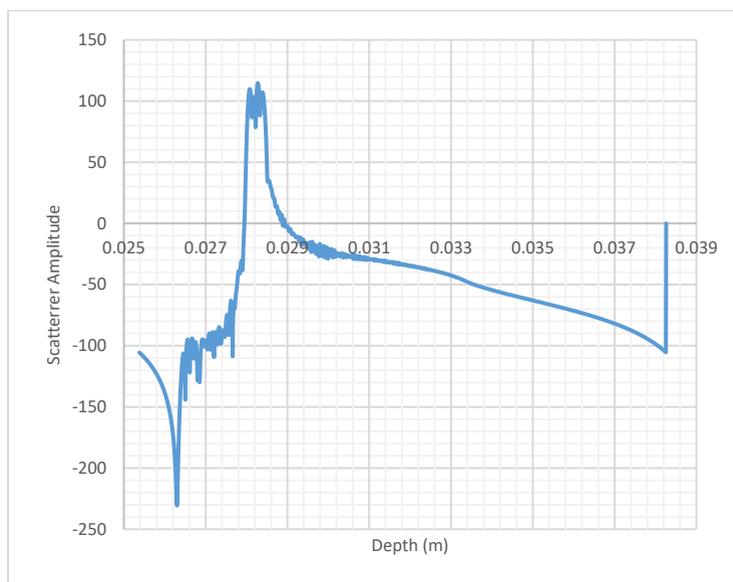

(a) Reference beam.

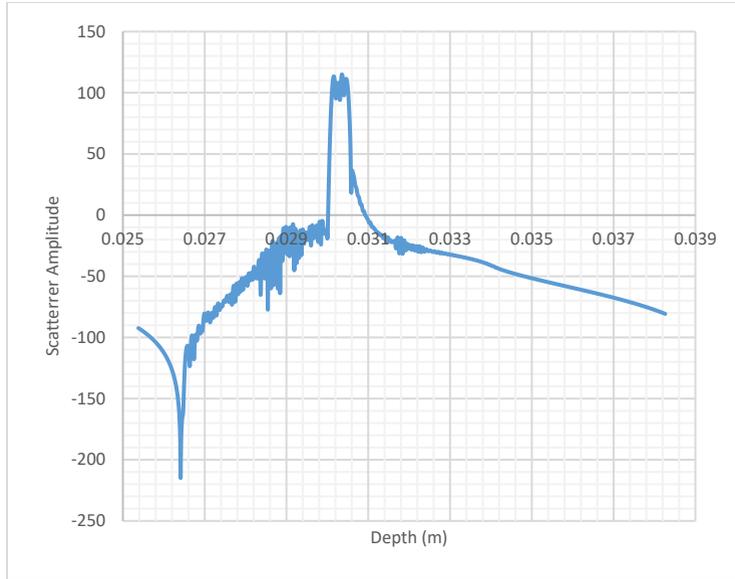

(b) 140 KPa beam.

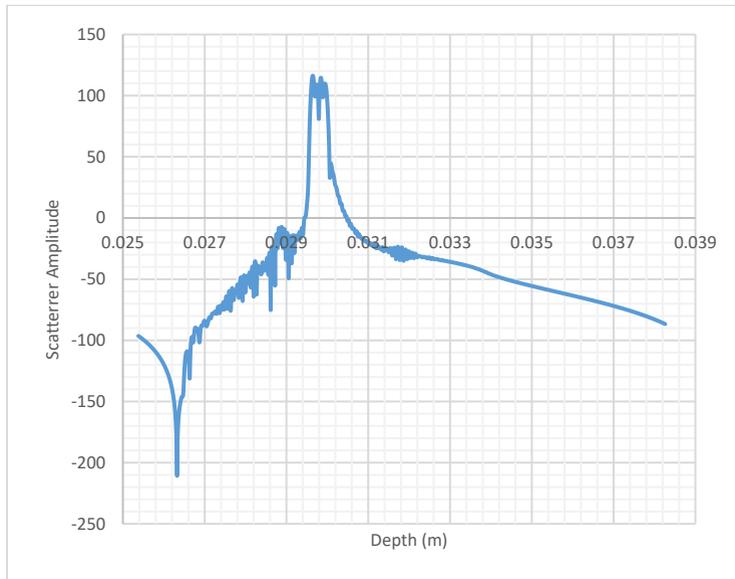

(c) 300 KPa beam.

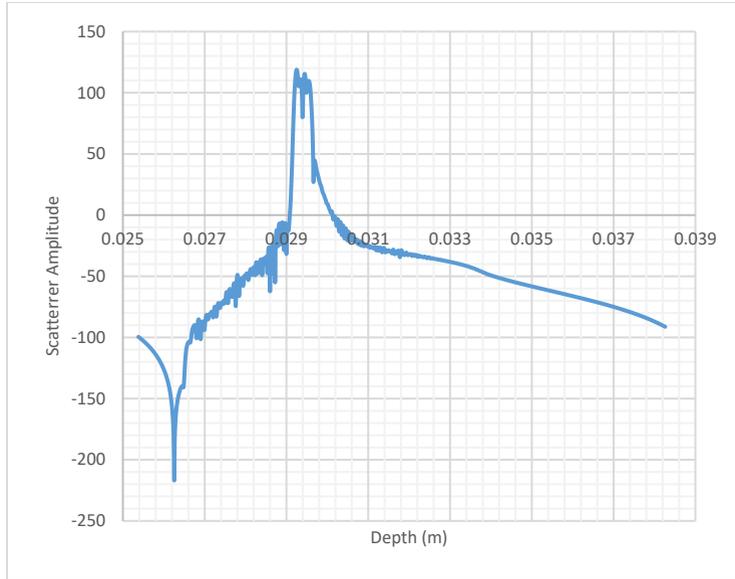

(d) 600 KPa beam.

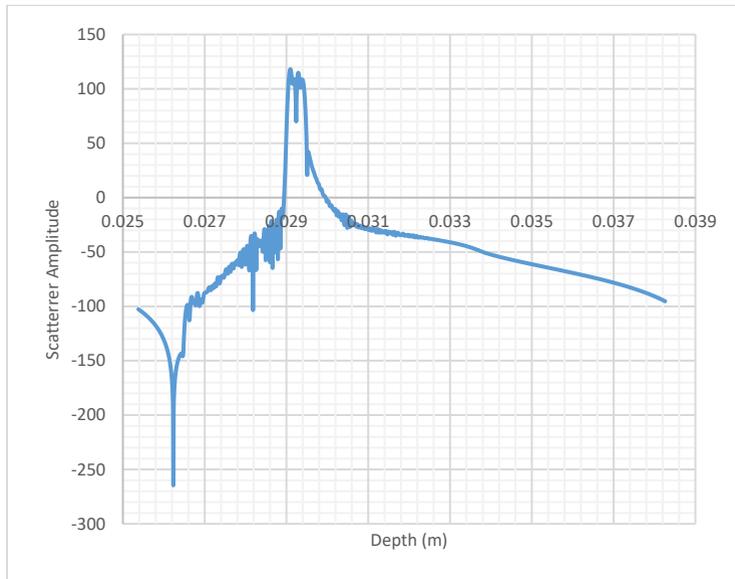

(e) 800 KPa beam.

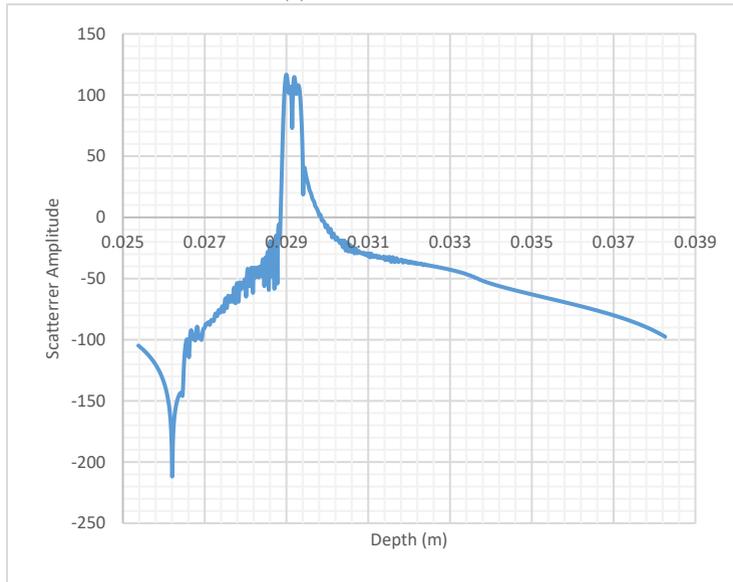

(f) 1 MPa beam.

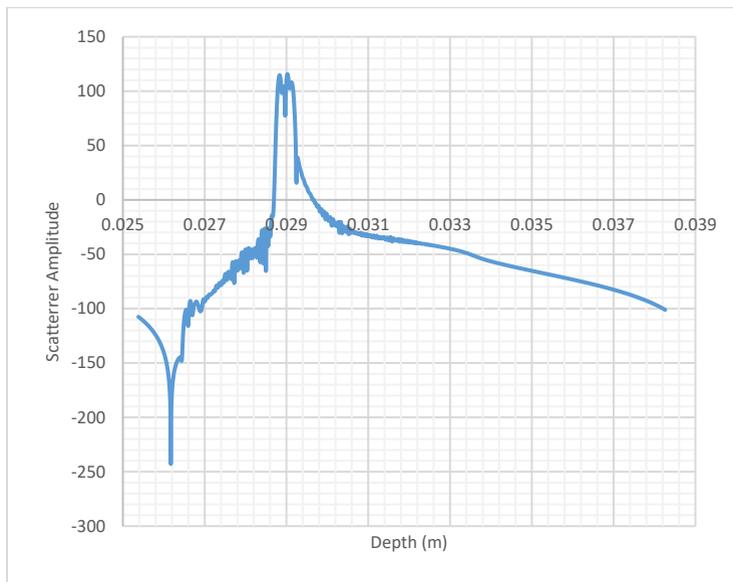

(g) 1.5 MPa beam.

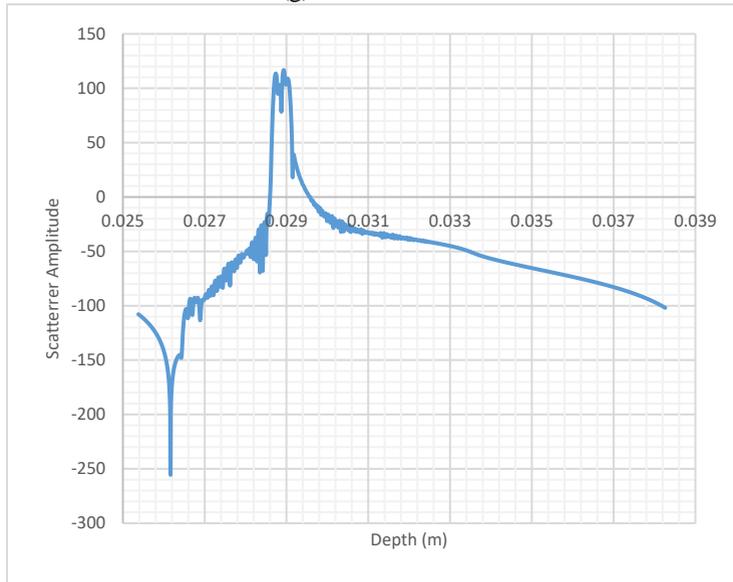

(h) 2 MPa beam.

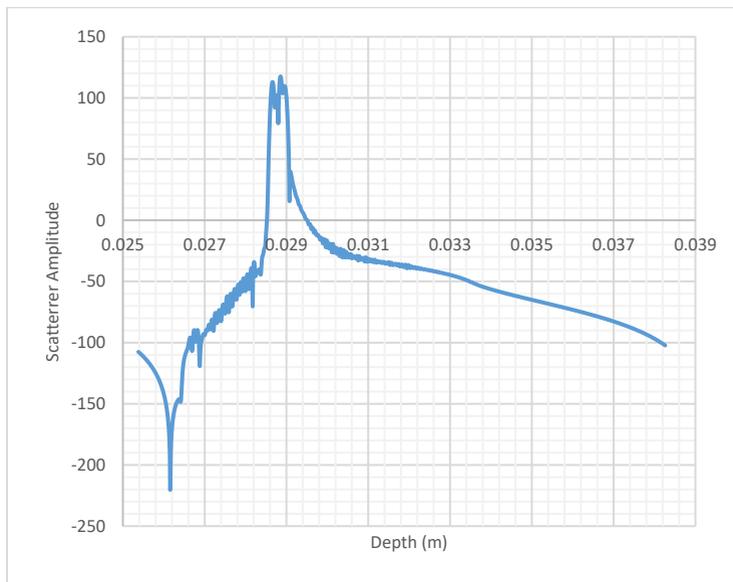

(i) 2.5 MPa beam.

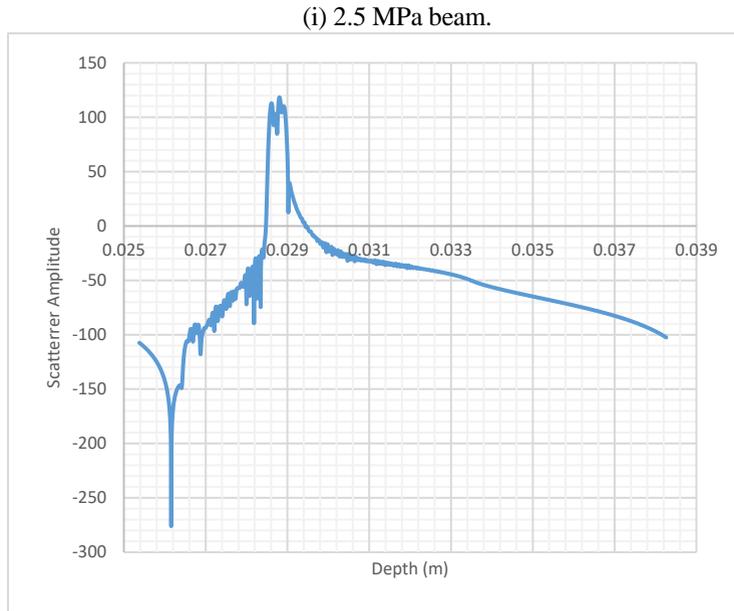

(d) 3 MPa beam.

Fig. 5.14 Reference beam along with all deformed beams for each cornea model.

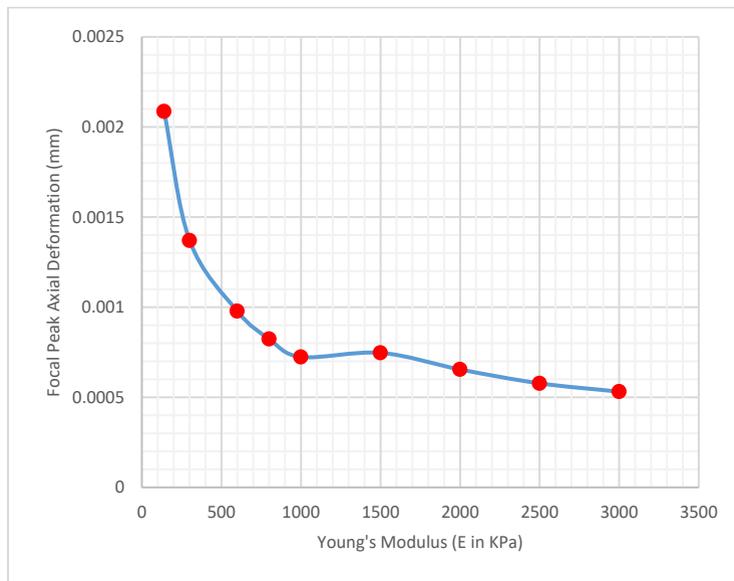

Fig. 5.15 Deformation amplitude for each cornea model.

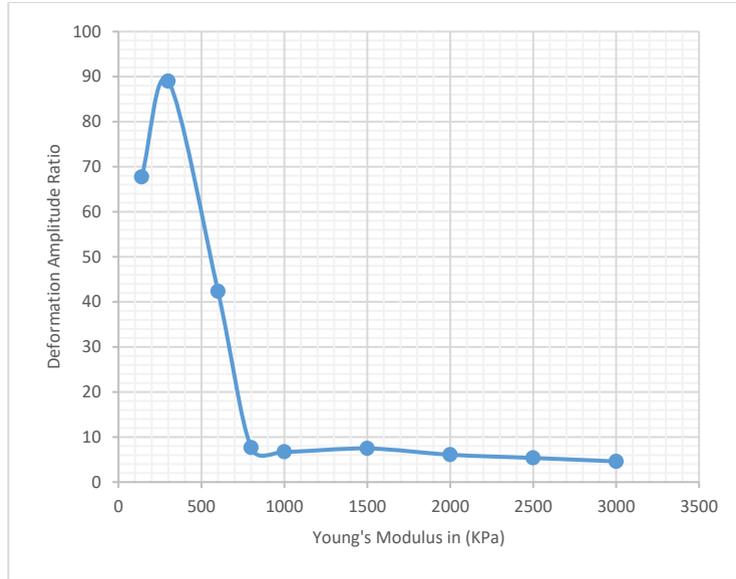

Fig. 5.16 Deformation amplitude ratio for each cornea model.

## 5.6. DISCUSSION

From Fig. 5.10. it is observed that corneal tissue experiences one complete cycle of wave propagation within 10 msec. at 140 KPa. The higher the elastic modulus for cornea tissue the higher the number of cycles of wave propagation. It is observed also that 140 KPa reaches its peak deformation at t3 when applying ARFI. The tissue starts by no deformation at the beginning of the simulation and deformation increases gradually till its peak value and then dampens again till reaching the no deformation state again. Many periods of ARFI application can be performed to obtain an average tissue behavior and more accurate results for biomechanical parameters.

TTP values for focal peak axial displacements shows a TTP value of 6.9 msec. at 140 KPa and 5.3 msec. at 3 MPa with time difference in TTPs between 140 KPa and 3 MPa of 1.6 msec. This narrow time difference makes it nearly impossible to estimate the corneal tissue biomechanics with ordinary ultrasound transducers operating with nearly several tens or even hundreds of frame rate. Yet, it is possible to achieve optimum temporal resolution with transducers operating with several thousand or even hundreds of thousands frame rate as in this study. In this study, 100 KHz frame rate gives about 160 frames for the 1.6 msec. time difference between TTPs of 140 KPa and 3 MPa, which is optimum to differentiate between them. TTP decrease as elasticity of tissue increases as shown in Fig. 5.12.

From Fig. 5.11. and Fig. 5.13. it is observed that force is dampened rapidly in temporal domain and spatial domain, where complete B-mode frame shows high localization of the force around the point of application, and deformation curves shows the rapid dampening in time for force after time of application.

From Fig. 5.15. We observed that deformation amplitudes decrease as the Young's modulus increase for corneal models. As elastic moduli shift to MPa range the difference between two consecutive deformation amplitudes becomes smaller. Differentiation between these deformation amplitudes is subjected to transducer axial resolution. Lateral resolution affects only the shear wave tracking process in the lateral direction, where the probing nodes are fixed with beam width. As beam width determines lateral resolution of the transducer, hence determining the lateral distance between the probing nodes. Smaller lateral distance yields worse temporal resolution between deformation curves of the two probing nodes of the wave speed, while larger lateral distance yields better temporal resolution.

We conclude from Fig. 5.14. that transducer's axial resolution is capable of differentiating between four of the major corneal layers.

Our estimation accuracy of SWS was maximum at 800 KPa with value of 99.8% and minimum at 2 MPa with value of 97.6% with respect to the theoretical SWS calculated from Eqn. (5.1).

As shown in Fig. 5.16. DA ratio at 2 mm represents how cornea apex deforms with respect to paracentral regions. It is observed that cornea models with low elastic modulus have high DA ratio values while DA ratio value decreases as the elastic modulus increases which is matching with experimental results obtained in relevant studies. DA ratio gives an objective information about how whole corneal tissue deforms in response to ARFI or any external force. This means that for corneal pre-refractive surgery; where its stiffness is considered to be high; DA ration at 2 mm is supposed to be small. While cornea post-refractive surgery is supposed to have high DA ratio at 2 mm as its stiffness is considered to be decreased by the surgery.

If we depend only on one of the estimated parameters for estimating the corneal biomechanics is not recommended. This is due to the misleading estimation as in the case of 300 KPa cornea model, where the estimated deformation amplitude ratio at 2mm value is supposed to be smaller than that of 140 KPa, while the estimated value is observed to be higher. Depending only on deformation amplitude ratio at 2 mm value supposes that the estimated corneal elastic modulus should be lower than that of 140 KPa. However, depending on all the estimated parameters values assessing corneal biomechanics leads to accurate estimation of corneal biomechanics and accurate assessment of post-refractive surgery.

## 5.7 SUMMARY

In this chapter, shear wave speed, deformation amplitude, Time-To-Peak deformation and deformation amplitude ratio for different elastic moduli cornea models are estimated respectively. Two models are used in parallel to study the behavior of cornea biomechanics pre- and post-refractive surgery, namely, FEM in conjunction with scatterer model. Third ultrasound transducer model is used to simulate the behavior of ultrasound transducer while imaging corneal tissue undergoing transient elastography. Also, the transducer model is used to simulate the transducer behavior while generating acoustic radiation force impulse that is used to excite corneal tissue. Nine FEMs are used to represent cornea in different biomechanical states pre- and post-refractive, 140 KPa, 300 KPa, 600 KPa and 800 KPa as post-refractive surgery corneal models and 1 MPa, 1.5 MPa, 2 MPa, 2.5 MPa and 3 MPa as pre-refractive surgery corneal models.

ARFI is applied transiently to each of the nine cornea models to induce deformation wave propagation. This wave of tissue deformation is tracked using B-mode imaging procedure that yields a video stack for each cornea model representing the wave propagation. The B-mode video stack including deformation wave is tracked by a 100 KHz frame rate. Resulting wave speed is tracked through two fixed probing points assigned on two fixed ultrasound beams, one point is at cornea apex, and the other is distal from cornea apex in either lateral direction. Number of probing points is subjected to the lateral resolution of the transducer, where involving more probing points inside the B-mode image frame means finer lateral resolution of the utilized transducer.

Focal TTP deformation values are estimated from TTP deformation curves for the focal apical probing point. It is observed that increasing corneal stiffness yields decrease in TTP deformation values.

Focal peak axial deformations are estimated from focal ultrasound beams by estimating corneal tissue axial location in both the reference frame and all peak deformation frames for each cornea model. Deformation amplitude values are observed to be decreasing by increase in corneal stiffness.

Similarly, paracentral axial deformations are estimated which enabled the estimation of deformation amplitude ration at 2 mm. Deformation amplitude ratios at 2 mm are shown to be decreasing with increasing the elastic modulus of the corneal tissue.

Shear wave speed values are estimated using the TTP deformation curves for both probing points. Lateral distance between the probing points divided by peaks arrival times gives an estimate to the corresponding shear wave speed of the cornea model under study. Elastic moduli can be estimated from the resulting shear wave speeds as they are related by Eqn. (4).

It is recommended to depend on each of the estimated parameters values of TTP deformation, deformation amplitude, deformation amplitude ratio at 2 mm and shear wave speed respectively when assessing cornea biomechanics post refractive surgery. Depending only on one of these values may be misleading and vague. As in the case of 300 KPa cornea model.

# Chapter 6 | Conclusion and Future Work

## 6.1. CONCLUSION

In this research, novel technique for estimating corneal biomechanics non-invasively in-situ is proposed and developed. This technique is investigated by using FEM simulations along with B-mode ultrasound simulations together. Ultrasonic methods are better than other methods in terms of feasibility, accuracy and it is non-destructive.

The new technique is developed over two major stages; the first stage is deriving an equivalent mathematical model to predict the corneal tissue deformation behavior in response to two different loading states, constant and transient state. Standard linear solid model is found to be optimum for describing the human corneal behavior in response to constant and transient load more efficiently. This model also describes the viscous damping behavior of the corneal tissue efficiently. This mathematical model concludes the optimum force values and the impact of the intra-ocular pressure over the deformation behavior of the corneal tissue.

The second major stage is to implement FEM simulation to predict the corneal tissue deformation as a 3D model, this enables the prediction of the entire human cornea behavior in response to a transient force. Eleven different elastic moduli FEMs are constructed to simulate human cornea behavior. FEM simulations are carried out for two major cases, the pre-refractive case and the post-refractive case, in order to reveal the impact of the refractive surgeries over the corneal biomechanics.

In terms of the ultrasound transducer modelling, the resulting deformation behavior from FEM is fed to scatterers model so as to simulate the ultrasound transducer behavior during B-mode imaging procedure. The transducer model is simulated to generate ARFI as well in order to induce tissue deformation.

Hence, the ultrasound transducer is modelled to simulate ARFI to induce corneal tissue deformation at first, and then the same ultrasound transducer is modelled to simulate the B-mode procedure of the resulting propagating deformation wave. The use of the FEM comes after generating the ARFI, where the generated ARFI map is applied to the generated 3D FEM corneal model to simulate its mechanical response to that ARFI. The resulting deformation maps are fed back again to the transducer model to simulate its equivalent ultrasound B-mode images.

In this research too, two corneal biomechanics estimation methods are developed and simulated. Both methods are depending on resulting deformation of the corneal tissue.

These two methods are, the FPAD method and the rSWS method respectively. The reason for developing these two methods is that we aim to make the corneal biomechanics estimation method applicable for a wide range of ultrasound devices in the market. The FPAD method has the advantage of estimating only qualitative corneal biomechanics properties using low market-available frequency ultrasound transducers. The rSWS method has the advantage of estimating complete quantitative corneal biomechanics properties maps using high frequency ultrasound transducers that are a bit expensive compared to the FPAD method. Hence, there is a tradeoff in estimation, for only qualitative estimation the FPAD method is considered to be the optimum solution, while for quantitative estimation the rSWS is considered to be the optimum method.

In terms of B-mode imaging frequency, two frame rates are investigated as well, in order to study the accuracy of estimation for the propagating deformation wave. These two frame rate are 10 KHz and 100 KHz respectively. The optimum frame rate for estimation at high-frequency ultrasound transducer is found to 100 KHz, where its accuracy curve is observed to be stable for all involved FEMs. At 10 KHz, the accuracy curve is observed to be fluctuating and not stable with involved FEMs which led to its exclusion.

Parameters for assessment of corneal biomechanics are utilized in this research as well. These parameters are; shear wave speed, deformation amplitude, Time-To-Peak deformation and deformation amplitude ratio respectively. It is recommended to depend on each of these parameters' values when assessing cornea biomechanics post refractive surgery.

## 6.2. STUDY LIMITATION AND FUTURE WORK

We would like to finish this research by highlighting the possible future directions that can enhance the search in this field while investigating them in more details. These directions are listed as follows:

a. Implementing the FPAD method on a conventional ultrasound probe to evaluate its efficiency in real-time deployment.
b. Implementing the rSWS method on a high-frequency ultrasound probe to evaluate its efficiency in real-time deployment.
c. Calculating the estimation accuracy and error in real deployment.
d. Investigating the thermal effect of ARFI on corneal tissue by a FEM to assess its side-effects.
e. Comparing the obtained results from this method with those obtained from ORA device in order to validate the proposed method.
f. Calculating the mechanical index experimentally and comparing it with FDA regulation value.